\documentclass[PAPER, atlasdraft=false,cernpreprint,UKenglish, texmf, orcidlogo]{atlasdoc}
\usepackage{atlaspackage}
\usepackage{atlasbiblatex}

\usepackage{atlasphysics}
\graphicspath{{logos/}{figures/}}

\usepackage{ANA-STDM-2020-07-PAPER-defs}

\AtlasTitle{Measurement of the $W$-boson angular coefficients and transverse momentum in $p\!p$ collisions at $\sqrt{s}=13$~TeV with the ATLAS detector}

\AtlasAbstract{%
The angular distributions of Drell--Yan lepton pairs provide sensitive probes of the underlying dynamics of quantum chromodynamics (QCD) effects in vector-boson production. This paper presents for the first time the measurement of the full set of angular coefficients together with the differential cross-section as a function of the transverse momentum of the $W$ boson, in the full phase space of the decay leptons. The measurements are performed separately for the $W^-$ and $W^+$ channels.
The analysis uses proton--proton collision data recorded by the ATLAS experiment at the Large Hadron Collider in 2017 and 2018, during special low-luminosity runs with a reduced number of interactions per bunch crossings (pile-up). The data correspond to an integrated luminosity of 338~pb$^{-1}$ at a centre-of-mass energy of $\sqrt{s} = 13$~TeV.
The low pile-up environment provides excellent experimental conditions for high-precision measurements of $W$-boson production. All results agree with theoretical predictions incorporating finite-order QCD corrections up to order $\alpha_S^2$.}
\AtlasRefCode{STDM-2020-07}

\PreprintIdNumber{CERN-EP-2025-188}

\AtlasJournal{EPJC}
\AtlasJournalRef{Eur. Phys. J. C 86 (2026) 483}
\AtlasDOI{10.1140/epjc/s10052-025-15264-w}

\AtlasCoverEgroupAnalysisTeam{atlas-stdm-2020-07-analysis-team@cern.ch}

\hypersetup{pdftitle={ATLAS document},pdfauthor={The ATLAS Collaboration}}

\begin{document}

\maketitle

\tableofcontents

%
%

%

\section{Introduction and motivation}

The angular distributions of lepton pairs produced in hadron collisions via the Drell--Yan process are governed by the spin correlation effects between the initial-state partons and the final-state leptons, mediated by a spin-1 intermediate state, such as in the case of the process $p_1 + p_2 \rightarrow \Wboson  \rightarrow \ell\nu$, where $p_1$ and $p_2$, are the incoming partons~\cite{Collins:1977prd,arXiv9406381}.
These spin correlations are described by a set of nine helicity density matrix elements, which can be calculated within the context of the parton model using perturbative quantum chromodynamics (QCD).  The full differential cross-section describing the kinematics of the two leptons arising from the decay of a \Wboson boson can be decomposed as a sum of harmonic polynomials, which depend on the lepton polar and azimuthal angles, denoted by $\theta$ and $\phi$, multiplied by the corresponding helicity-dependent cross-sections which depend on the \Wboson-boson transverse momentum, \ptw, invariant mass, \mw, and rapidity, \yw. It is a standard convention to factorise out the unpolarised cross-section, denoted by $\sigma^{U+L}$, and to present this differential cross-section as an expansion into the nine harmonic polynomials  %
and eight dimensionless angular coefficients $\Ai$~\cite{Mirkes:1992hu, Mirkes94PhysRevD, Mirkes94prep}, which represent ratios of helicity cross-sections relative to the unpolarised one, $\sigma^{U+L}$.

All eight \Zboson-boson angular coefficients were measured at a centre-of-mass energy of $\sqrt{s}=8~\TeV$ by the \atlas as a function of both the \Zboson-boson transverse momentum, \ptz, and rapidity, \yz, using \Zll events, where $\ell=\left\{e,\mu\right\}$~\cite{STDM-2014-10,STDM-2018-05}. Other measurements of the \Zboson-boson angular coefficients were also extracted as a function of \ptz from \Zmm events by the \lhcb~\cite{LHCb:2022tbc} and the \cms~\cite{CMS-SMP-13-010}, resulting in the extraction of four and five of the eight coefficients, respectively. Prior to these measurements at the Large Hadron Collider (LHC)~\cite{Evans:2008zzb}, four of the eight coefficients as a function of \ptz were measured by the \cdf at the Tevatron from \Zee events~\cite{CDF:ZAi}.

Direct measurements of only two of these angular coefficients for the \Wboson boson have previously been performed. The $A_2$ and $A_3$ angular coefficients were extracted as a function of \ptw by the \cdf from $W \rightarrow \ell\nu$ events, where $\ell=\left\{e,\mu\right\}$~\cite{CDF:WAi}. However, other measurements of the \Wboson-boson polarisation were previously done by the \atlas~\cite{STDM-2011-37} and by the \cms~\cite{CMS-EWK-10-014,CMS-SMP-18-012}. These polarisation fraction measurements are related to the $A_0$ and $A_4$ coefficients and so can be used to probe their values.

The angular coefficients of the \Wboson boson are related to the nature of electroweak processes, the \Wboson-boson polarisation, and the presence of QCD effects. They are important for modelling QCD effects in the production of \Wboson bosons and have a direct impact on the precision of other measurements, such as the determination of the \Wboson-boson mass~\cite{Pellen_2022}. Currently, the validation of the modelling of these coefficients for the \Wboson-mass measurements at the LHC relies on extrapolations from the predictions and measurements of the angular coefficients of the \Zboson boson, thereby contributing to larger uncertainties~\cite{STDM-2019-24,LHCb:2021bjt,CMS-SMP-23-002}.
Precise determinations of the \Wboson-boson transverse momentum spectrum and its angular coefficients offer direct experimental insights into the proton structure and parton density and consequently to the production dynamics of the \Wboson and enable stringent comparisons with state-of-the-art QCD predictions. This model-independent approach exploits the spin-one nature of the gauge boson and the spin-1/2 nature of the decay leptons, constrained by angular momentum conservation. This method bypasses the need for detailed modelling of the gauge-boson polarisation and decay, and instead isolates the production dynamics. Consequently, it mitigates theory uncertainties associated with spin correlations and the resummation of fiducial power corrections, which can be sizeable~\cite{Ebert:2020dfc,Glazov:2020gza,Salam:2021tbm,Alekhin:2021xcu}.

This paper presents, for the first time, simultaneous measurements of the full set of \Wboson-boson angular coefficients \Ai\ and the differential cross-section \dsdpT\ as functions of \ptw, performed in the full phase space of the decay leptons using $\Wpm \rightarrow \ell^{\pm}\nu$ events, where $\ell=\left\{e,\mu\right\}$.
This represents a complementary approach to  previous precise measurements of \Wboson-boson differential cross-section~\cite{STDM-2018-17,STDM-2018-26},  which were limited to the fiducial phase space.
The measurements are based on dedicated proton-proton ($pp$) collision data samples collected by the ATLAS experiment at the LHC with low instantaneous luminosity at a centre-of-mass energy of $\sqrt{s} = 13~\TeV$, corresponding to 338\,pb$^{-1}$ of integrated luminosity. The mean number of additional $pp$ interactions per bunch crossing (pile-up events) in the dataset is approximately two. This low pile-up dataset provides excellent experimental conditions for high-precision measurements of \Wboson-boson production.

While this analysis is based on the same dataset and systematic uncertainty evaluations as in Refs.~\cite{STDM-2018-17,STDM-2018-26}, the use of differential angular distributions introduces additional sensitivity to certain experimental systematic effects. Notably, these angular observables are not directly involved in the detector calibration procedures, thereby providing complementary and independent constraints.

The structure of the paper is as follows: Section~\ref{sec:detector} details the ATLAS detector and Section~\ref{sec:procedure} describes the methodology used to extract the angular coefficients and the differential cross-section in the full phase space of the decay leptons. Section~\ref{sec:analysis} outlines the experimental set-up, including the dataset, event selection, and background estimate. The impact of systematic uncertainties in the final results is discussed in Section~\ref{sec:Systematic_uncertainties}. Finally, the results are presented in Section~\ref{sec:Results}.


%

\section{ATLAS detector}
\label{sec:detector}

\newcommand{\AtlasCoordFootnote}{%
ATLAS uses a right-handed coordinate system with its origin at the nominal interaction point (IP)
in the centre of the detector and the \(z\)-axis along the beam pipe.
The \(x\)-axis points from the IP to the centre of the LHC ring,
and the \(y\)-axis points upwards.
Polar coordinates \((r,\phi)\) are used in the transverse plane,
\(\phi\) being the azimuthal angle around the \(z\)-axis.
The pseudorapidity is defined in terms of the polar angle \(\theta\) as \(\eta = -\ln \tan(\theta/2)\) and is equal to the rapidity
$ y = \frac{1}{2} \ln \left( \frac{E + p_z}{E - p_z} \right) $ in the relativistic limit.
Angular distance is measured in units of \(\Delta R \equiv \sqrt{(\Delta y)^{2} + (\Delta\phi)^{2}}\).}

The ATLAS experiment~\cite{PERF-2007-01} at the LHC is a multipurpose particle detector
with a forward--backward symmetric cylindrical geometry and a near \(4\pi\) coverage in
solid angle.\footnote{\AtlasCoordFootnote}
It consists of an inner tracking detector surrounded by a thin superconducting solenoid
providing a \qty{2}{\tesla} axial magnetic field, electromagnetic and hadronic calorimeters, and a muon spectrometer.
The inner tracking detector covers the pseudorapidity range \(|\eta| < 2.5\).
It consists of silicon pixel, silicon microstrip, and transition radiation tracking detectors.
Lead/liquid-argon (LAr) sampling calorimeters provide electromagnetic (EM) energy measurements
with high granularity within the region \(|\eta|< 3.2\).
A steel/scintillator-tile hadronic calorimeter covers the central pseudorapidity range (\(|\eta| < 1.7\)).
The endcap and forward regions are instrumented with LAr calorimeters
for EM and hadronic energy measurements up to \(|\eta| = 4.9\).
The muon spectrometer surrounds the calorimeters and is based on
three large superconducting air-core toroidal magnets with eight coils each.
The field integral of the toroids ranges between \num{2.0} and \qty{6.0}{\tesla\metre}
across most of the detector.
The muon spectrometer includes a system of precision tracking chambers up to \(|\eta| = 2.7\) and fast detectors for
triggering up to \(|\eta| = 2.4\).
The luminosity is measured mainly by the LUCID--2~\cite{LUCID2} detector, which is located close to the beampipe.
A two-level trigger system is used to select events~\cite{TRIG-2016-01}.
The first-level trigger is implemented in hardware and uses a subset of the detector information
to accept events at a rate close to \qty{100}{\kHz}.
This is followed by a software-based trigger that
reduces the accepted rate of complete events to \qty{1.25}{\kHz} on average
depending on the data-taking conditions.
A software suite~\cite{SOFT-2022-02}  is used in data simulation, in the reconstruction
and analysis of real and simulated data, in detector operations, and in the trigger and data acquisition
systems of the experiment.


%

\section{Measurement methodology}
\label{sec:procedure}
This methodology used for this analysis relies on the same formalism already developed and published for the ATLAS extraction of the \Zboson-boson angular coefficients and full phase-space cross-section~\cite{STDM-2018-05}.
The full five-dimensional differential cross-section describing the kinematics of the two Born-level leptons from the \Wboson-boson decay  can be decomposed as:
\begin{align}
\label{eq:master2}
\frac{ \mathrm{d}\sigma}{\mathrm{d}\ptw~\mathrm{d}\yw~\mathrm{d}\mw~\mathrm{d}\cos\theta~\mathrm{d}\phi} =&
\frac{3}{ 16 \pi} \frac{\mathrm{d}\sigma ^{ U+ L }}{\mathrm{d}\ptw~\mathrm{d}\yw~\mathrm{d}\mw}  \nonumber \\
\bigg\{&(1 + \cos^2\theta) + \frac{1}{2}\ A_0 (1 - 3 \cos^2\theta) + A_1 \ \sin 2 \theta\ \cos\phi \\
+& \frac{1}{2}\ A_2 \ \sin^2\theta\ \cos 2 \phi +  A_3 \ \sin \theta\ \cos \phi +     A_4 \ \cos\theta  \nonumber \\
+&     A_5 \ \sin^2\theta \ \sin 2 \phi +     A_6 \ \sin 2 \theta\ \sin \phi +     A_7 \ \sin \theta \ \sin \phi \bigg\}\nonumber.
\end{align}
The dependence of the differential cross-section on the lepton angular $\cos\theta$ and $\phi$ distributions in the Collins-Soper (CS) frame~\cite{Collins:1977prd} is analytical and fully contained in the nine harmonic polynomials~$P_i(\cos \theta, \phi)$ given in the equation. On the other hand, the dependence on \ptw, \yw\ and \mw is entirely contained in the unpolarised cross-section, $\sigma^{U+L}$, and in the \Arangesub~angular coefficients. Therefore, all hadronic dynamics from the production mechanism are factorised from the decay kinematics in the \Wboson-boson rest frame.

If the \Wboson boson is produced with no initial-state radiation, it is polarised along the beam axis, due to the vector minus axial vector, V-A, nature of the weak interaction and helicity conservation. In that case, $A_4$ is the only non-zero coefficient.   If the \Wboson boson is produced with non-negligible transverse momentum balanced by the associated production of one or more
jets, the $A_{0-3}$ angular coefficients become non-zero and hence the cross-section depends on the azimuthal angle, $\phi$, as well.
The remaining three angular coefficients, $A_{5-7}$, are non-zero only if gluon loops of $\mathcal{O}(\alpha_s^2)$ are present in the production of the \Wboson boson.

The Lam-Tung relation~\cite{LamTung:1978,LamTung:1979,LamTung:1980}, which predicts that $A_0 - A_2 = 0$ is a consequence of the spin-1 nature of the gluon, is expected to hold up to $\mathcal{O}(\alpha_s)$, but can be violated at higher orders. Such violations were experimentally confirmed in \Zboson-boson events~\cite{STDM-2014-10}, while this has yet to be confirmed in \Wboson-boson events.

The \Wboson-boson kinematic dependence of the coefficients varies strongly with the choice of the spin quantisation axis in
the \Wboson-boson rest frame ($z$-axis). In the CS reference frame adopted for this analysis, the $z$-axis is defined in
the \Wboson-boson rest frame as the external bisector of the angle between the momenta of the two protons, as is
described in more detail in Ref.~\cite{STDM-2014-10}. The positive direction of the $z$-axis is defined by the direction of positive longitudinal \Wboson-boson momentum in the laboratory frame.
At detector level, the directions of the leptons in the CS frame are defined by two angles, \thetacs and \phics. The angle \thetacs is defined as the angle between the momentum of the negatively charged lepton and the $z$-axis. In the case of \Wplus, there is no negatively charged lepton and instead the direction of the neutrino is used~\cite{Mirkes:1992hu}.

For the leptonic decay modes, \Wboson-bosons have the disadvantage relative to \Zboson-bosons that the \lnu decay
kinematics cannot be completely reconstructed due to the  inability to fully reconstruct the longitudinal
momentum of the neutrino in a hadron collider.
The neutrino transverse components $p_{x}^{\nu}$ and $p_{y}^{\nu}$ are determined from measurable laboratory-frame quantities, while the longitudinal component  \pzn  is determined only up to a sign. This twofold sign ambiguity in the reconstruction of the neutrino momentum in the CS frame translates into an ambiguity on the sign of \coscs, while \phics is determined. This effect dilutes the sensitivity of the measurement to the coefficients that directly depend on \coscs namely \Azero and  \Afour .

\subsection{Inferring the neutrino \pzn information}
\label{sec:neutrino}

\begin{figure}[!h]
\centering
\subfloat[]{\includegraphics[width=0.45\textwidth]{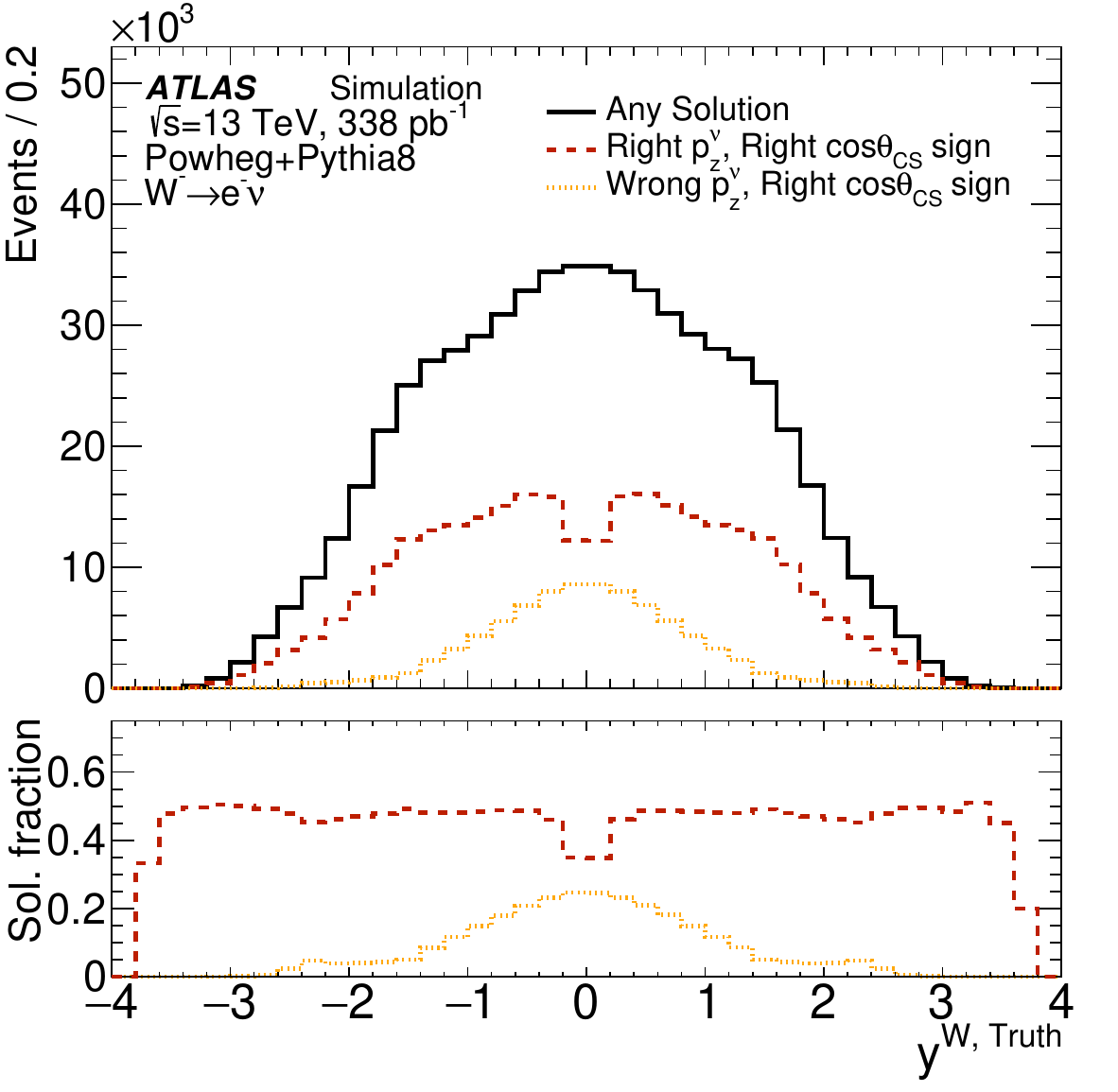}}
\subfloat[]{\includegraphics[width=0.45\textwidth]{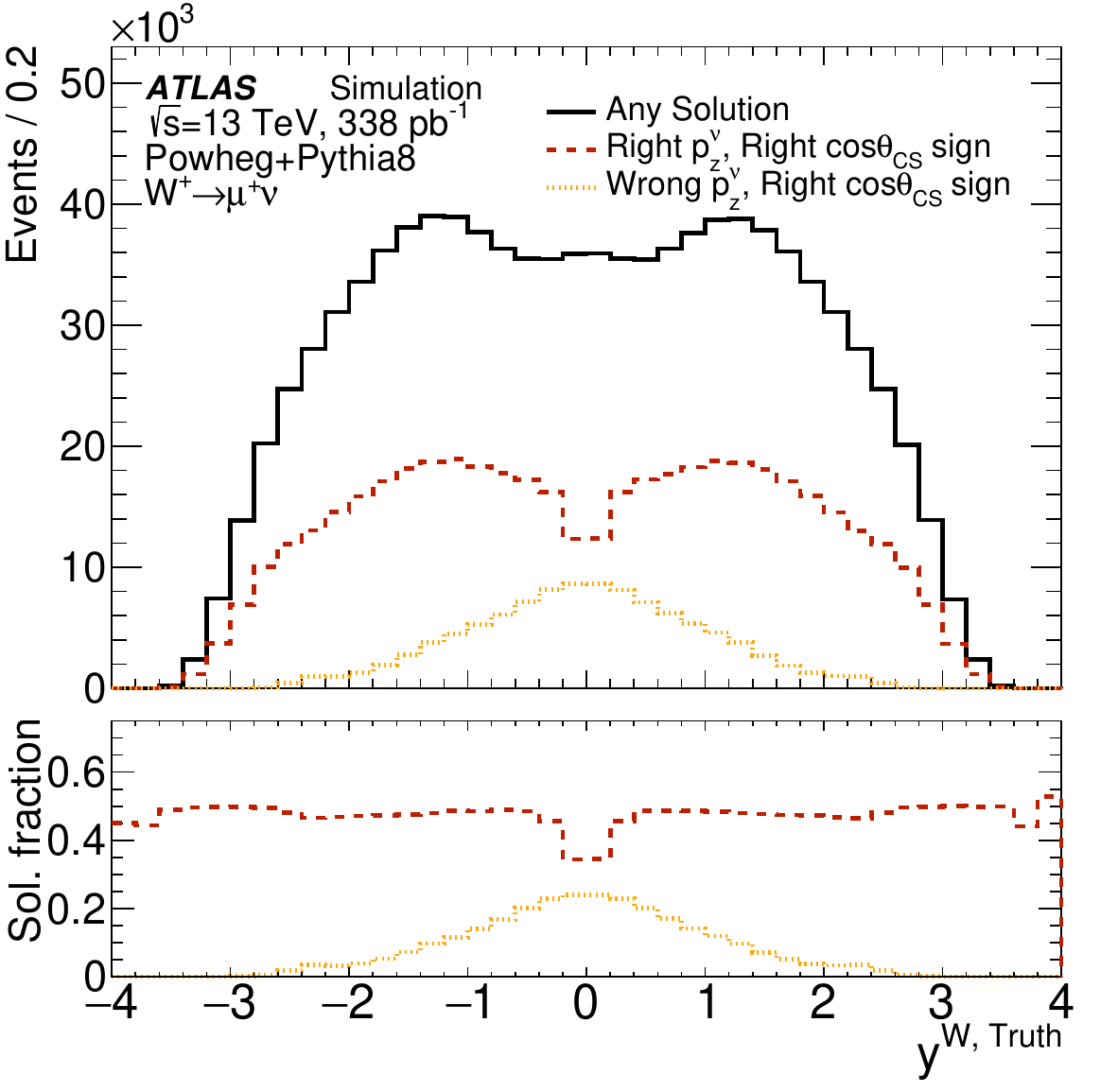}}
\caption{ \Wboson-boson rapidity distributions for events reconstructed with the correct or the wrong sign of \(\coscs\) for the  (a) \Wmenu and the (b) \Wpmunu channels.
Events with the chosen solution closest to the true \pzn are labelled as \emph{right \pzn}.
Events with an unambiguous determination of \(\coscs\) are labelled as \emph{right \coscs\ sign}.
The differences between the electron and muon distributions reflect both the different efficiencies for reconstructing and selecting the two lepton flavours, and the intrinsic kinematic differences between $W^-$ and $W^+$-boson events. The ratio panel shows the fraction of events with an unambiguous  determination of \(\coscs\) relative to all events with a real-valued solution for \pzn.	 }
\label{fig:pz}
\end{figure}

The unmeasured longitudinal momentum of the neutrino is inferred by assuming that the dilepton system from the Drell--Yan  process has an invariant mass equal to the measured mass of the \Wboson boson~\cite{Richter_Was_2017}. In the laboratory frame, both the charged lepton and the neutrino are treated as massless particles. By imposing the \mW constraint and expressing the resulting condition in terms of the neutrino's unknown longitudinal component, a quadratic equation is obtained. Solving this equation yields two possible real solutions for the neutrino's longitudinal momentum, \pzn.
For this analysis,  one of the two solutions for the neutrino longitudinal momentum is chosen randomly, with equal probability.
As an illustration, the resulting rapidity distributions obtained with this random choice are shown in Figure~\ref{fig:pz} for the \Wmenu and \Wpmunu channels.%

As mentioned already in Section~\ref{sec:procedure}, the \(z\)-axis of the CS frame is defined along the direction of the \Wboson-boson's longitudinal momentum,
leading to the transformation \(\coscs \to \coscs \times \mathrm{sign}(y^{\ell\nu})\), where  \(y^{\ell\nu}\) is the \Wboson-boson rapidity. Under this convention, events for which the two real \pzn solutions yield opposite signs for \(y^{\ell\nu}\) correspond to cases where the ambiguity in the sign of \(\coscs\) cancels, resulting in an unambiguous determination of \(\coscs\). Nevertheless, an ambiguity in the sign of the \Wboson-boson rapidity remains. These events, which predominantly occur at central rapidities (see Figure~\ref{fig:pz}), allow for the resolution of the \(\coscs\) sign ambiguity in an additional 25\% of the dataset, enabling a direct measurement of the forward--backward asymmetry, associated with the \(A_4\) coefficient, in \Wboson-boson production.%

Events for which the quadratic equation yields no real solution are still included in the analysis. In these cases, the reconstructed neutrino kinematics are adjusted by applying a momentum imbalance constraint, which modifies the input to ensure a unique, real-valued solution. This approach follows a methodology similar to that described in Ref.~\cite{BETCHART2014169}.
The fraction of events with no real solution increases with the reconstructed transverse momentum of the \Wboson boson, \ptln. It is approximately 15\% at low \ptln and rises to around 60\% for \ptln values exceeding 250~\GeV. This behaviour is primarily driven by the limited resolution in the reconstruction of the neutrino transverse momentum at detector level, which degrades with increasing \ptln.

\subsection{Extraction of angular coefficients and differential cross-sections }
\label{sec:methodology}

The angular coefficients, \Ai, and the differential cross-sections as a function of \ptw, \dsdpT, are extracted from the data by performing a template fit to the polynomial terms $P_i(\cos \theta, \phi)$ defined in Eq.~\ref{eq:master2} to the two-dimensional reconstructed angular distributions, using a binning of $10 \times 8$ in the (\coscs, \phics) plane.
This procedure is performed separately for \Wminus and \Wplus candidates.
The particle-level distributions are related to the reconstructed angular distributions using a folding procedure, which accounts for detector effects and bin migrations in the (\coscs, \phics) plane.
Compared to the methodology employed in Refs.~\cite{STDM-2014-10,STDM-2018-05}, the binning in \coscs is modified to account for events without a real solution for $\pzn$. Events with a real solution, as described in Section~\ref{sec:neutrino}, are assigned to one of eight same-size \coscs bins, distributed over the range $[-1, 1]$, while events without a real solution are included as overflow in the tenth bin in \coscs. The ninth \coscs bin serves as an empty separator. This approach ensures that events lacking a real $\pzn$ solution contribute to the fit, thereby preserving sensitivity to the \phics  distribution and avoiding distortions in the  \coscs\ shape.

Each template is normalised by a free parameter for its corresponding polynomial coefficient, \Arangesub, and by a common parameter representing the unpolarised cross-section, $\sigma^{U+L}$. The polynomial $1+\cos^2\theta$ in Eq.~\ref{eq:master2} is normalised only by the parameter for the unpolarised cross-section. All of the angular coefficients together with the corresponding unpolarised cross-section parameters are measured in each of the analysis bins in the \ptln space.
The binning in \ptln is chosen based on the experimental resolution at low \ptln and on the number of events at
high \ptln, and has the following boundaries:

\begin{equation}
\begin{aligned}
\label{eq:binning}
\ptln &= [0, 8, 17, 27, 40, 55, 75, 110, 150, 210, 600]~\GeV. \\
\end{aligned}
\end{equation}

In the same way as described in Ref.~\cite{STDM-2018-05}, a likelihood is built as a product of Poisson probabilities across all the bins in the reconstructed \coscs, \phics, and \ptln distributions and of auxiliary constraints for each nuisance parameter.
A profile likelihood ratio method is used to extract the best fit values of the parameters of interest~(POIs) and their uncertainties.
The POIs include the angular coefficients, \Aij, and the cross-sections, \sigmaj$^{U+L}$, in each $j$-th measurement bin defined in Eq~\ref{eq:binning}.
%


%

\section{Analysis}
\label{sec:analysis}

The analysis is performed with leptonically decaying \Wboson-boson events, $W \rightarrow \ell\nu$, where here the charged lepton refers to electrons and muons, $\ell = {e,\mu}$. The analysis is carried out separately for positively and negatively charged leptons.
The hadronic recoil, $\vec{u}_\mathrm{T}$, reconstructed in the plane transverse to the beam using charged-particle tracks and energy deposits is used to infer the transverse momentum of the decay neutrino, \ptn.
The extraction of the angular coefficients and the differential cross-section as a function of the \Wboson-boson transverse momentum are first performed separately for the electron and muon channels. After confirming the compatibility of the results, the electron and muon channels are combined, taking into account correlations between systematic uncertainties.

In this analysis, the full set of angular coefficients is measured, along with the difference  $A_0 - A_2$.
Given the level of statistical precision provided by the data, only the coefficients $A_0$, $A_2$, $A_3$, and $A_4$ are measured to be significantly different from zero. The sensitivity to $A_1$, the higher-order coefficients $\Ahighrng$, and the difference $A_0 - A_2$ is limited. Nevertheless, these coefficients are included in the likelihood fit, and the results are reported.

\subsection{Data and simulated-event samples}
\label{sec:samples}

The data were collected by the ATLAS detector in 2017 and 2018 at a centre-of-mass energy of $\sqrt{s} =13$\,TeV, and correspond to a total integrated luminosity of $338.1 \pm 3.1$\,pb$^{-1}$~\cite{DAPR-2021-01}. These data were collected during dedicated LHC low pile-up runs, where the average number of additional $pp$ interactions was approximately two, which is about a factor 20 smaller than that of the nominal LHC Run~2 13\,TeV operation between 2015 and 2018. To fully exploit these low pile-up conditions, the thresholds applied to suppress noise in the reconstruction of clusters of energy in the calorimeters were lowered, thereby optimising the reconstruction of the hadronic recoil, as is discussed in Ref.~\cite{STDM-2018-17}.

Samples of Monte Carlo (MC) simulated events are used to model the signal and all background processes except for QCD multijet production. All of the samples were processed with the $\GEANT$-based simulation~\cite{Agostinelli:2002hh} of the ATLAS detector~\cite{SOFT-2010-01} using settings specific to the low pile-up run conditions. The effects of pile-up collisions in the same or neighbouring bunch crossings were included in the MC simulation by overlaying inelastic $pp$ interactions produced using
\PYTHIAV{8}~\cite{Sjostrand:2007gs}  with the NNPDF2.3LO set of parton distribution functions (PDFs)~\cite{Ball:2012cx} and the A3 set of parameters (tune)~\cite{ATL-PHYS-PUB-2016-017}.

The main signal event samples for \Wboson-boson and for background \Zboson-boson production were generated
using the \Powheg\ event generator at next-to-leading-order (NLO) in QCD~\cite{Nason:2004rx,Frixione:2007vw,Alioli:2008gx,Alioli:2010xd}
using the CT10 PDFs~\cite{Lai:2010vv}, interfaced to \PYTHIAV{8}~\cite{Sj_strand_2015} using the
AZNLO tune~\cite{STDM-2012-23}. These \Powheg+\PYTHIAV{8} samples were interfaced
to \Photos~\cite{Golonka:2005pn} to simulate the effect of final-state quantum electrodynamics
(QED) radiation. The effective sample size is typically $10$--$20$ times larger than
the data, to minimise the impact of MC statistical uncertainties.

Alternative signal samples were prepared with \Sherpa\ 2.2.1~\cite{Bothmann:2019yzt} using the NNPDF3.0 next-to-next-to-leading-order (NNLO) PDFs~\cite{Ball:2014uwa} and merging matrix element calculations from \textsc{Comix}~\cite{Gleisberg:2008fv} and \OPENLOOPS~\cite{Buccioni:2019sur,Cascioli:2011va,Denner:2016kdg} for vector-boson $V+0,1,2$ partons at NLO accuracy with $V+3,4$ partons at leading-order (LO) accuracy in the MEPS@NLO scheme~\cite{Schumann:2007mg,Hoeche:2011fd,Hoeche:2012yf,Catani:2001cc,Hoeche:2009rj}. These samples are used to evaluate the uncertainty arising from the choice of the model used to simulate the signal events.

The \Wboson-boson samples were normalised to NNLO calculations using the \textsc{DYTurbo}~\cite{Camarda:2019zyx,Camarda:2021ict,Camarda:2021jsw,Camarda:2023dqn} program with the MMHT2014 NNLO PDF set~\cite{Harland-Lang:2014zoa}. A total uncertainty of $\pm 5\%$ was taken on the predictions normalised using these cross-sections, and accounts for uncertainties in the choice of PDF set and corresponding variations ($3-4\%$), as well as variations in the value of the strong coupling constant $\alpha_S$ ($1-2\%$), and missing higher order corrections (1\%).

Backgrounds arising from top-antitop-quark-pair (\ttbar) production as well as single-top-quark production ($Wt$ associated production, $t$-channel and $s$-channel) were generated with \Powheg+\PYTHIAV{8}~\cite{ATL-PHYS-PUB-2020-023} and normalised to the NNLO predictions with resummation at next-to-next-to-leading-logarithmic (NNLL) accuracy~\cite{Beneke:2011mq,Cacciari:2011hy,Baernreuther:2012ws,Czakon:2012zr,Czakon:2012pz,Czakon:2013goa,Czakon:2011xx}.
Diboson production $VV (V=W,Z)$ was generated with \Sherpa\ in all decay channels with at least one real lepton in the final state and is considered as background in this analysis~\cite{ATL-PHYS-PUB-2017-005}.

In addition, two auxiliary MC samples are used to model backgrounds from non-prompt leptons. The first, generated with \PYTHIAV{8} and \textsc{EvtGen}1.2.0~\cite{evtgen}, simulates $b$- and $c$-hadron decays. The second models backgrounds to prompt photon and electron production and includes all tree-level $2 \to 2$ QCD processes, top-quark pair, and weak vector-boson production, generated with \PYTHIAV{8} using the A14 tune~\cite{Skands:2014pea} and the NNPDF2.3LO PDF set~\cite{Ball:2012cx}.

\subsection{Analysis overview and event selection}

The data were collected with triggers that require at least one electron or one muon with transverse momentum $\pt^e > 15\,\gev$ or $\pt^\mu > 14\,\gev$, respectively~\cite{TRIG-2016-01, TRIG-2018-05,TRIG-2018-01}. Loose electron and muon identification criteria are applied at the trigger level. Events are selected if they include exactly one electron or muon candidate matched to a corresponding trigger-lepton candidate.

Events are also required to have at least one reconstructed collision vertex with two or more charged-particle tracks~\cite{ATL-PHYS-PUB-2015-026}. The vertex with the largest sum of squared transverse momenta of its associated tracks is taken as the primary vertex.

Electron candidates are reconstructed from clusters of energy deposited in the electromagnetic calorimeter and associated with at
least one track in the inner detector~\cite{EGAM-2018-01}. Electron candidates are required to fall within the coverage of the inner detector and the precision region of the EM calorimeter, namely  $|\eta| < 2.47$. Electrons in the transition region between the barrel and endcap calorimeters, $1.37<|\eta|<1.52$, are excluded. Electron candidates are required to have $p_\mathrm{T} > 25$\,GeV and satisfy the \textit{Tight} likelihood identification requirements~\cite{EGAM-2018-01}.

Muon reconstruction is first performed independently in the inner detector and in the muon spectrometer. A muon candidate is then  formed using the combined information from both detectors~\cite{MUON-2018-03}. The muon candidates are required to have an absolute pseudorapidity of $|\eta| < 2.4$, a transverse momentum of $p_\mathrm{T} > 25$\,GeV, and to satisfy the \textit{Medium} identification criteria~\cite{MUON-2018-03}.

Both electron and muon candidates are also required to be isolated from nearby activity, as measured by tracks in a cone of size $\Delta R < 0.2$ around the candidate lepton. The scalar sum of the $p_\mathrm{T}$ of these tracks, $p_\mathrm{T}^\mathrm{cone20}$, is required to be less than $10\%$ of the lepton $p_\mathrm{T}$ and may not exceed 5\,GeV for leptons with $p_\mathrm{T}  > 50$\,GeV, i.e. $p_\mathrm{T}^\mathrm{cone20}/\min(p_\mathrm{T}, 50\,\mathrm{GeV}) < 0.1$.

Dedicated lepton calibrations and efficiency corrections are applied to the reconstructed electron and muon candidates as described in Ref.~\cite{STDM-2018-17}.

The charged-lepton candidates are required to originate from the primary vertex. The track transverse impact parameter significance, $|d_0/\sigma_{d_{0}}|$, calculated relative to the beam line, must be smaller than three for muons and five for electrons. Furthermore, the longitudinal impact parameter, $z_0$, which is the difference between the $z$-coordinate of the point on the track at which $d_0$ is defined and the longitudinal position of the primary vertex, is required to satisfy $|z_0 \sin\theta| <$ 0.5~mm.

The missing transverse momentum, $\vec{p}_\mathrm{T}^\mathrm{miss}$, with its magnitude \met,  represents a measure of the transverse momentum of the neutrino.  It is defined as $\vec{p}_\mathrm{T}^\mathrm{miss} = - \left(\vec{p}_\mathrm{T} +\vec{u}_\mathrm{T}\right)$ using the lepton $\vec{p}_\mathrm{T}$ and hadronic recoil $\vec{u}_\mathrm{T}$.
The hadronic recoil is reconstructed in the plane transverse to the beam using particle-flow objects  (PFOs), that combine information from charged-particle tracks in the inner detector and energy deposits in the calorimeter~\cite{PERF-2015-09,JETM-2018-05}.
The hadronic recoil is calibrated using $Z\to \ell\ell$ events by comparing $u_\mathrm{T}$ with $p_\mathrm{T}^{\ell\ell}$, the transverse momentum of the dilepton system~\cite{STDM-2018-17}.  The low pile-up datasets used in this analysis significantly improve the resolution of the recoil measurement and therefore the overall resolution of the detector-level measurement of the neutrino-related observables and the reconstructed \ptw, referred to in the following as \ptln.

The background from QCD multijet events is reduced by the requirement $\met > 25$\,GeV. Furthermore, the \Wboson-boson transverse mass $m_\mathrm{T} = \sqrt{2 p_\mathrm{T}  \met  (1-\cos\Delta\phi_{\ell\nu})}$ is required to exceed 50\,GeV, where $\Delta\phi_{\ell\nu}$ is the azimuthal angle between the lepton $\vec{p}_\mathrm{T}$ and $\vec{p}_\mathrm{T}^\mathrm{miss}$.

After all selections, the numbers of $W\to e\nu$ and $W\to \mu\nu$ candidates used in this analysis are approximately $1.95\times 10^{6}$ and $2.21\times 10^{6}$, respectively.

In each of the analysis \ptln bins defined in Eq.~\ref{eq:binning}, two-dimensional (\coscs, \phics) angular distributions are measured separately for positively and negatively charged leptons. Those two-dimensional distributions serve as the basis for the simultaneous extraction of all the angular coefficients and of the differential cross-sections, following the methodology described in~Section~\ref{sec:methodology}.

%

\subsection{Background estimates}
\label{sec:background}

\begin{figure}[tbp]
\centering
\subfloat[]{\includegraphics[width=0.45\linewidth]{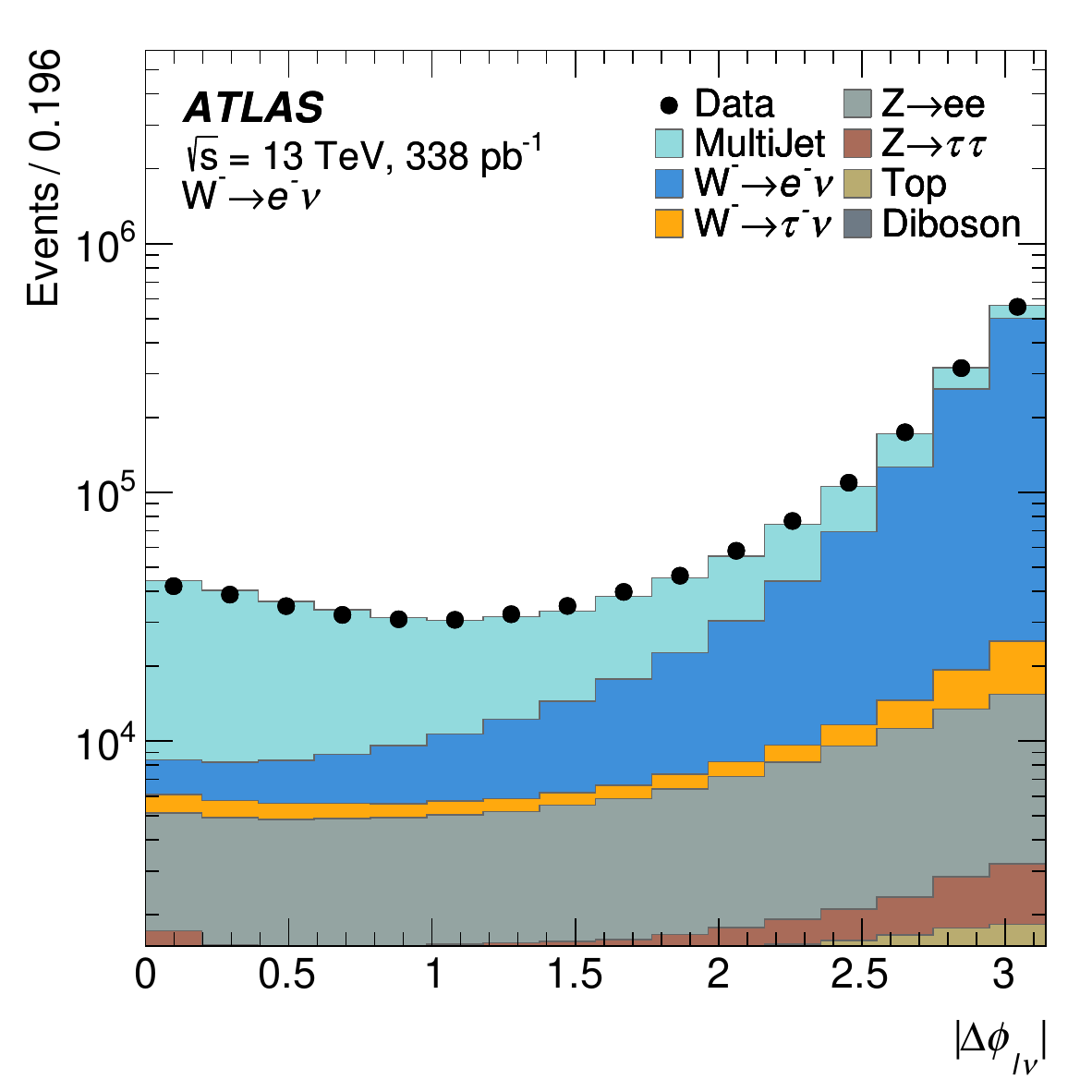}}
\subfloat[]{ \includegraphics[width=0.45\linewidth]{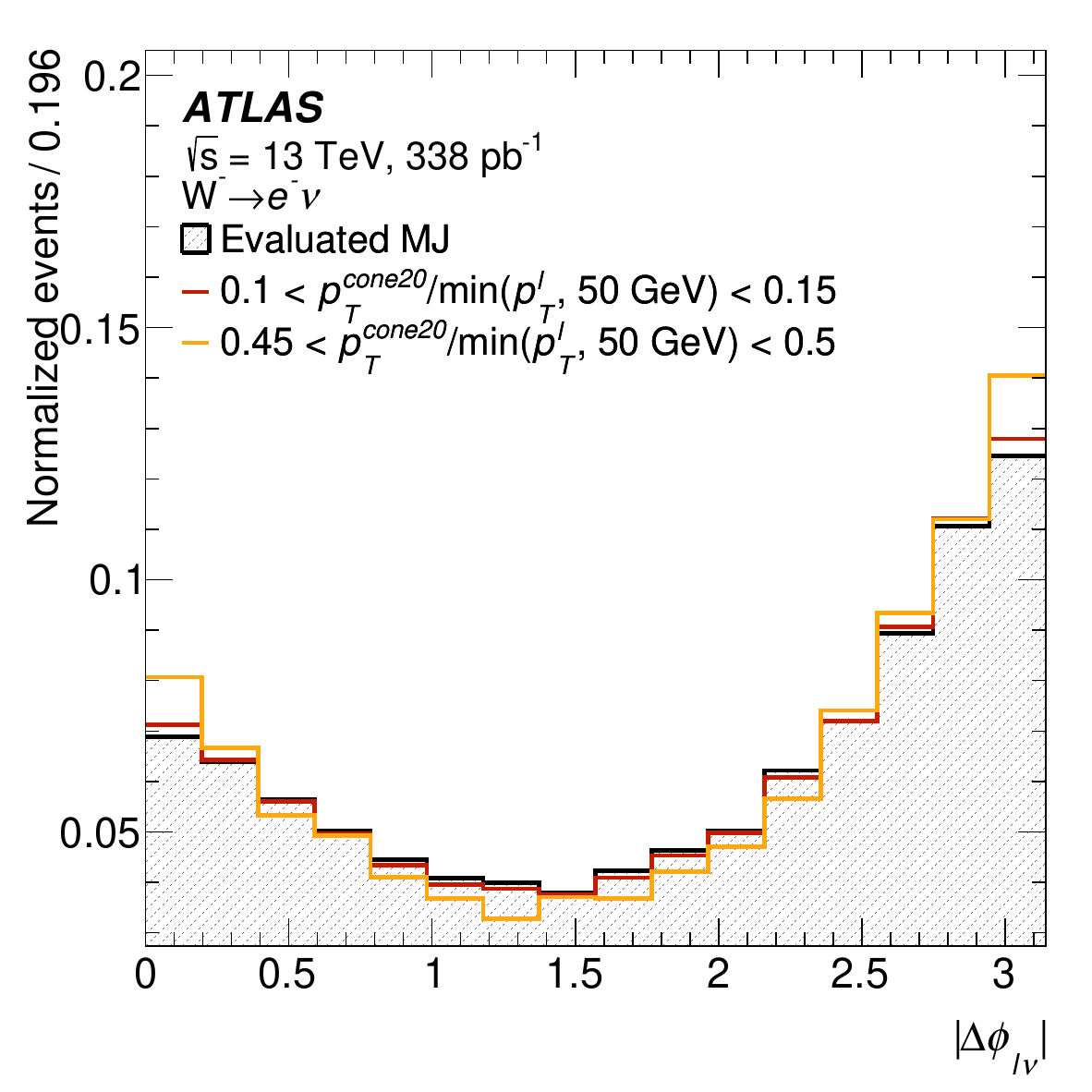}}
\\
\subfloat[]{\includegraphics[width=0.45\linewidth]{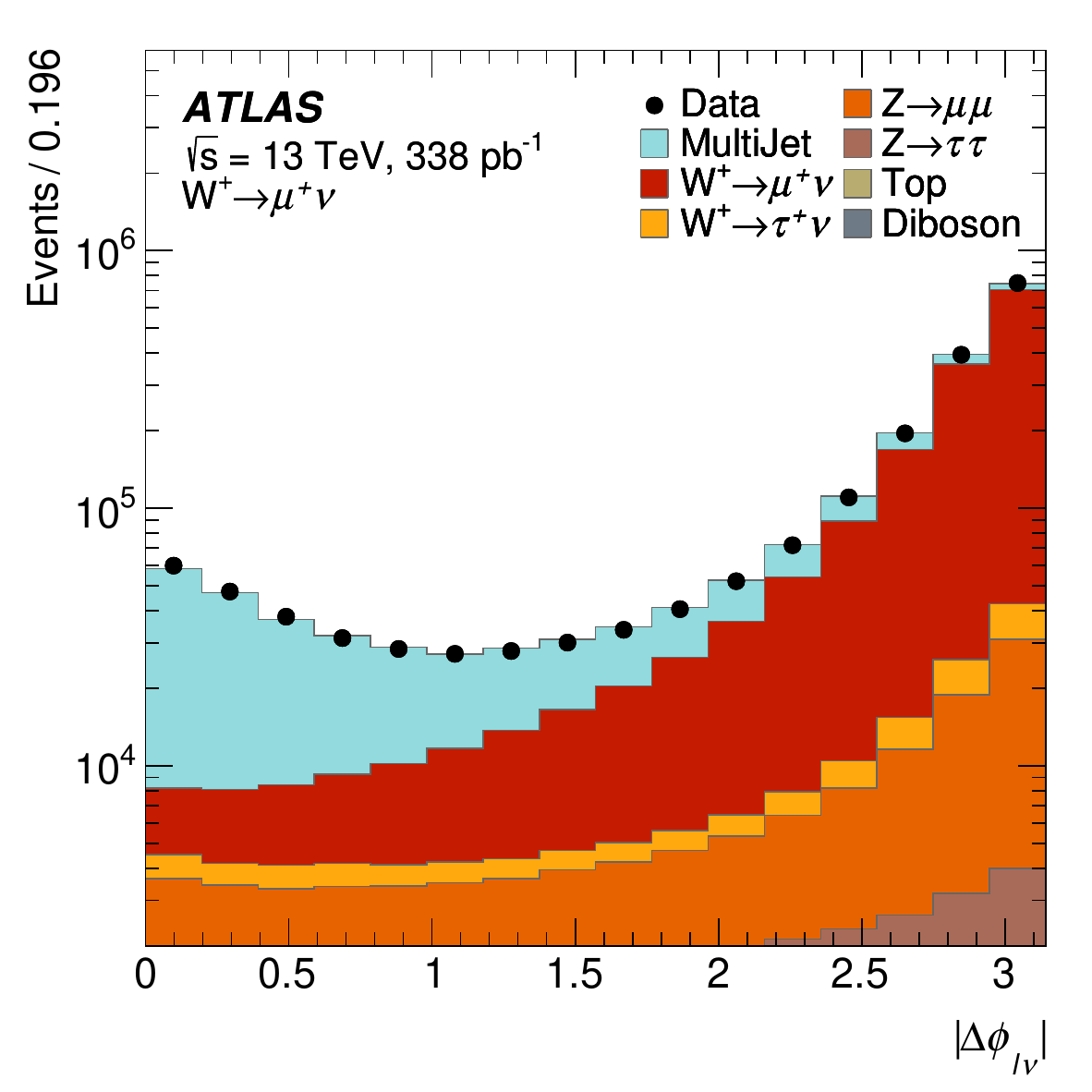}}
\subfloat[]{ \includegraphics[width=0.45\linewidth]{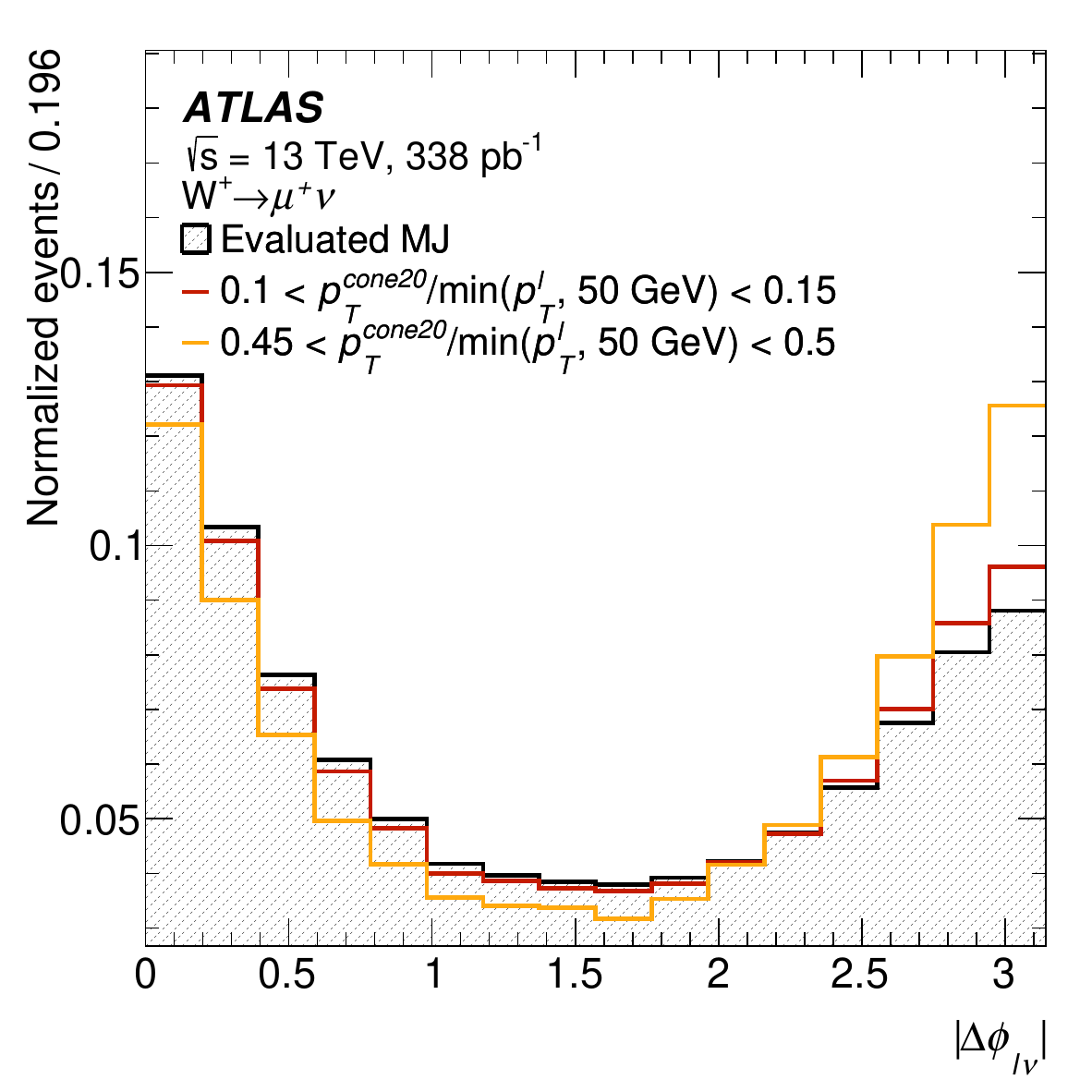}}
\caption{$|\Delta\phi_{\ell\nu}|$ distribution in the fit region for data, multijet background and for signal and background from simulation in the  (a) \Wmenu and (c) \Wpmunu channel.  The multijet template shown in the plots is the one obtained from data in the first anti-isolation interval, as an example.
Multijet templates for the $|\Delta\phi_{\ell\nu}|$ distribution in the signal region in the (b) \Wmenu  and (d) \Wpmunu channel, derived from two anti-isolation intervals and compared with the final multijet template shown in black. }
\label{fig:multijetShape}
\end{figure}

Background contributions from electroweak processes such as single-boson (in this analysis, \Wtaunu and \Zboson-boson production  \Ztau and \Zll are treated as backgrounds) and diboson production,
as well as background arising from processes involving the top-quark, such as \ttbar and single-top-quark production,
are directly estimated from the MC simulation samples described in Section~\ref{sec:samples}.

The background from QCD multijet (MJ) production cannot be reliably simulated and so is derived from data.
Depending on the lepton flavour, the MJ background has significant contributions from
leptons produced in semileptonic decays of heavy quarks (beauty or charm), pion and kaon decays, and in the case of electron candidates  photon conversions and hadronic activity faking a lepton.
\begin{figure}[tbp]
\centering
\subfloat[]{ 	\includegraphics[width=.49\textwidth]{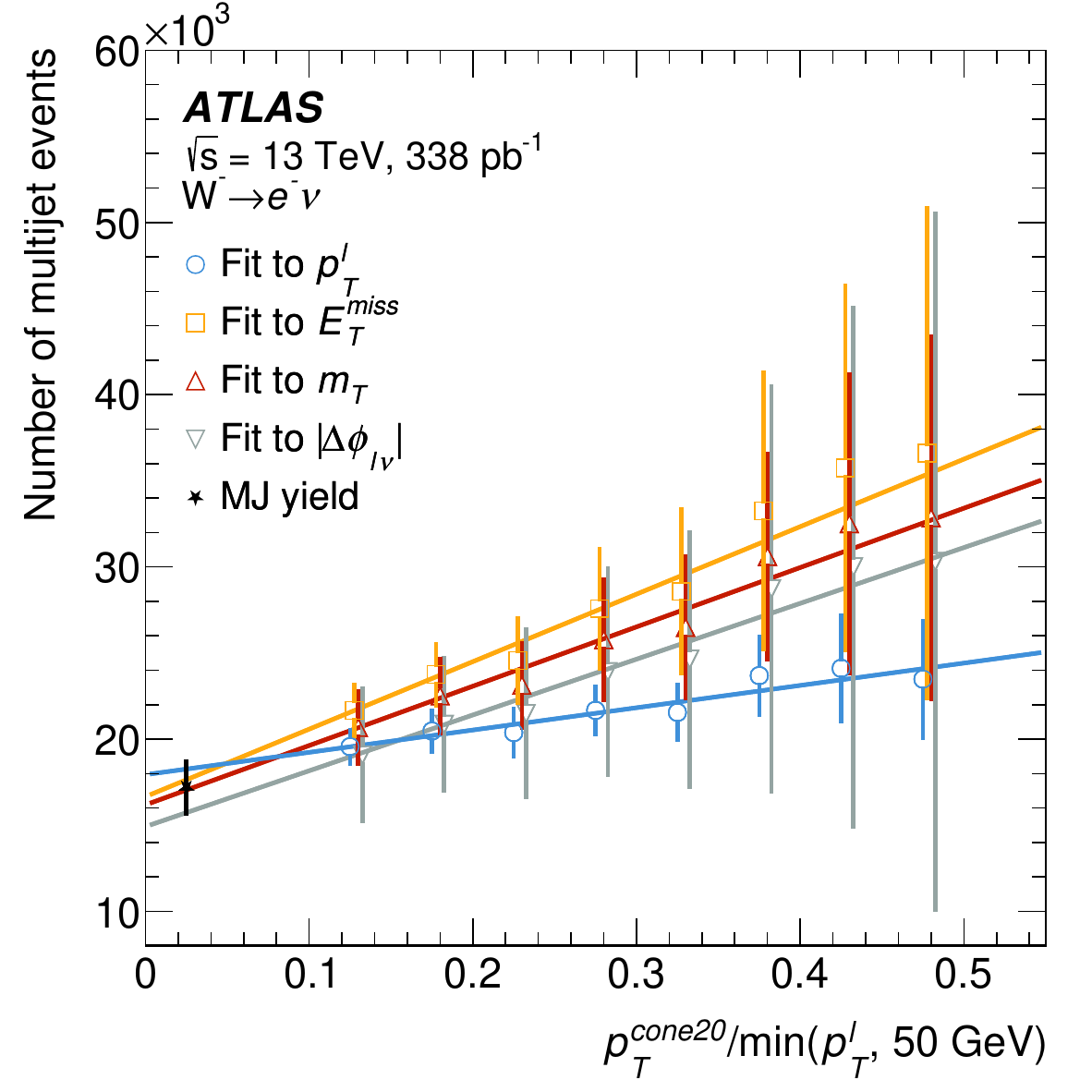}}
\subfloat[]{ 	\includegraphics[width=.49\textwidth]{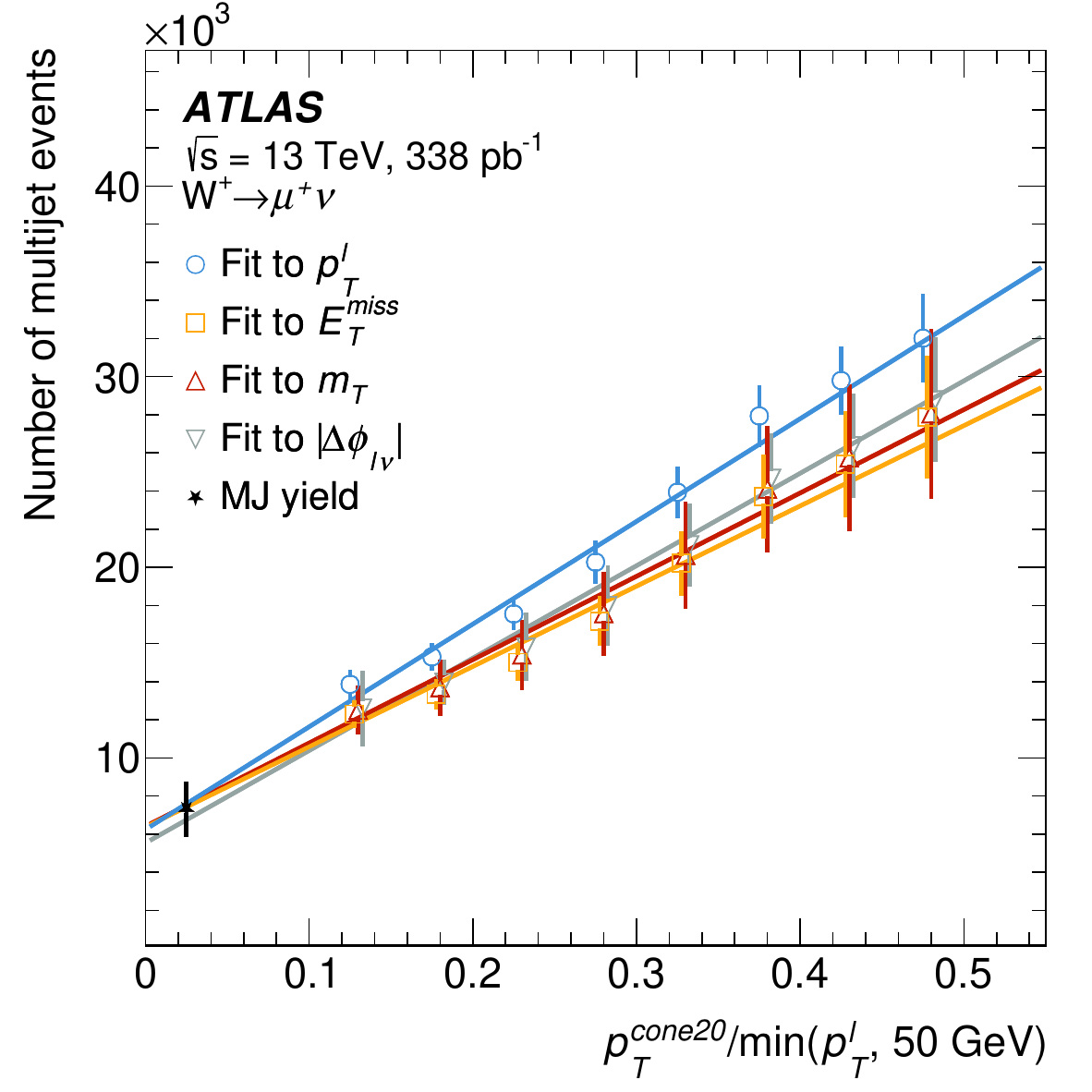}}

\caption{
Examples of multijet event yields extraction procedure from the template fit described in the text as a function of the different anti-isolation regions in $p_\mathrm{T}^\mathrm{cone20}/\mathrm{min}(p_\mathrm{T}, 50\,\mathrm{GeV})$ are shown for the (a)  \Wmenu and the (b) \Wpmunu  channels. The legend indicates each kinematic distributions used in the fit, as detailed in the text. The black vertical error bar on the extrapolated multijet yield in the isolated region represents the total uncertainty in the signal region. }
\label{fig:multijet}
\end{figure}

The multijet yields are estimated following an approach similar to that described in Ref.~\cite{STDM-2018-17}, performed separately in the electron and muon channels, as well as for negatively and positively charged \Wboson-boson candidates. Differences relative the previous method are discussed below.

As multijet production mainly arises at lower values of \ptl,  \met, and \mt\ as compared with the signal, this background
is estimated by using a template fit to these three kinematic distributions in the so-called fit region (FR), which is defined as the signal region with the \met\ and \mt\ selections removed.
As this analysis involves angular observables, the $|\Delta\phi_{\ell\nu}|$ variable is also included.
As an example, the distribution of $|\Delta\phi_{\ell\nu}|$ in the fit region is shown in Figure~\ref{fig:multijetShape} (a) for \Wmenu and in Figure~\ref{fig:multijetShape} (c) for \Wpmunu events  where the shape difference between signal and multijet background is clearly visible and exploited in the fit.

The multijet kinematic distributions used in the template fit are derived from data in anti-isolation regions (i.e. regions where the isolation requirement is inverted), after correcting for contamination from the $W\to\ell\nu$ signal and other non-multijet backgrounds using simulation.
As shown in Figure~\ref{fig:multijetShape} (b,d), the shape of the multijet templates depends on the value of the isolation criterion.
Mutually exclusive intervals
in the isolation variable are defined to create statistically
independent multijet templates that  approach
progressively to the signal-candidate selection. Following the procedure described in more detail in Ref.~\cite{STDM-2018-17}, the soft-energy activity in the anti-isolated regions is further corrected to reduce biases in the multijet kinematics arising from the dependence on the anti-isolation requirement.

For each kinematic distribution, the template fit in the fit region is repeated multiple times, each time using a different multijet template. Each multijet yield extracted in this way is then scaled by a transfer factor that corrects for the acceptance of the  \met and \mt selections in the signal region.
As an example, Figure~\ref{fig:multijet} shows for the \Wmenu and \Wpmunu channels the variation of the extracted MJ yields as a function of the isolation interval used to define the template, together with a linear fit used to extrapolate to the average isolation value in the signal region. Alternative functional forms were tested for the extrapolation and found to give consistent results within the associated uncertainties in the yield.
The final MJ yield is determined by averaging the extrapolated values from the four kinematic variables.
The dominant uncertainties in this procedure arise from the linear extrapolation to the isolated region, the subtraction of signal and other backgrounds contamination in the anti-isolation regions, and the mismodelling of the soft energy activity in the anti-isolated regions.  The final uncertainty in the MJ yield is about 10\% in the electron channel and around 20\% in the muon channel.

\begin{figure}[tbp]
\centering
\subfloat[]{   \includegraphics[width=0.45\linewidth]{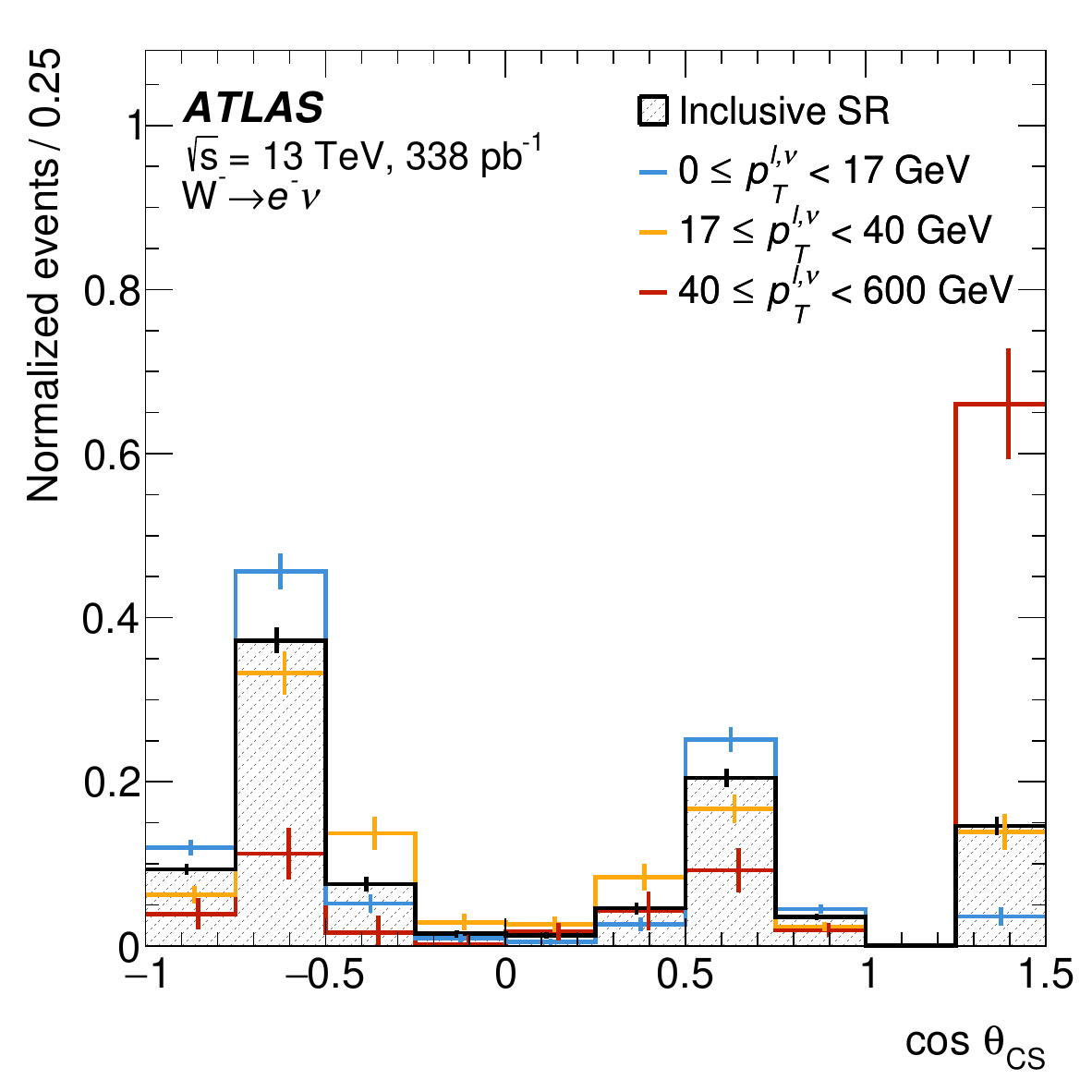} }
\subfloat[]{   \includegraphics[width=0.45\linewidth]{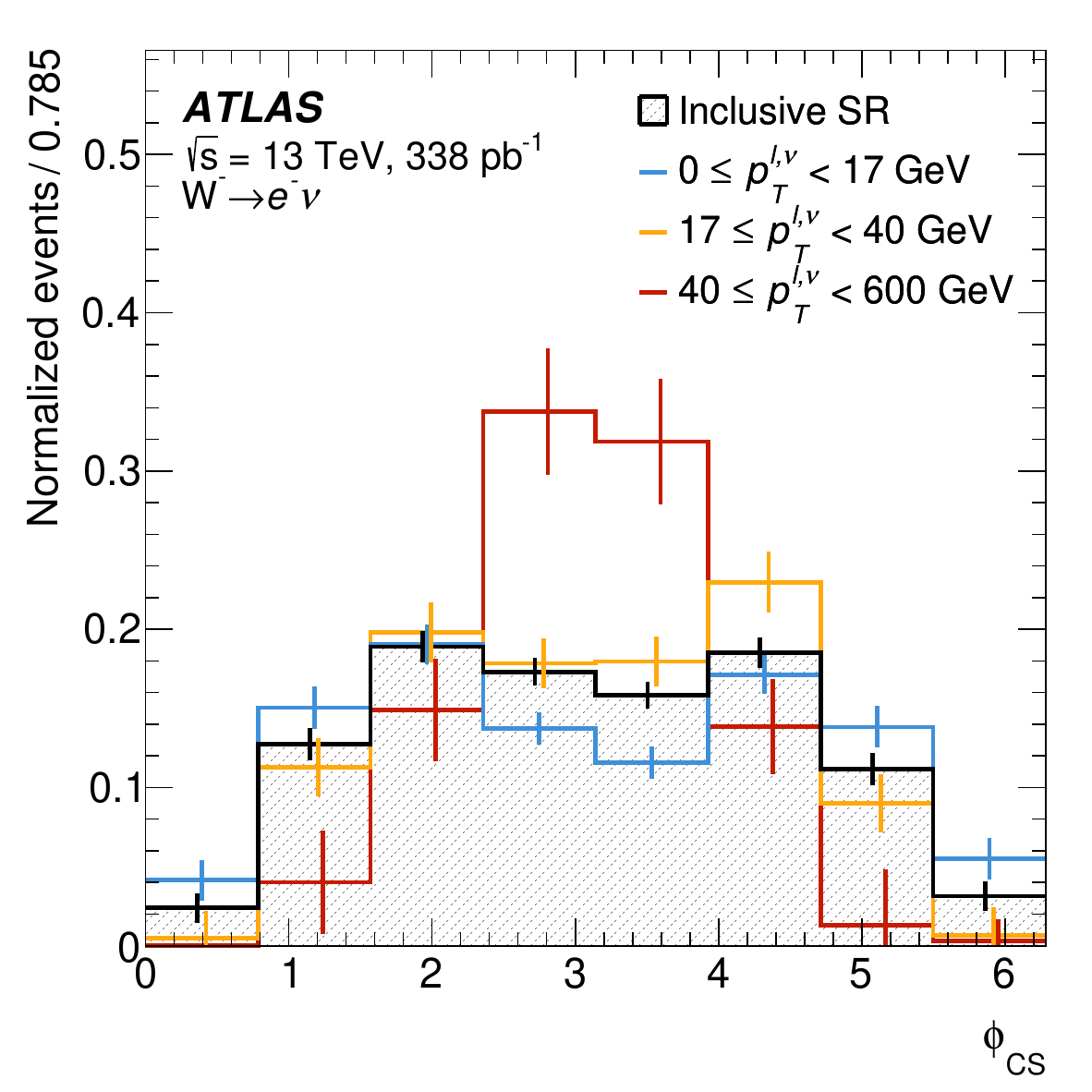}  }
\\
\subfloat[]{   \includegraphics[width=0.45\linewidth]{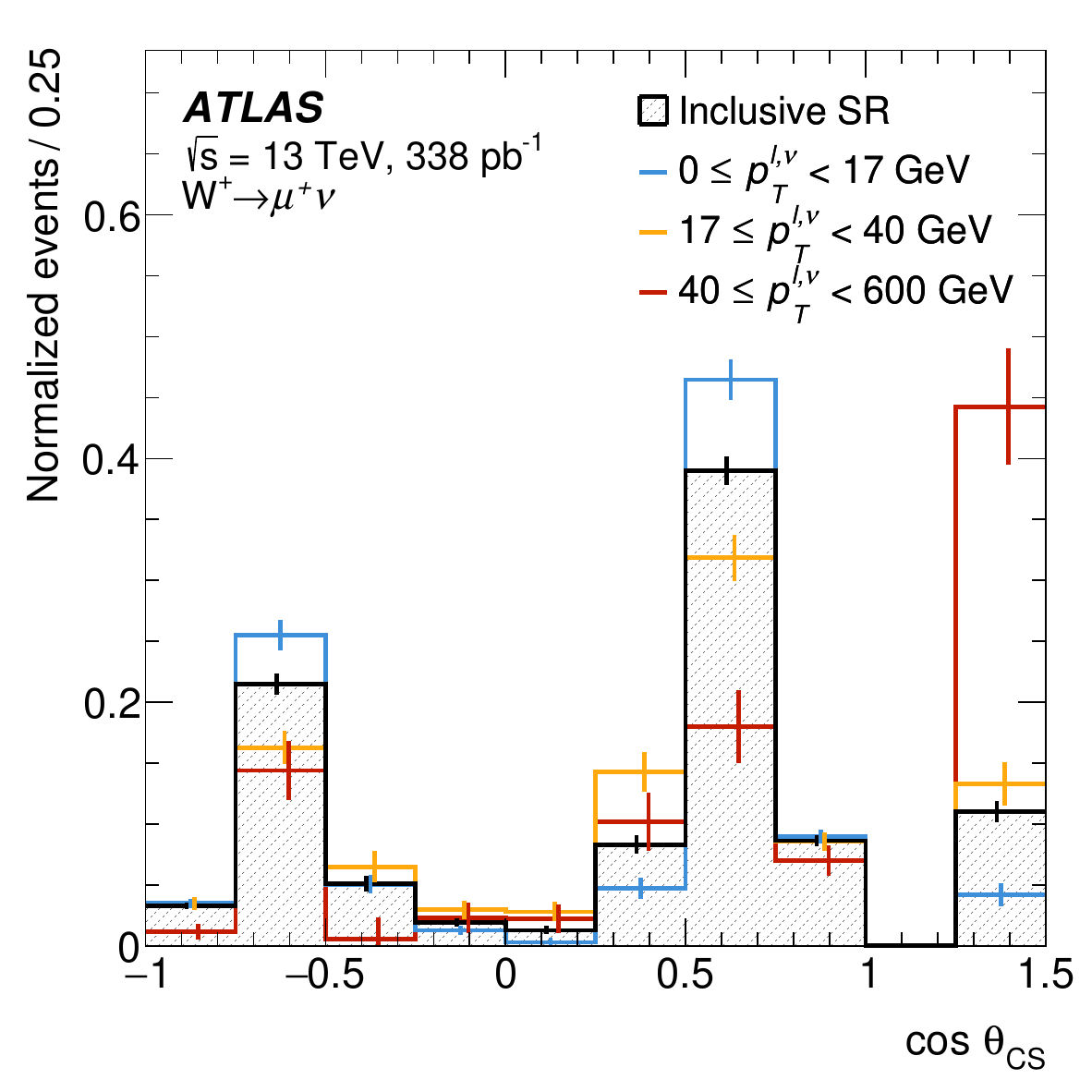} }
\subfloat[]{   \includegraphics[width=0.45\linewidth]{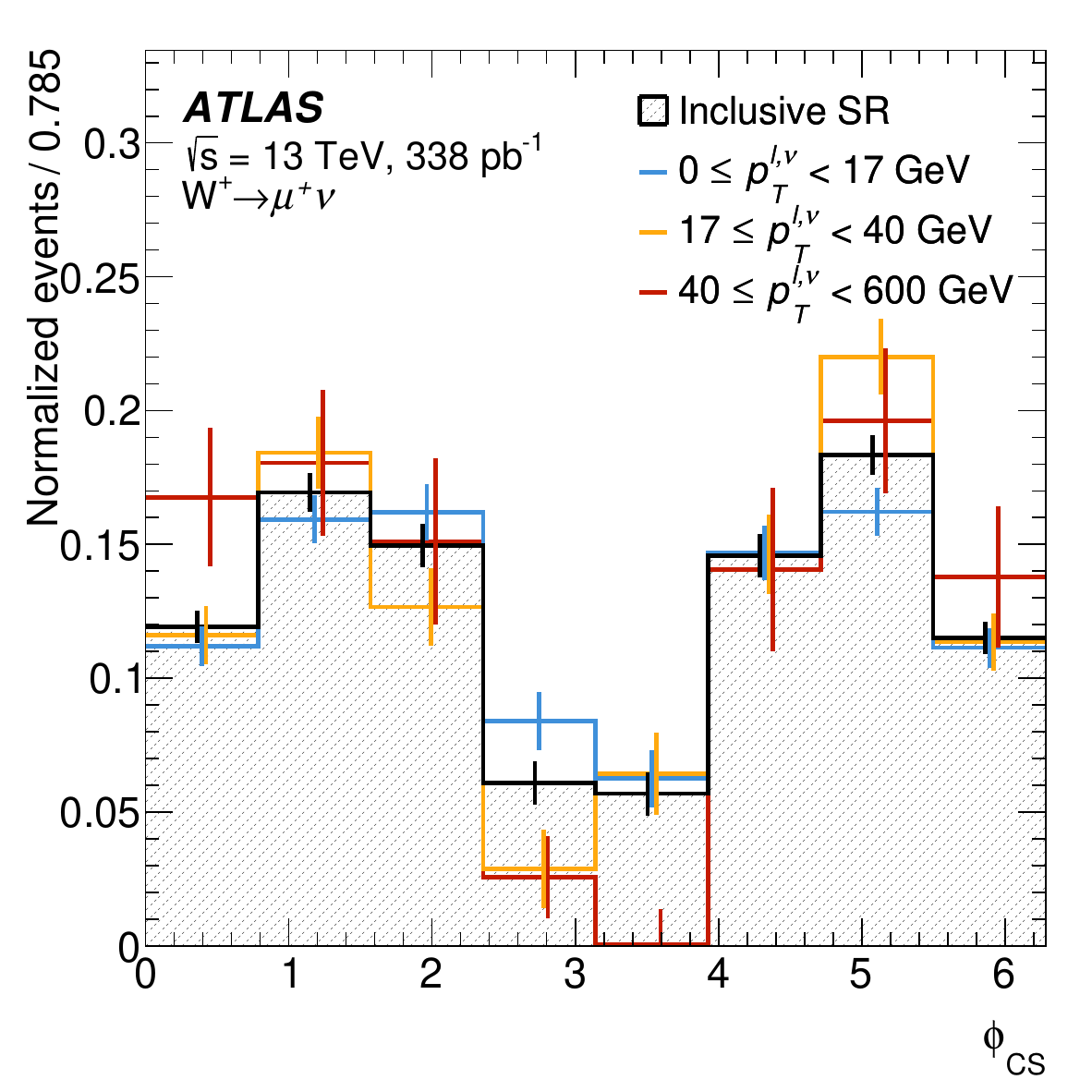} }
\caption{Examples of one dimensional projections of the normalised angular distributions \(\coscs\) and \(\phics\) of the multijet background for the (a) (b) \Wmenu and (c) (d) \Wpmunu  channels extracted in the \(\ptln\)-inclusive signal region (black curve)  and in three broad \(\ptln\) intervals $[0, 17)$~\text{GeV}, $[17, 40)$~\text{GeV}, and $[40, 600)$~\text{GeV} (coloured curves). As mentioned in Section~\ref{sec:methodology}, events without a real solution for $\pzn$ are included as overflow in the tenth bin in \coscs. }
\label{fig:mjAccCorr}
\end{figure}

Similarly to the yield extraction, the shapes of the MJ kinematic distributions in the signal region are determined by linearly extrapolating the shapes observed in the mutually exclusive anti-isolation intervals into the isolated region, as illustrated by the black histograms in Figure \ref{fig:multijetShape} (b,d).
Additional corrections to the shapes of the angular distributions of the MJ background in the signal region are applied to account for the strong dependence of the two-dimensional (\coscs, \phics)  distributions on \ptln, as shown in Figure~\ref{fig:mjAccCorr}. To derive these corrections, MJ background templates are constructed independently in three broad \(\ptln\) intervals: \(\ptln \in [0, 17)~\text{GeV}\), \([17, 40)~\text{GeV}\), and \([40, 600)~\text{GeV}\).
Within each $\ptln$ interval, acceptance corrections derived from MC simulation are applied to the MJ angular distributions to account for kinematic differences between the broad $\ptln$ intervals used in the estimate and the finer binning defined in Eq.~\ref{eq:binning} for the final measurement.   The dominant sources of uncertainty in the corrected multijet angular distributions arise from the limited data statistics in the anti-isolated regions of each $\ptln$ interval, the limited size of the MC samples used to derive acceptance corrections, and differences in the modelling of non-prompt lepton backgrounds between data and simulation.

The final multijet background yields, as well as the two-dimensional angular distributions in (\coscs, \phics), are provided for each \ptln bin as input for the likelihood fit described in~Section~\ref{sec:methodology}.
This information is given separately for \Wminus and \Wplus in each channel.
The corresponding uncertainties, both in normalisation and shape, are propagated to the subsequent stages of the analysis.

\subsection{Event yields and detector-level distributions}

\begin{figure}[tbp]
\centering
\subfloat[]{  \includegraphics[width=0.45\linewidth]{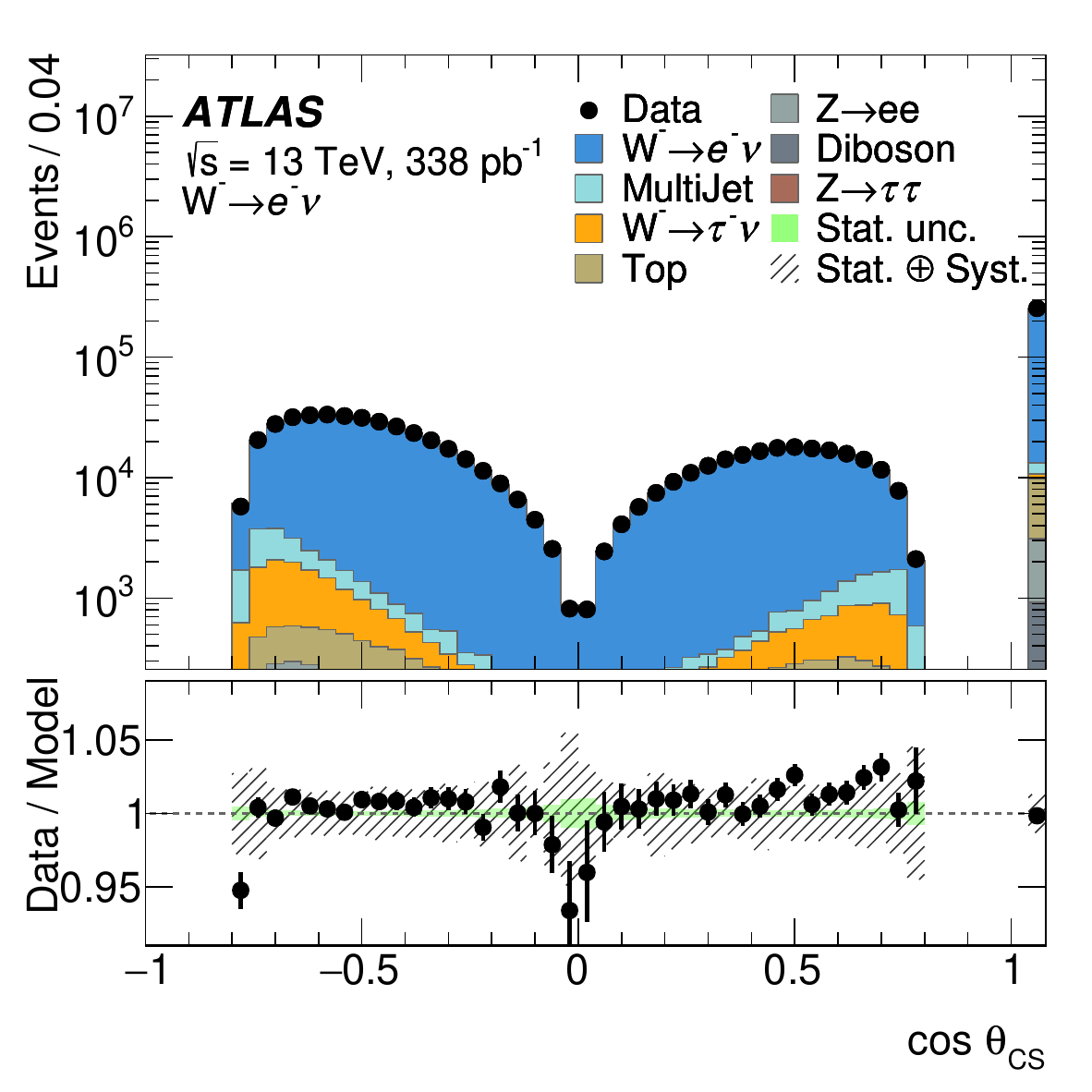} }
\subfloat[]{  \includegraphics[width=0.45\linewidth]{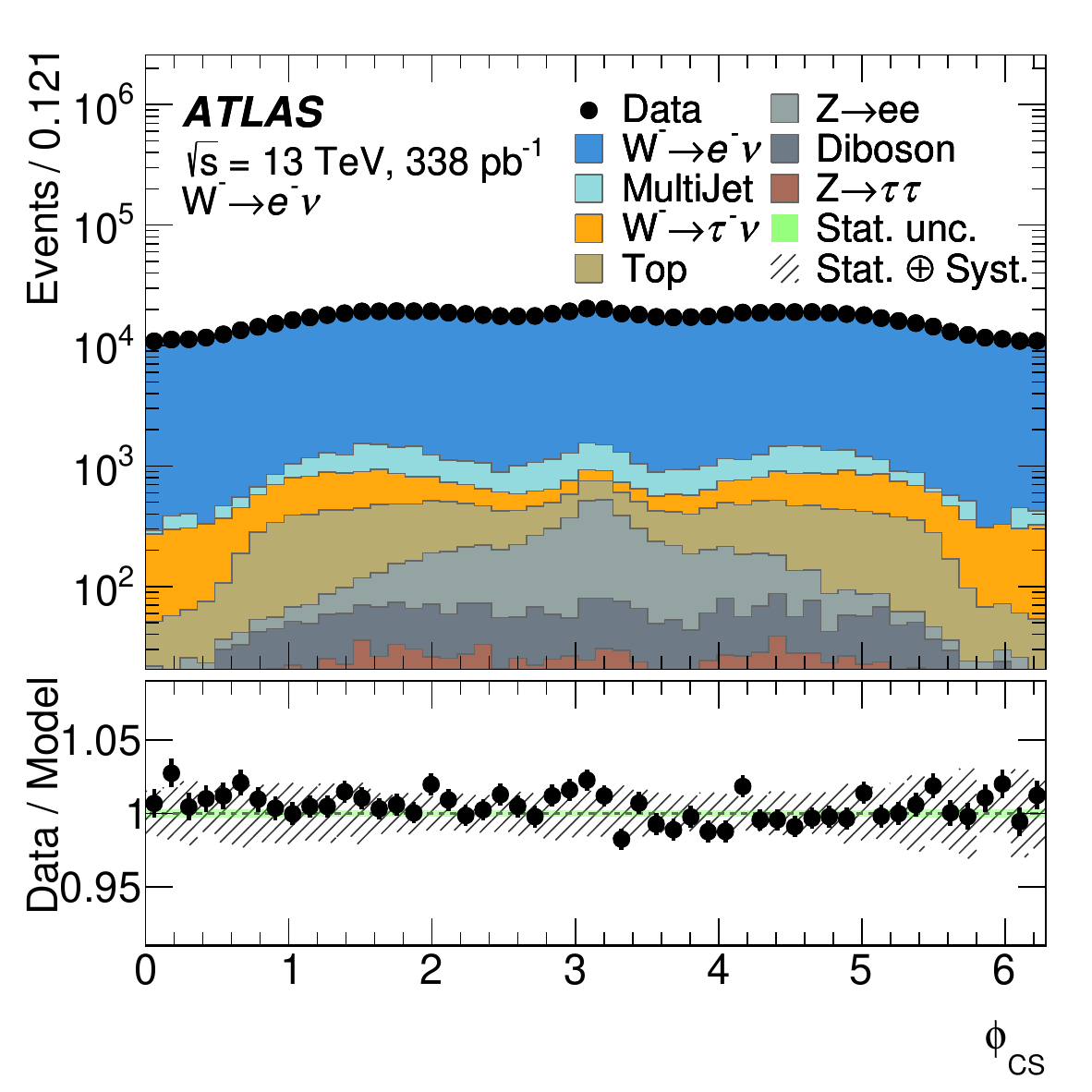} } \\
\subfloat[]{  \includegraphics[width=0.45\linewidth]{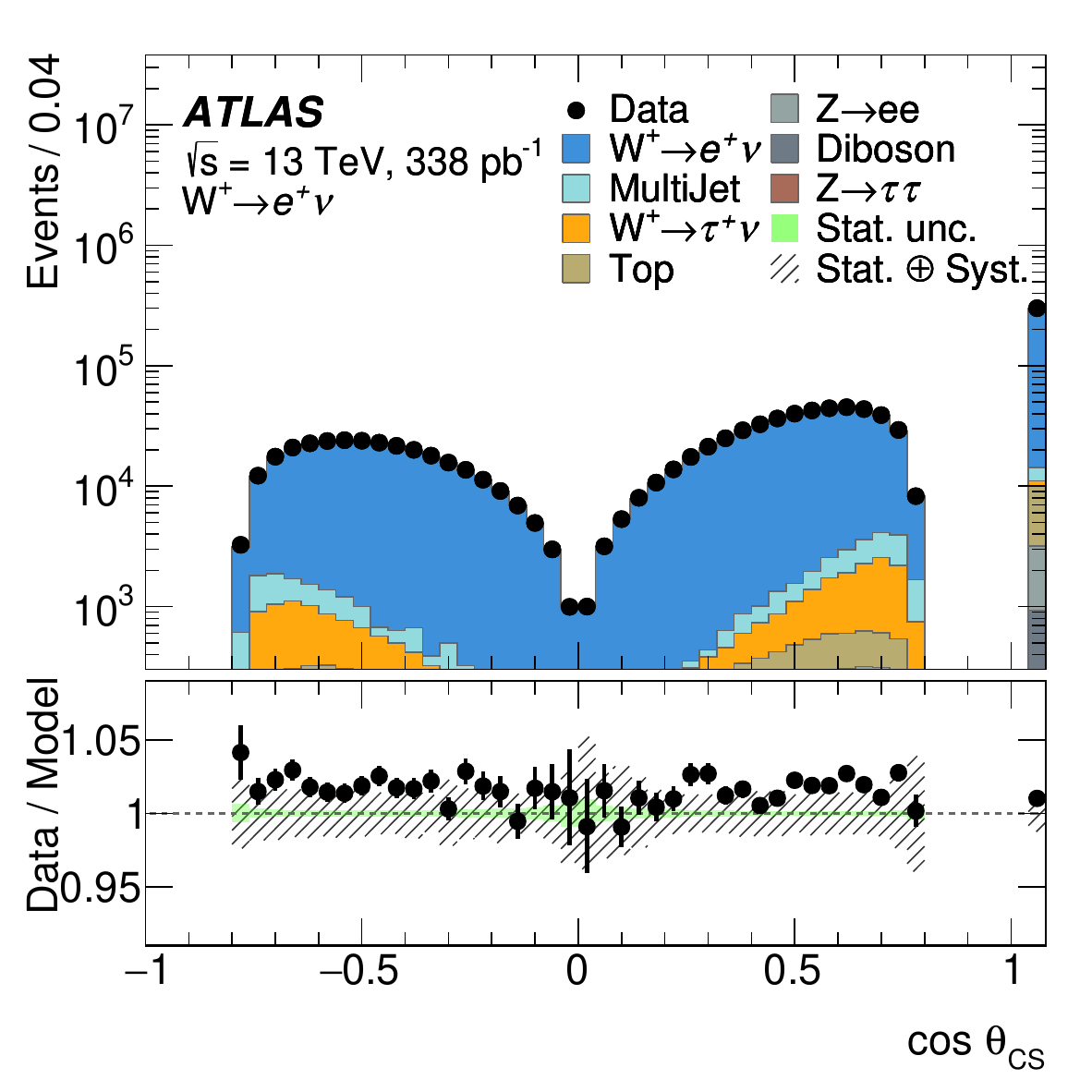} }
\subfloat[]{  \includegraphics[width=0.45\linewidth]{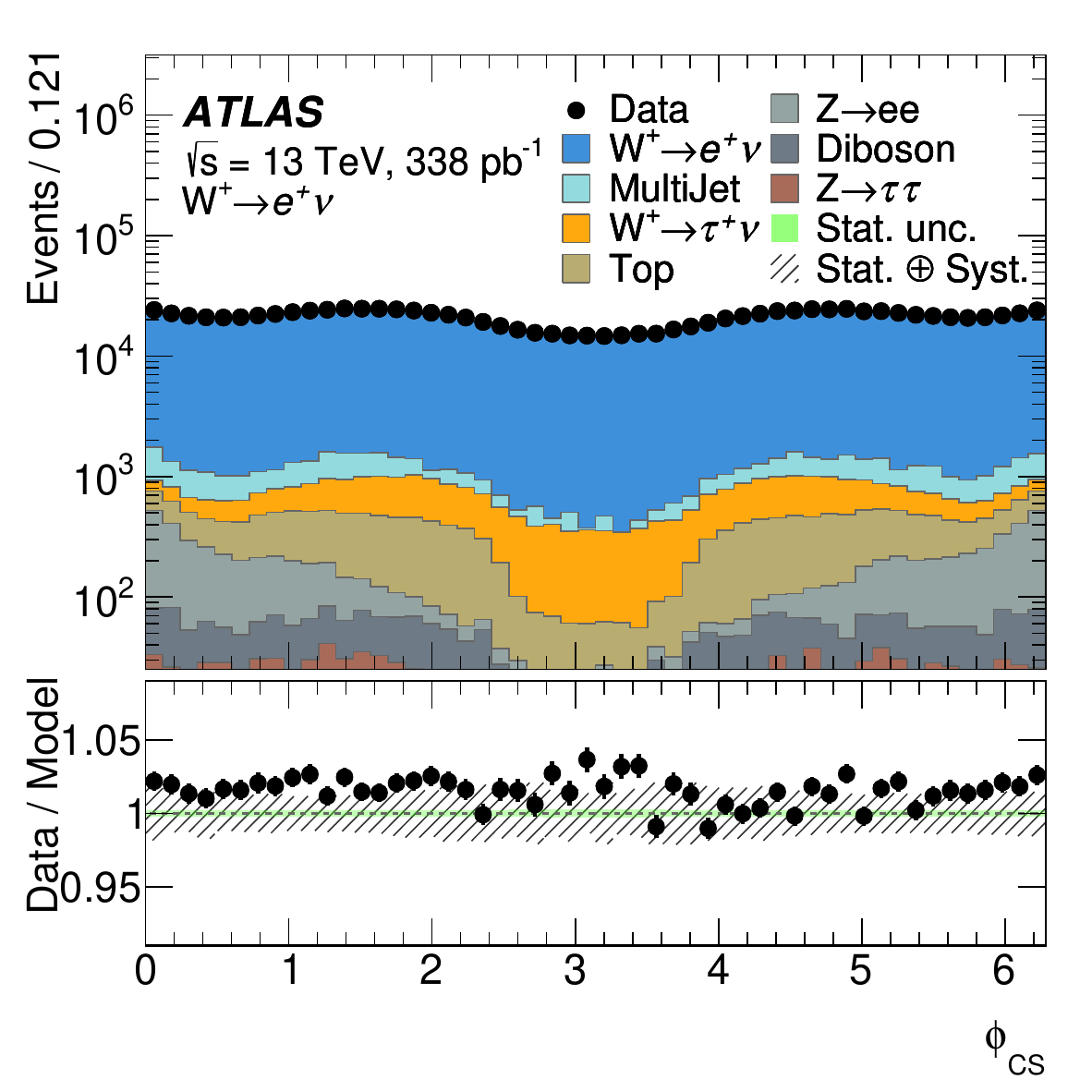} }
\caption{
The \coscs and \phics angular distributions, integrated over~\ptln for the \Wenu channel for  (a) (b) negative and (c) (d) positive  leptons. The expected distributions are shown separately for the signal and the different background sources contributing to each channel. The grey band represents the total experimental uncertainties. Events without a real solution for $\pzn$ are included as overflow bin in \coscs distribution.}
\label{fig:controlPlots_wenu}
\end{figure}

\begin{figure}[tbp]
\centering
\subfloat[]{   \includegraphics[width=0.45\linewidth]{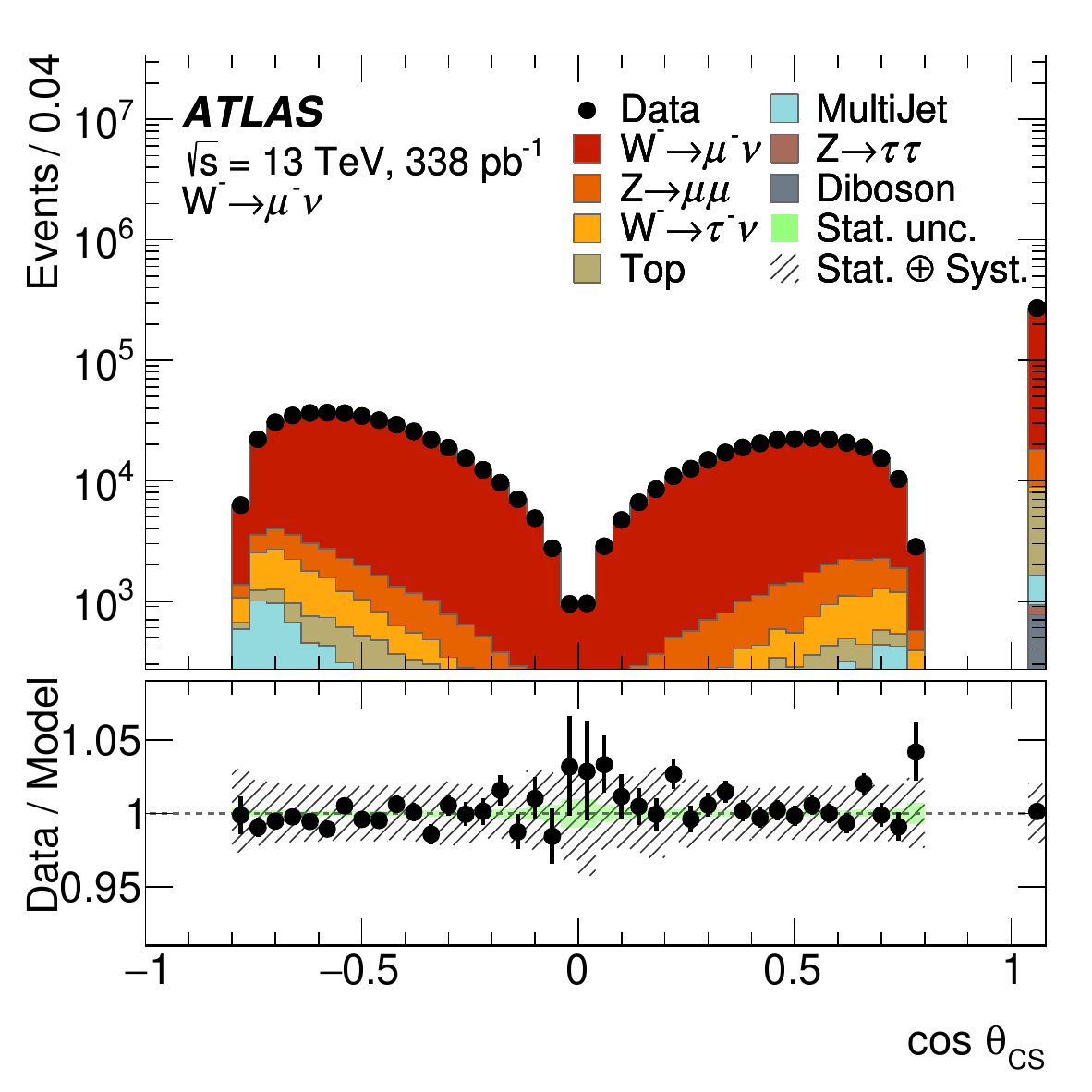} }
\subfloat[]{   \includegraphics[width=0.45\linewidth]{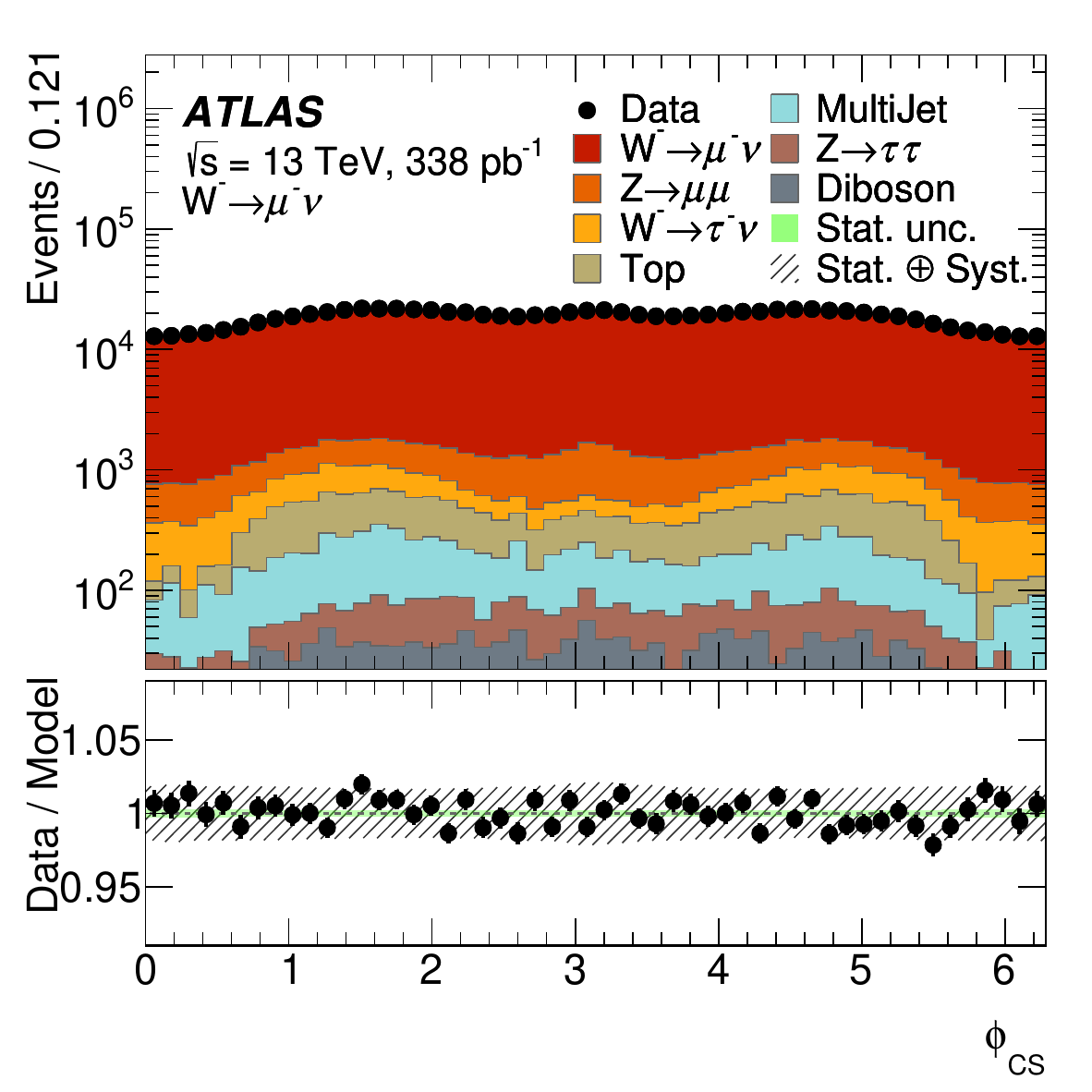} }
\\
\subfloat[]{  \includegraphics[width=0.45\linewidth]{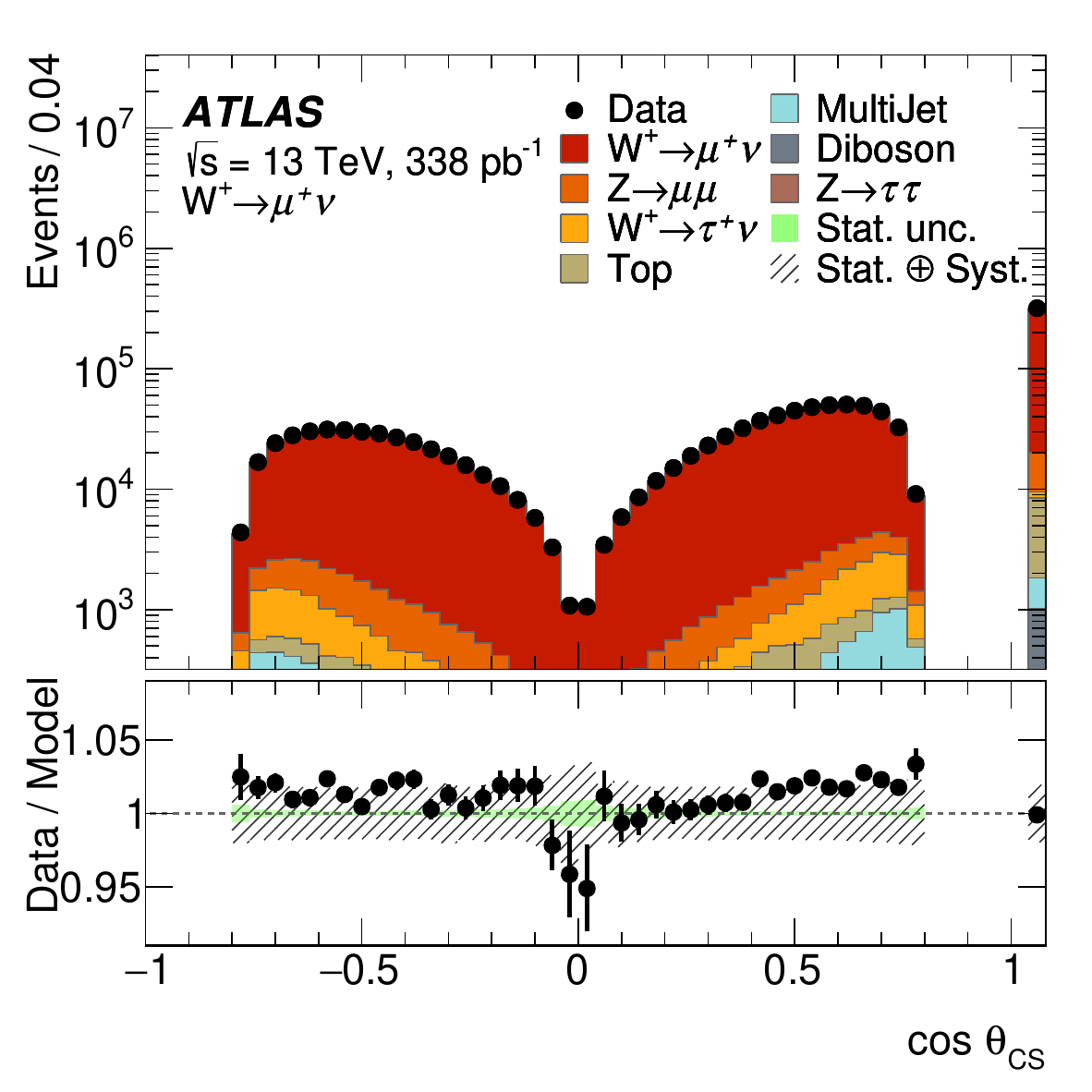} }
\subfloat[]{  \includegraphics[width=0.45\linewidth]{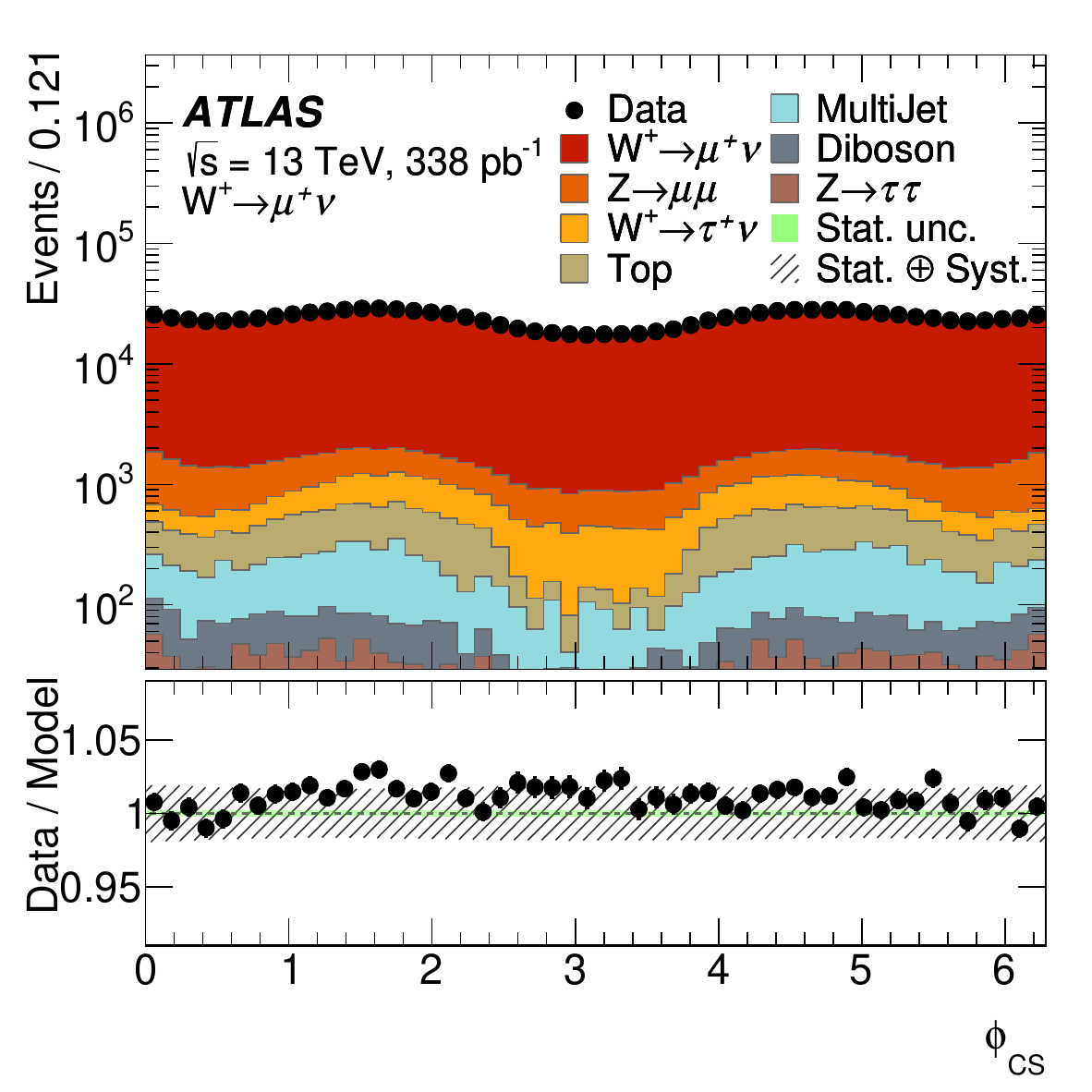} }
\caption{
The \coscs  and \phics angular distributions, integrated over~\ptln for the \Wmunu channel for (a) (b) negative and (c) (d) positive leptons. The expected distributions are shown separately for the signal and the different background sources contributing to each channel. The grey band represents the total experimental uncertainties. Events without a real solution for $\pzn$ are included as overflow bin in \coscs distribution.}
\label{fig:controlPlots_wmunu}
\end{figure}

Projections of the detector-level distributions of the angular distributions \coscs and \phics integrated over \ptln are shown for \Wminus and \Wplus in Figure~\ref{fig:controlPlots_wenu} for the electron channel and in Figure~\ref{fig:controlPlots_wmunu} for the muon channel.

The differences between the electron and muon distributions reflect the different efficiencies for reconstructing and selecting the two lepton flavours. Intrinsic kinematic differences between $W^-$-boson events and $W^+$-boson events at $\sqrt{s}=13$\,TeV are also observed when comparing the two charged channels.
Overall, good agreement is observed between the data and the expectations in all the kinematic distributions.
Small normalisation differences between the data and the MC distributions are observed at the level of a few percent, compatible with the combined uncertainties in the integrated luminosity and signal cross-section. The measurement of the angular coefficients is, however, independent of the normalisation between data and simulation in each measurement bin.

The expected event yields for signal and background processes are shown in Table~\ref{tab:event-yields}, along with the yields as measured in data.  The total MJ background yield and total uncertainty  described in Section~\ref{sec:background} are also presented  in Table~\ref{tab:event-yields}. The estimated multijet background fractions represent $2.0\%$ and $0.7\%$  of the total number of observed \Wminus candidate events in the electron and muon channels, respectively. The corresponding numbers for \Wplus candidate events are $1.6\%$ and $0.6\%$, respectively.

\begin{table}[tbp]
\caption{Observed and expected event yields for candidate \Wminus and \Wplus events in the electron and muon channels.
The uncertainty in the expected event yields includes only the statistical uncertainty of the simulation, whereas the total uncertainty is given for the multijet background.  }
\label{tab:event-yields}
\setlength\tabcolsep{12pt}
\resizebox{\textwidth}{!}{

\begin{tabular}{ l r@{\,} c@{\,} l r@{\,} c@{\,} l r@{\,} c@{\,} l r@{\,} c@{\,} l }
Channel & \multicolumn{3}{c}{$W^{-}\rightarrow e^{-}\nu$} & \multicolumn{3}{c}{$W^{+}\rightarrow e^{+}\nu$} & \multicolumn{3}{c}{$W^{-}\rightarrow \mu^{-}\nu$} & \multicolumn{3}{c}{$W^{+}\rightarrow \mu^{+}\nu$}\\
\hline \hline
\Wenu  & 802370 & $\pm$ & 260 & 1019420 & $\pm$ & 290 & & -- & & & -- & \\
\Wmunu & & -- & & & -- & & 893900 & $\pm$ & 270 & 1153450 & $\pm$ & 310 \\
$W\rightarrow \tau\nu$  & 14950 & $\pm$ & 110 & 18260 & $\pm$ & 130 & 14670 & $\pm$ & 110 & 18210 & $\pm$ & 130 \\
\Zee  & 4953 & $\pm$ & 17 & 5227 & $\pm$ & 17 & & -- & & & -- & \\
\Zmm  & & -- & &  & -- & & 33469 & $\pm$ & 42 & 37148 & $\pm$ & 44 \\
$Z\rightarrow \tau\tau$  & 1106 & $\pm$ & 32 & 1183 & $\pm$ & 32 & 1705 & $\pm$ & 51 & 1663 & $\pm$ & 49 \\
Top  & 11928 & $\pm$ & 47 & 12372 & $\pm$ & 49 & 11948 & $\pm$ & 47 & 12516 & $\pm$ & 49 \\
Diboson  & 1581 & $\pm$ & 43 & 1605 & $\pm$ & 46 & 1617 & $\pm$ & 43 & 1718 & $\pm$ & 44 \\
\hline
Multijet  & 17200 & $\pm$ & 1600 & 18600 & $\pm$ & 1900 & 6800 & $\pm$ & 1300 & 7400 & $\pm$ & 1400 \\
\hline
Total expected &
854000 & $\pm$ & 1700 &
1076600 & $\pm$ & 2000 &
964100 & $\pm$ & 1400 &
1232100 & $\pm$ & 1500 \\
\hline
\hline
Data  & 857807 & &   & 1093356 & &  & 964514 & &   & 1245755 & &   \\
\hline \hline
\end{tabular}

%
%
%
%
%
%
%
%
%
%
%
%
%
%
%
%
%
%
%
%
%
%
%
%


}
\end{table}



%

\section{Systematic uncertainties}%
\label{sec:Systematic_uncertainties}

\begin{figure}[ht]
\centering

\subfloat[]{\includegraphics[width=\half]{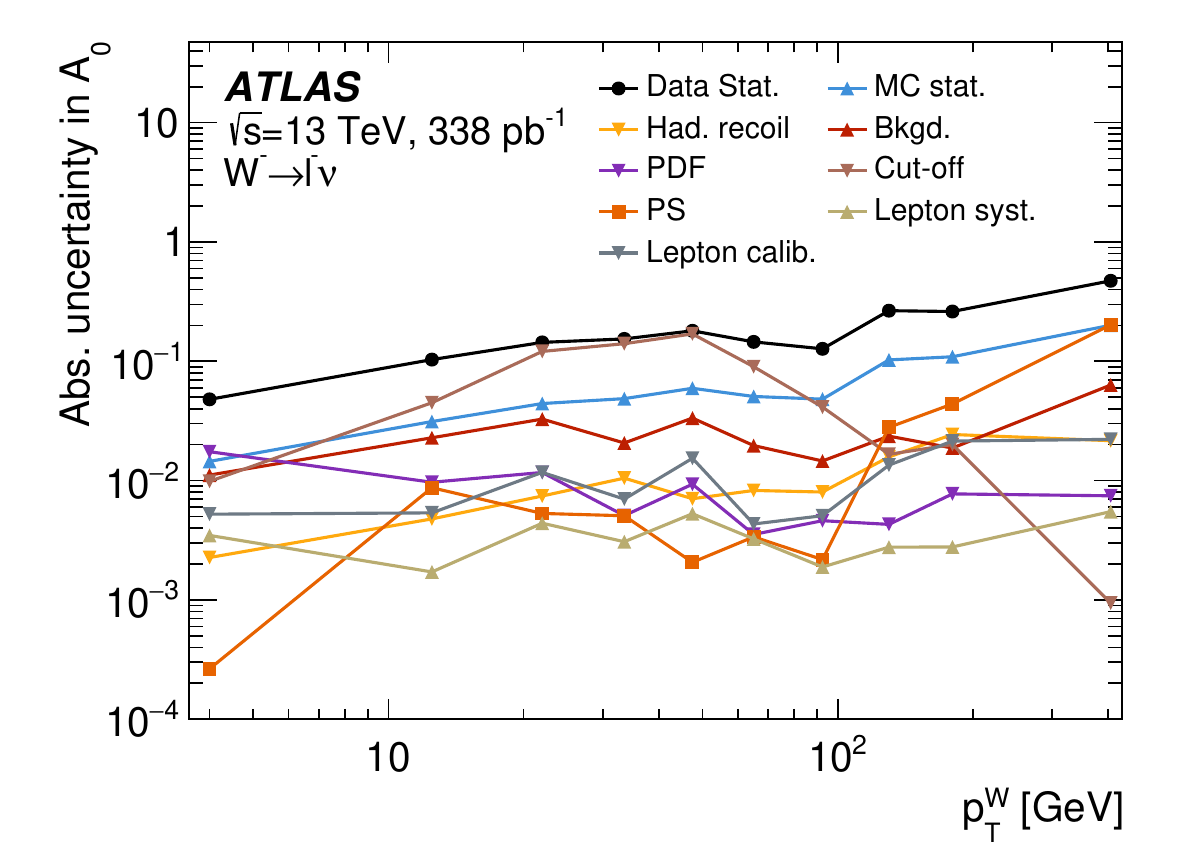}}
\subfloat[]{\includegraphics[width=\half]{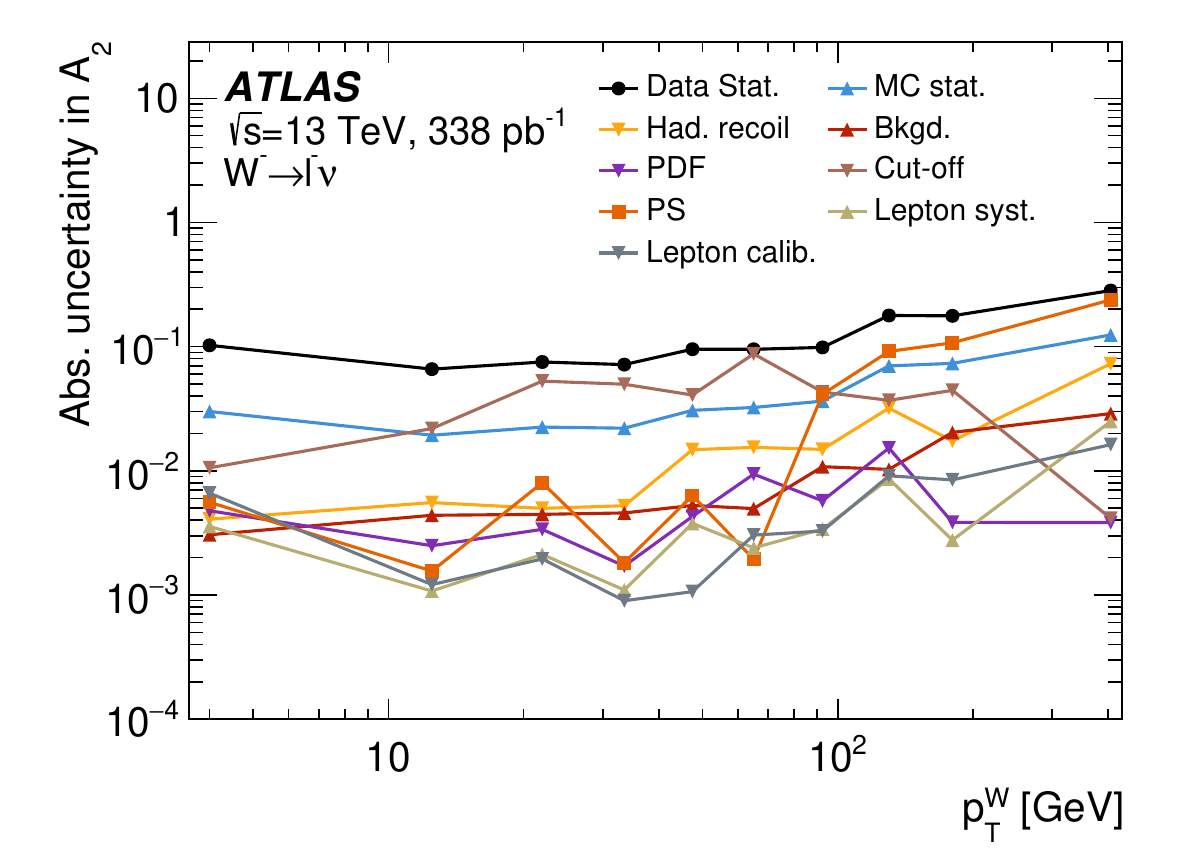}}\\
\subfloat[]{\includegraphics[width=\half]{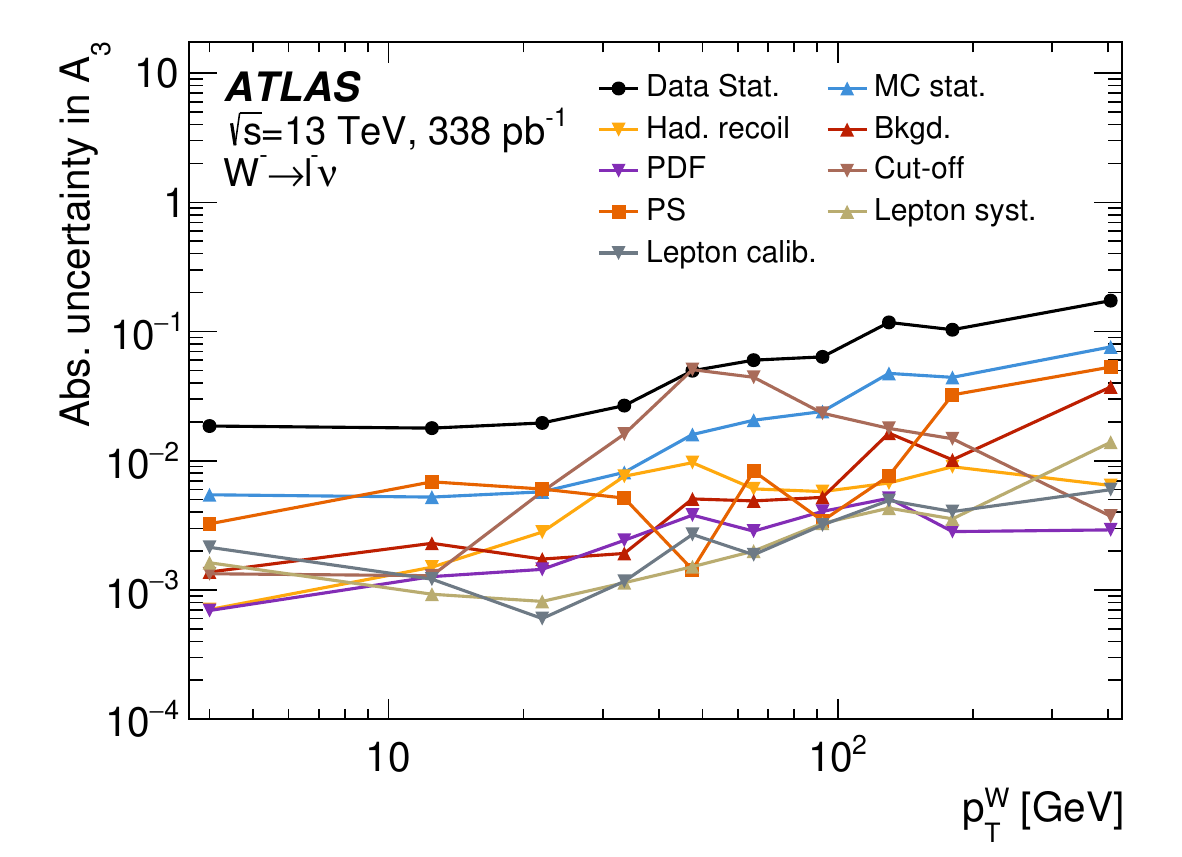}}
\subfloat[]{\includegraphics[width=\half]{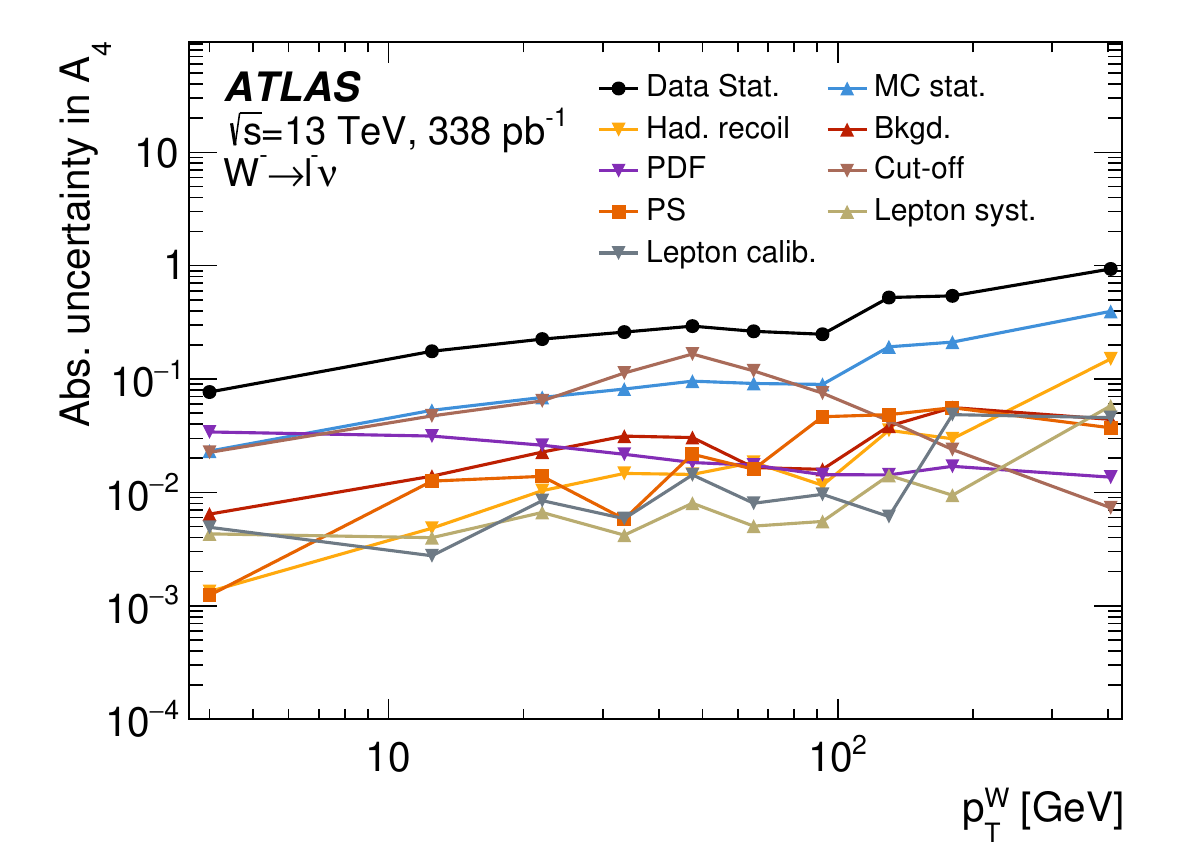}}
\caption{The breakdown into statistical and systematic components of all the uncertainties for the measured (a) \Azero, (b) \Atwo, (c) \Athree and (d) \Afour angular coefficients  as a function of \ptw in the \Wmlnu channel, as obtained from the full fit. The values are given as absolute uncertainty.  }
\label{fig:WMLL_pt_Summary}
\end{figure}

\begin{figure}[ht]
\centering
\subfloat[]{\includegraphics[width=\half]{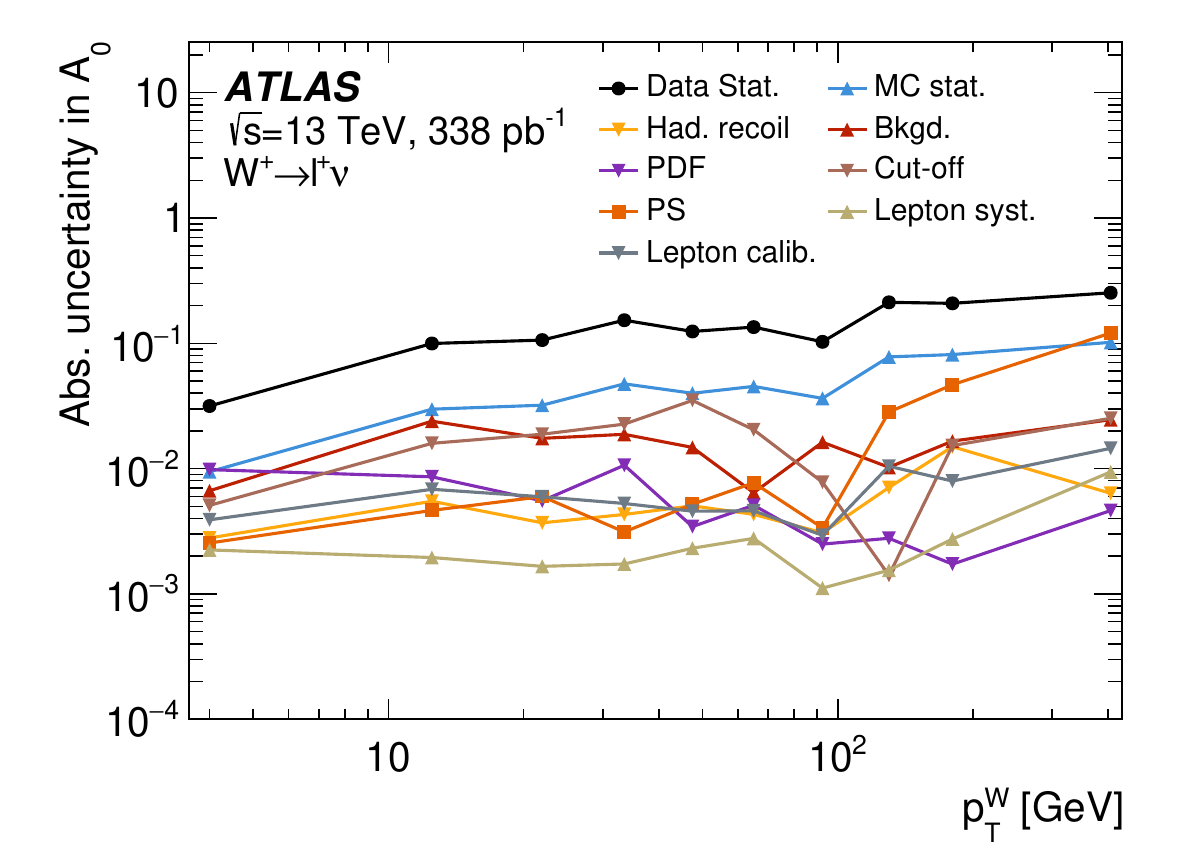}}
\subfloat[]{\includegraphics[width=\half]{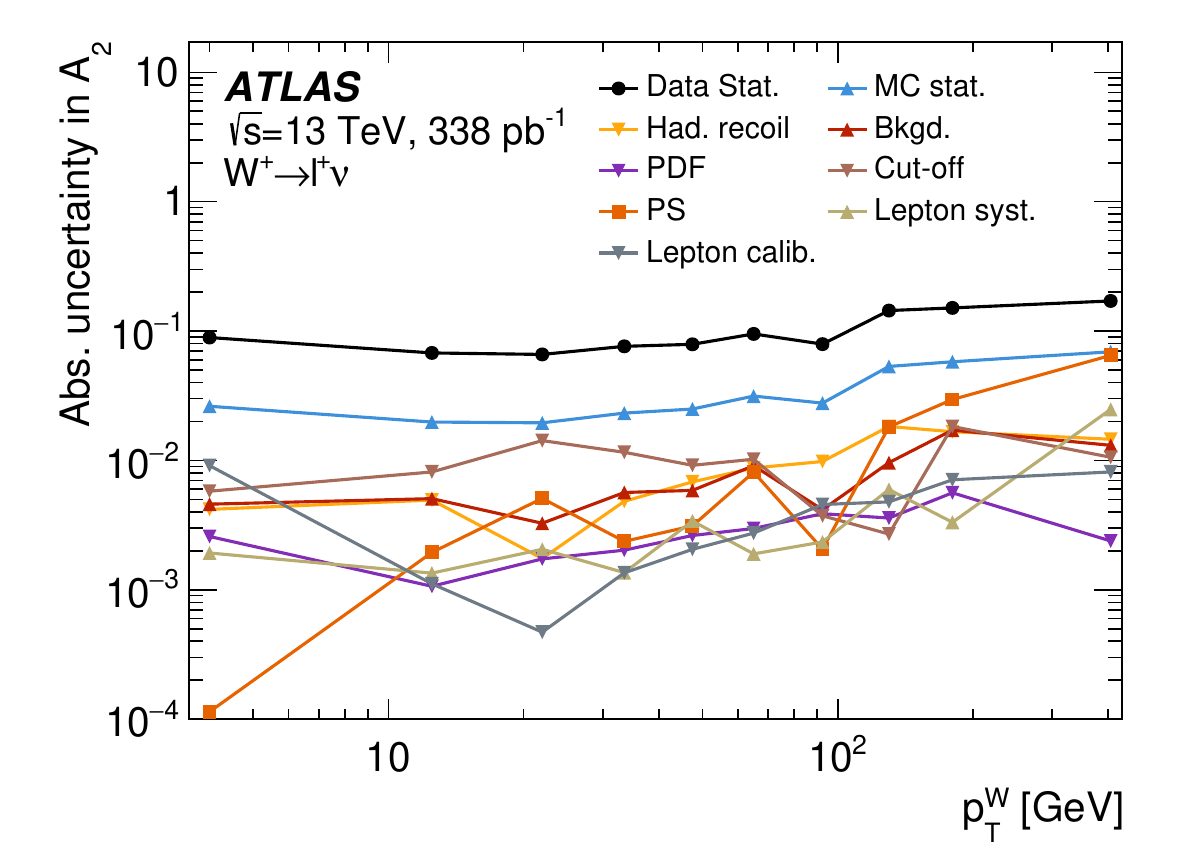}}\\
\subfloat[]{\includegraphics[width=\half]{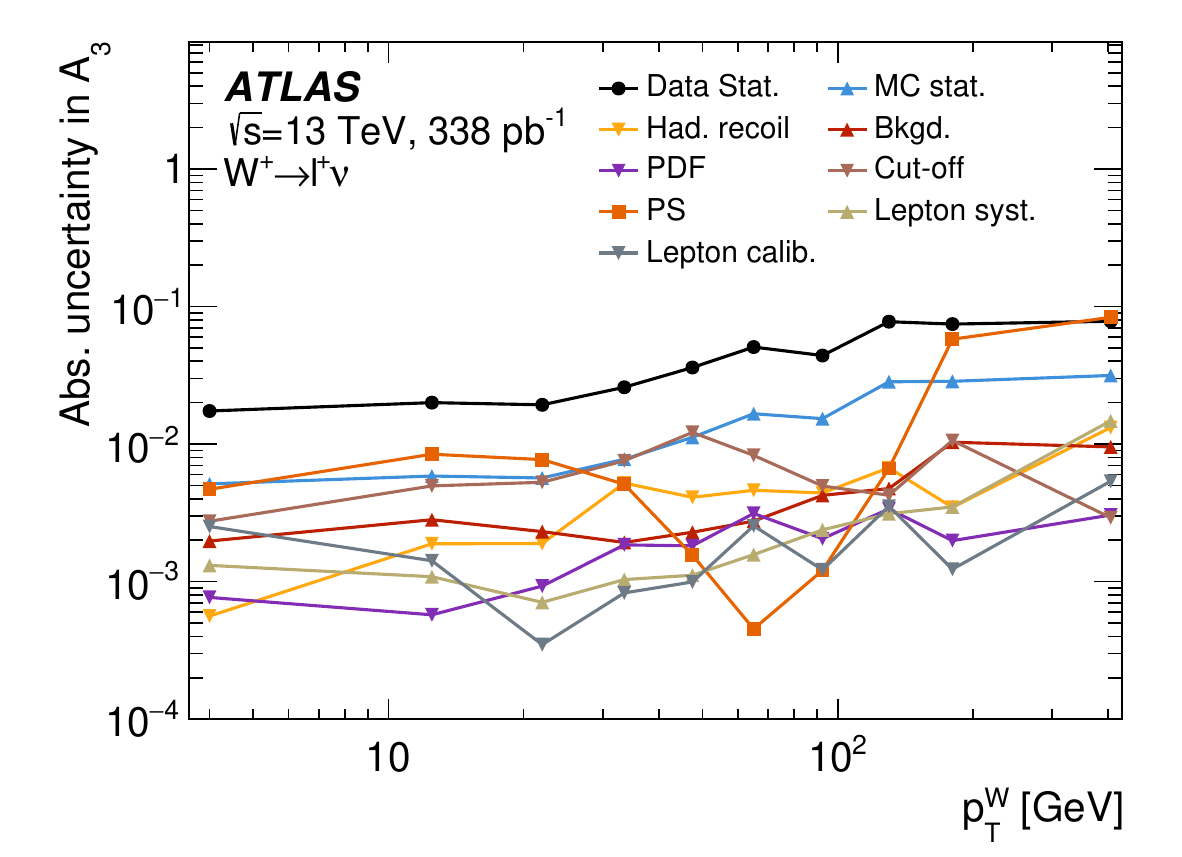}}
\subfloat[]{\includegraphics[width=\half]{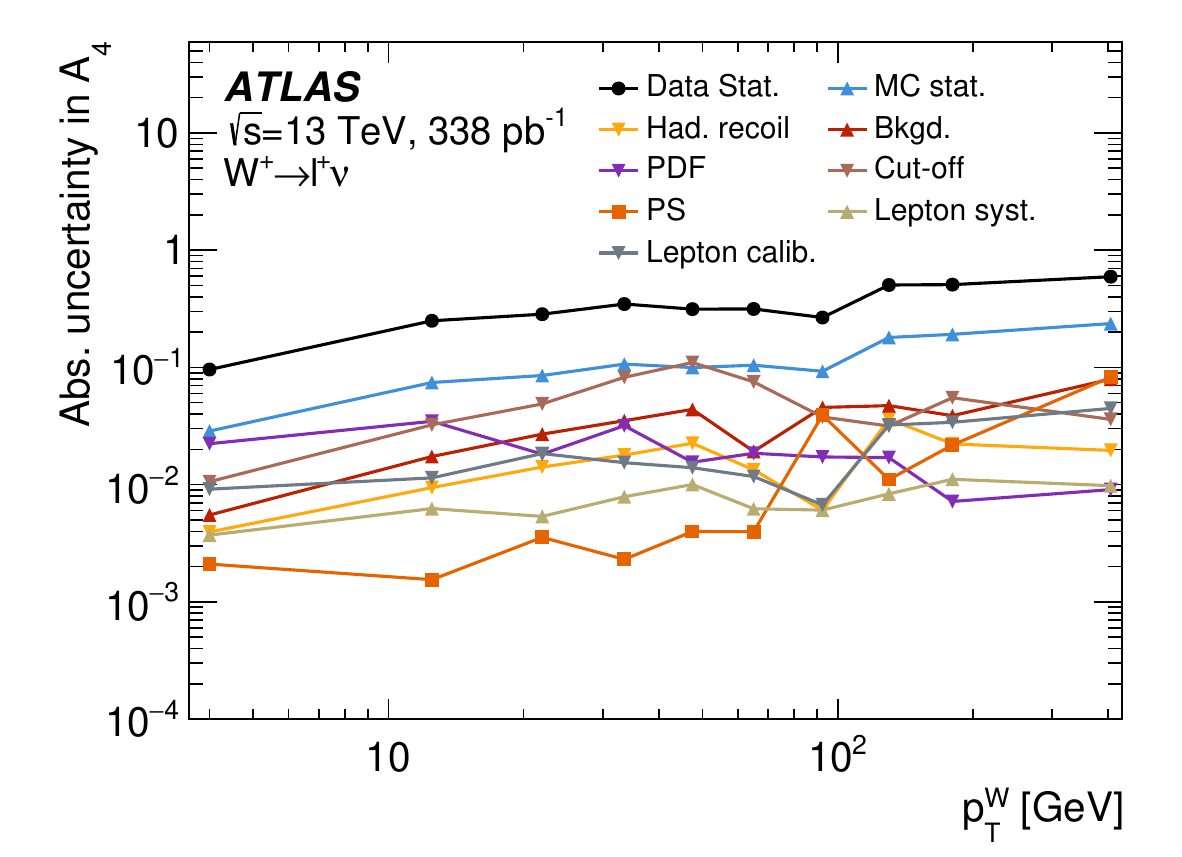}}
\caption{The breakdown into statistical and systematic components of all the uncertainties for the measured (a) \Azero, (b) \Atwo, (c) \Athree and (d) \Afour angular coefficients  as a function of \ptw in the \Wplnu channel, as obtained from the full fit. The values are given as absolute uncertainty. }
\label{fig:WPLL_pt_Summary}
\end{figure}

\begin{figure}[ht]
\centering
\subfloat[]{\includegraphics[width=\half]{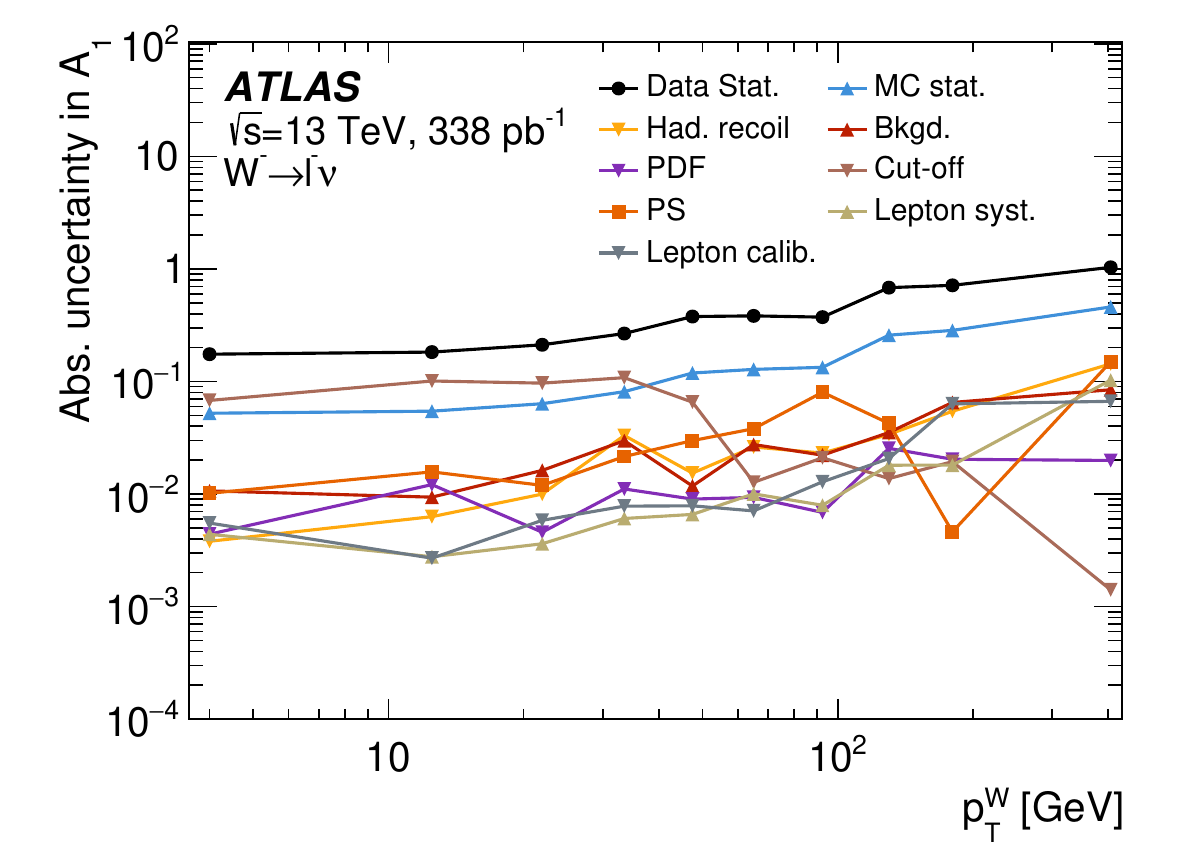}}
\subfloat[]{\includegraphics[width=\half]{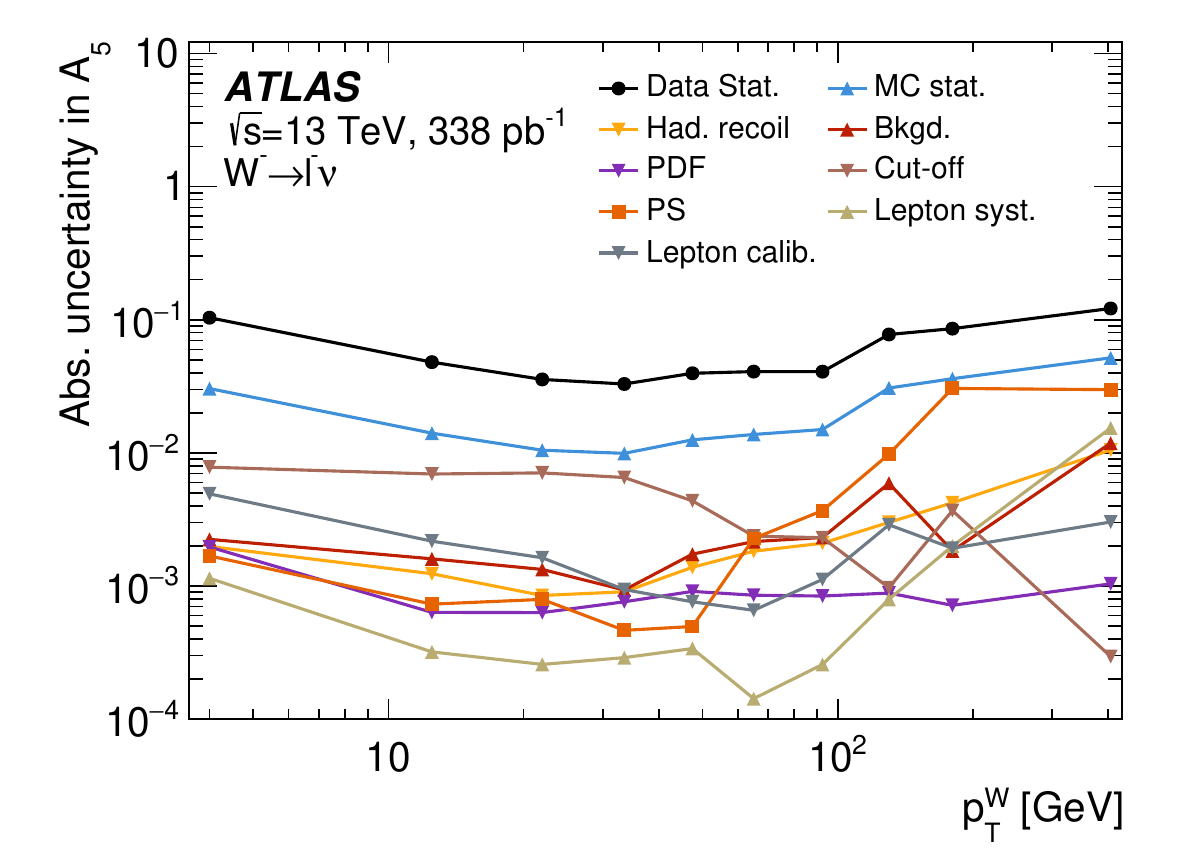}}\\
\subfloat[]{\includegraphics[width=\half]{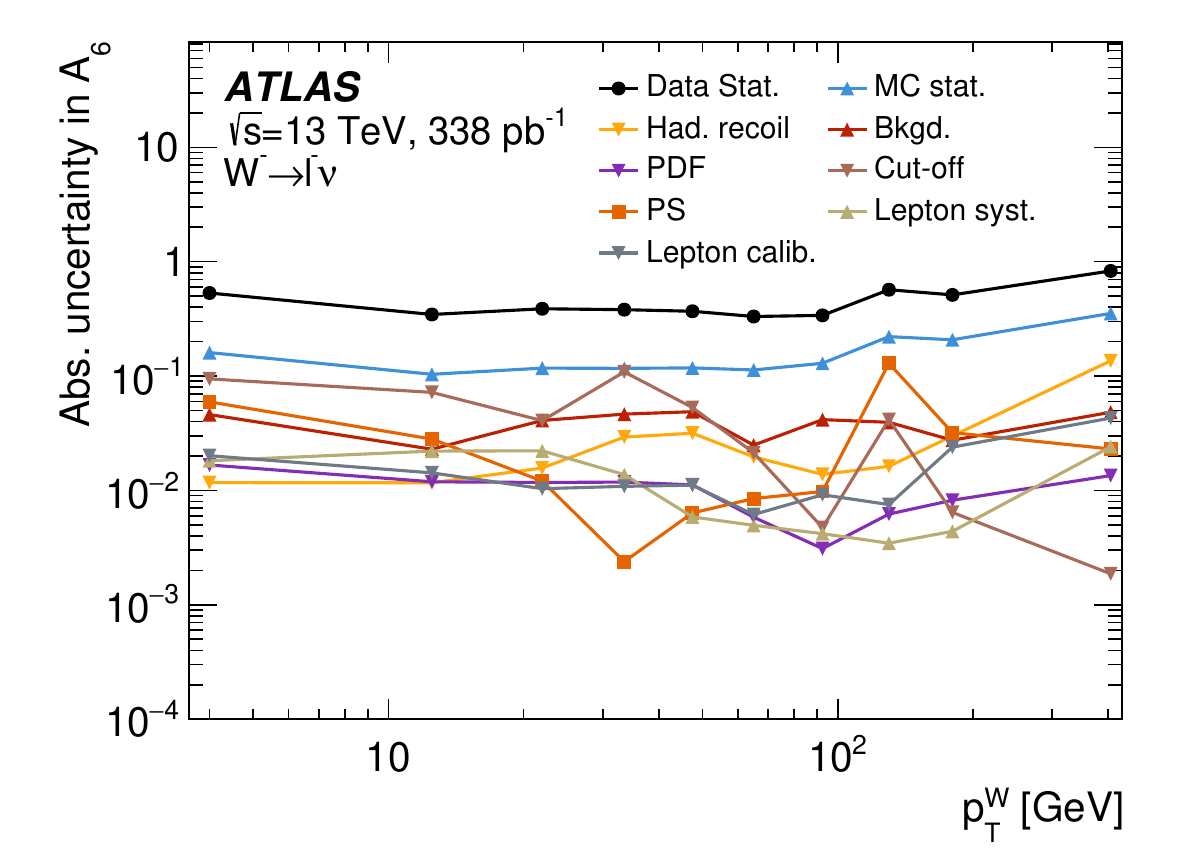}}
\subfloat[]{\includegraphics[width=\half]{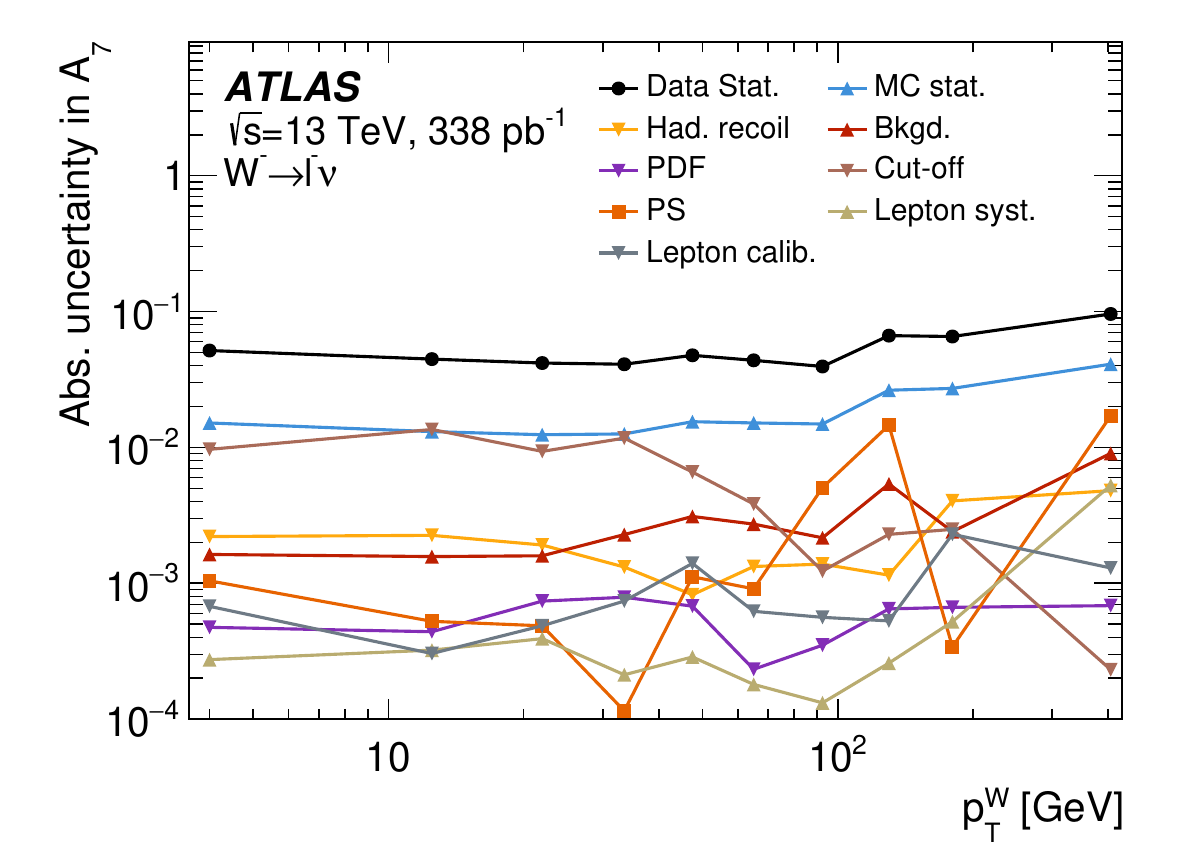}}
\caption{The breakdown into statistical and systematic components of all the uncertainties for the measured (a) \Aone, (b) \Afive, (c) \Asix and (d) \Aseven angular coefficients  as a function of \ptw in the \Wmlnu channel, as obtained from the full fit.  The values are given as absolute uncertainty. }
\label{AUXfig:WMLL_pt_Summary}
\end{figure}

\begin{figure}[ht]
\centering
\subfloat[]{\includegraphics[width=\half]{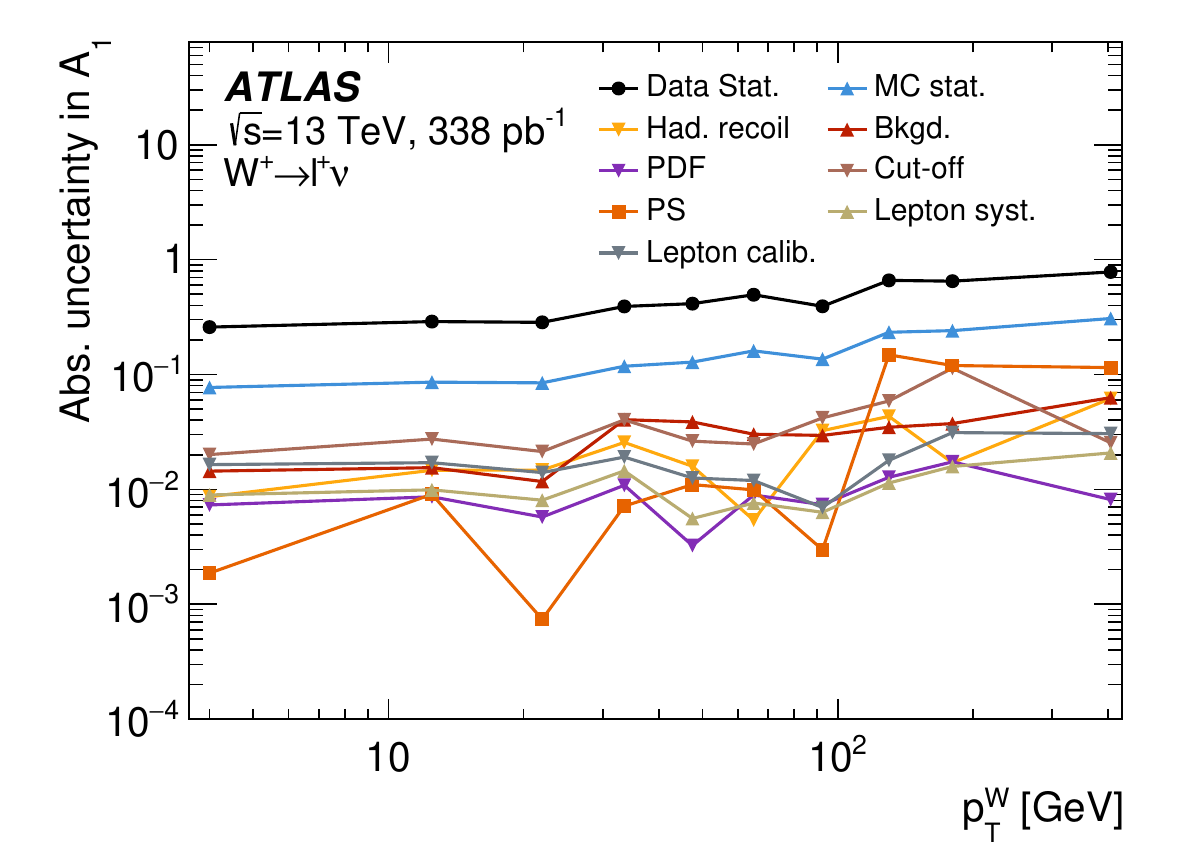}}
\subfloat[]{\includegraphics[width=\half]{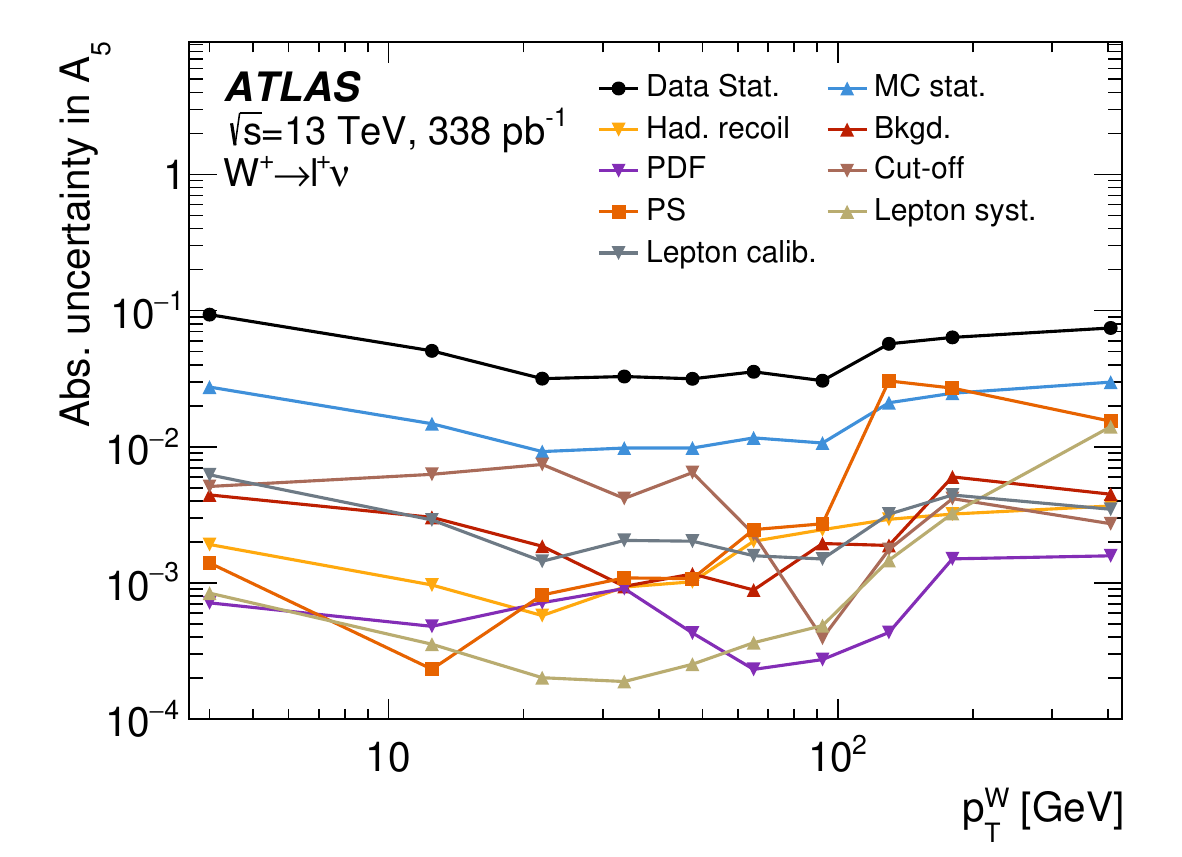}}\\
\subfloat[]{\includegraphics[width=\half]{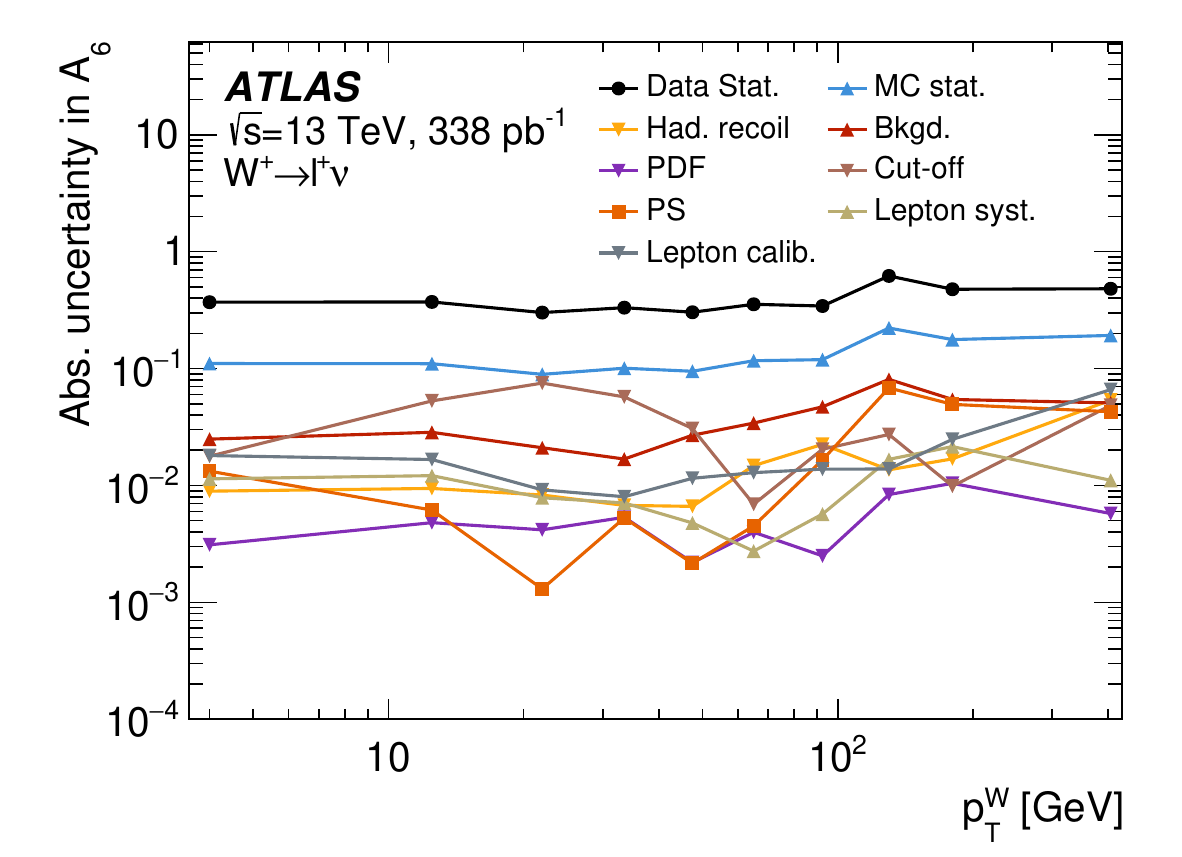}}
\subfloat[]{\includegraphics[width=\half]{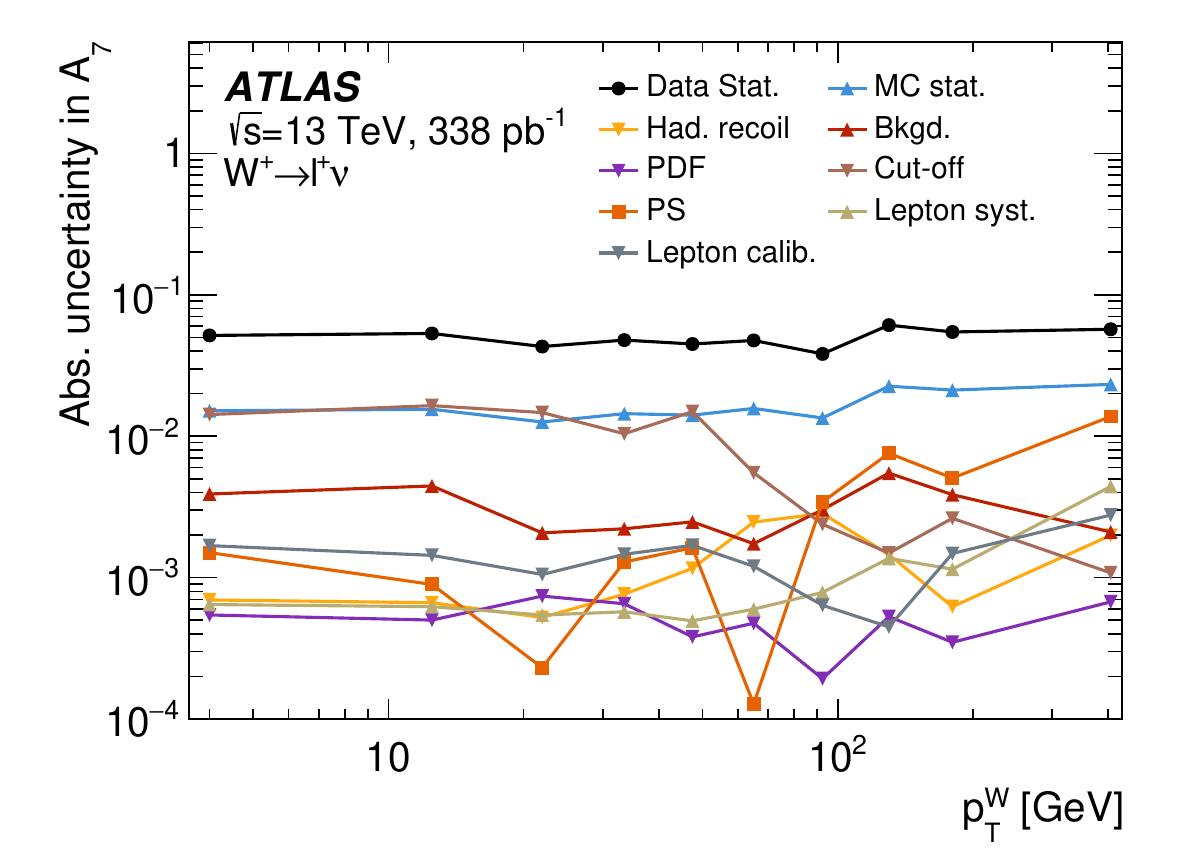}}
\caption{The breakdown into statistical and systematic components of all the uncertainties for the measured  (a) \Aone, (b) \Afive, (c) \Asix and (d) \Aseven angular coefficients  as a function of \ptw in the \Wplnu channel, as obtained from the full fit.  The values are given as absolute uncertainty. }
\label{AUXfig:WPLL_pt_Summary}
\end{figure}

\begin{figure}[!h]
\centering
\subfloat[]{\includegraphics[width=\half]{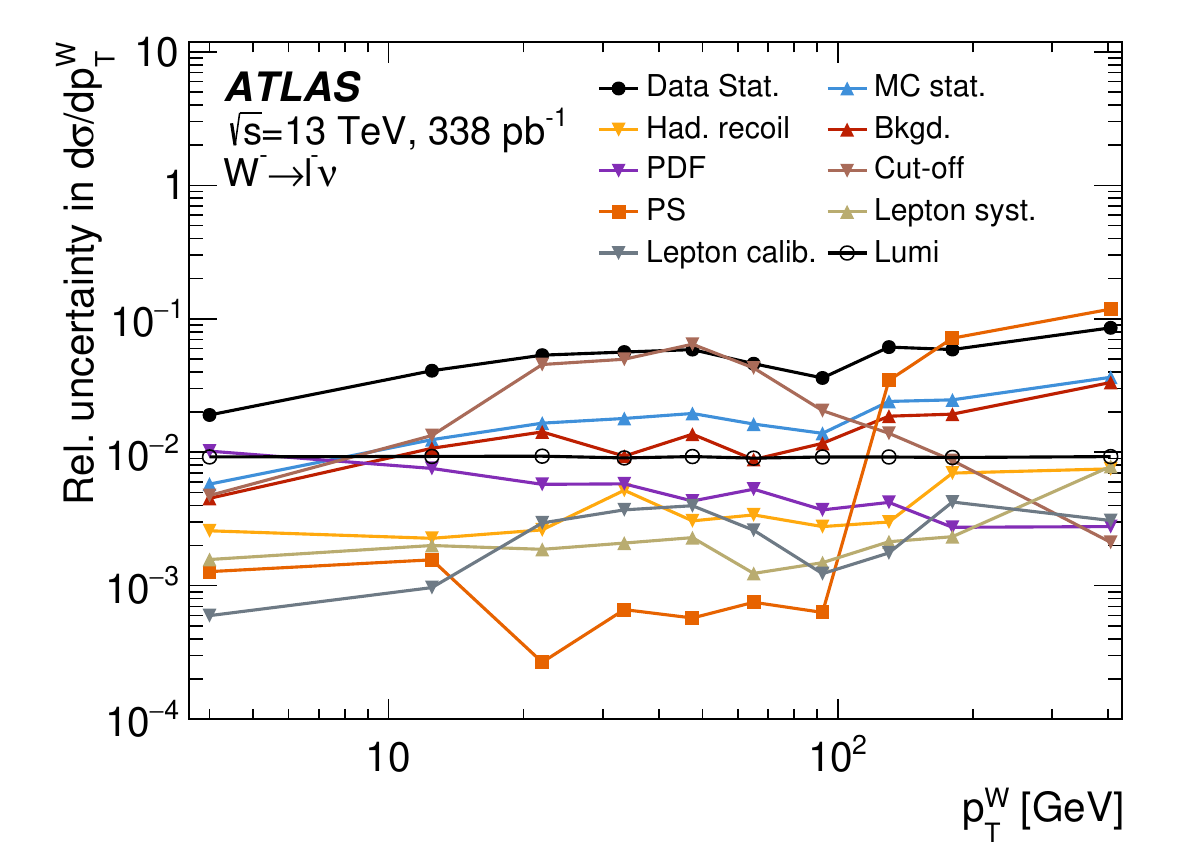}}
\subfloat[]{\includegraphics[width=\half]{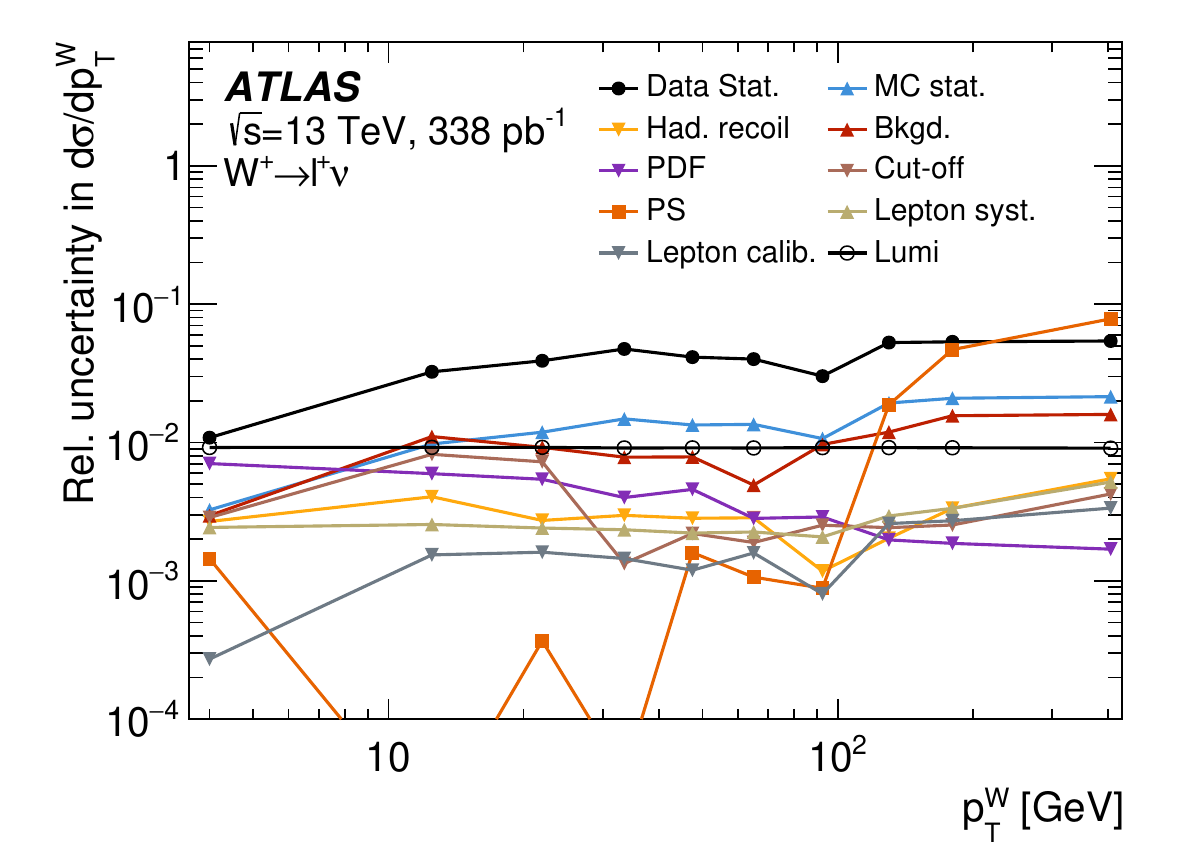}}
\caption{The breakdown into statistical and systematic components of all the uncertainties for the measured \dsdpT in the  (a)  \Wmlnu and (b) \Wplnu channel, as obtained from the full fit. The values are given as relative uncertainty. The uncertainty of $0.92\%$ in the integrated luminosity is included.  }
\label{fig:WMLL_Xs_Summary}
\end{figure}

Several sources of statistical and systematic uncertainty affect the extraction of the angular coefficients and differential cross-sections presented in this paper.
Systematic uncertainties are incorporated into the likelihood fit as constrained nuisance parameters, and their categorisation follows the methodology outlined in Ref.~\cite{STDM-2014-10}.  The overall impact of experimental and theory uncertainties in the values of \Asubrange\ is illustrated in Figures~\ref{fig:WMLL_pt_Summary} and~\ref{fig:WPLL_pt_Summary} for the $W^{-}$ and $W^{+}$ channels respectively, while the corresponding results for \Asamll\ are shown in Figures~\ref{AUXfig:WMLL_pt_Summary} and~\ref{AUXfig:WPLL_pt_Summary}.
The uncertainties in the \Asamll\ coefficients are generally larger than those for \Asubrange, due to their suppressed contribution.
The dependence of the uncertainties in the differential cross-section, \dsdpT, as a function of \ptw, is also shown in Figure~\ref{fig:WMLL_Xs_Summary} separately for the $W^{-}$ and $W^{+}$ channels.
The impact of all uncertainty sources on the final results is evaluated after combining the electron and muon channels (see Section~\ref{sec:Results} for a discussion of their compatibility). A detailed discussion of these uncertainties is presented in this section.

The dominant contribution arises from the statistical uncertainty associated with the limited size of the dataset.
Although the harmonic polynomials used in this methodology are completely orthogonal in the full phase space, detector-level resolution and acceptance effects lead to non-zero correlations between the measured coefficients.  Due to the limited resolution and the intrinsic twofold ambiguity in the reconstruction of the neutrino momentum, the templates used to extract the cross-sections and angular distributions measured in bins of (reconstructed) $\ptln$ receive contributions from a broad range of (true) $\ptw$ values.
This migration leads to correlations between the measured coefficients which increase their statistical uncertainties.
For the cross-section measurements, the number of data events available is the largest source of uncertainty at large \ptw where data statistics are limited, but otherwise the dominant source of uncertainty arises from bin-to-bin migration, making it the leading contributor to the overall statistical uncertainty.

Monte Carlo simulation is used to build the folding templates for the likelihood fit, as described in Section~\ref{sec:methodology}, and for modelling event migration between truth and reconstructed bins.
The statistical uncertainty from MC simulation is treated as uncorrelated between each bin of
the three-dimensional (\ptln, \coscs, \phics) distribution and represents a non-negligible, though subdominant, contribution towards the total uncertainty.
To ensure the statistical stability of the folding procedure, a threshold is applied such that templates contributing less than $2\%$ of the expected number of events in a given truth bin are excluded. This \emph{cut-off} value is determined by generating pseudo-data through random fluctuations of the expected MC yields and performing a series of likelihood fits with varying threshold values; the optimal cut-off is defined as the point where the variation in the $\chi^2$ between successive fits is minimised.
The systematic uncertainty associated with this choice is quantified separately for the $W^{-}$ and $W^{+}$ channels, by varying  the threshold up and down by $50\%$ . The lower variation ($1\%$) corresponds to the minimum value of the cut-off required to remove potential biases in the folded results, while the upper variation ($3\%$) accounts for additional uncertainties arising from the reduced event information in the folding migrations. Both the direct impact on the fit results and the induced uncertainty due to event migration are taken into account. This uncertainty is added post-fit to the total systematic uncertainty.
The $W^{-}$ channel exhibits larger sensitivity to this variation, resulting in a systematic uncertainty that becomes comparable to the statistical uncertainty in the cross-section measurement at intermediate values of \ptw.

Experimental systematic uncertainties affect both the \coscs  and \phics  distributions used to construct the templates and the event weights applied to the simulation.
The main experimental uncertainties affecting the measurements arise from the hadronic recoil calibration.   The uncertainties in the scale and resolution of the hadronic recoil are discussed in greater detail in Ref.~\cite{STDM-2018-17}.
In this analysis, the hadronic recoil systematic uncertainty remains below $1\%$ across the entire \ptw range.
Lepton-related systematic uncertainties due to energy/momentum scale and resolution as well as reconstruction and selection efficiency corrections are propagated considering the statistical and systematic uncertainties in the procedures, as detailed in Ref.~\cite{STDM-2018-17}. They have only a small impact on both the \Ai and cross-section measurements.

Uncertainties arising from the use of MC simulation to estimate the backgrounds are obtained by independently varying their corresponding cross-sections.
An additional uncertainty in the $t\overline{t}$ background is evaluated using alternative MC samples, where the modelling parameter associated with high-$p_\mathrm{T}$ radiation is varied. This variation impacts both the normalisation and the shape of the $t\overline{t}$ background contribution.
Systematic uncertainties affecting both the yield and shape of the multijet background are evaluated separately for $W^{-}$ and $W^{+}$ candidate events, as detailed in Section~\ref{sec:background}, and are treated as uncorrelated between the electron and muon channels. This choice is motivated by the use of independent MC samples to correct the acceptance of the angular distribution templates in $\pt^{\ell\nu}$ bins for each lepton flavour.
The overall impact of background-related uncertainties in the $\Ai$ coefficients is subdominant but for the differential cross-sections, the uncertainty reaches up to 3\% for $W^{-}$ and 2\% for $W^{+}$ at high \ptw.

The overall uncertainty in the integrated luminosity of the 13\,TeV low pile-up datasets is $0.92\%$~\cite{DAPR-2021-01}. The integrated luminosity uncertainty affects the absolute cross-section measurements, as shown in Figure~\ref{fig:WMLL_Xs_Summary}, but it is negligible in the \Ai measurements because the coefficients are effectively cross-section ratios.

Theory systematic uncertainties due to the choice of parton distribution functions, parton shower modelling and the underlying event are also considered. Other theory uncertainties due to QCD renormalisation and factorisation scales, MC event generator modelling, and QED/EW (electroweak) corrections are not considered in this measurement as they were found to contribute negligibly in the previous, more precise, ATLAS measurement of \Zboson angular coefficients~\cite{STDM-2014-10}.
The PDF-related uncertainties are evaluated using the CT10 eigenvector variations after rescaling them by a factor of 1.64 to obtain 68\% confidence level (CL) variations. The resulting impact on the angular coefficients is subdominant, and the effect on the differential cross-section remains below $\sim1\%$.
Systematic uncertainties related to the underlying event and parton shower modelling (PS) are evaluated using a dedicated \Sherpa~sample reweighted to match the $W$-boson kinematics of the nominal \Powheg\ prediction. Folding templates derived from the \Sherpa\ sample are used to extract the angular coefficients and differential cross-sections, and the absolute difference relative to the nominal result defines the uncertainty. This uncertainty is evaluated per channel and included post-fit in the total systematic uncertainty. The impact is generally subdominant, except in the highest $p_{T}^{\ell\nu}$ bin, where it becomes comparable to or exceeds the statistical uncertainty.


%

\section{Results}%
\label{sec:Results}

\begin{figure}[ht]
\centering
\subfloat[]{\includegraphics[width=\half]{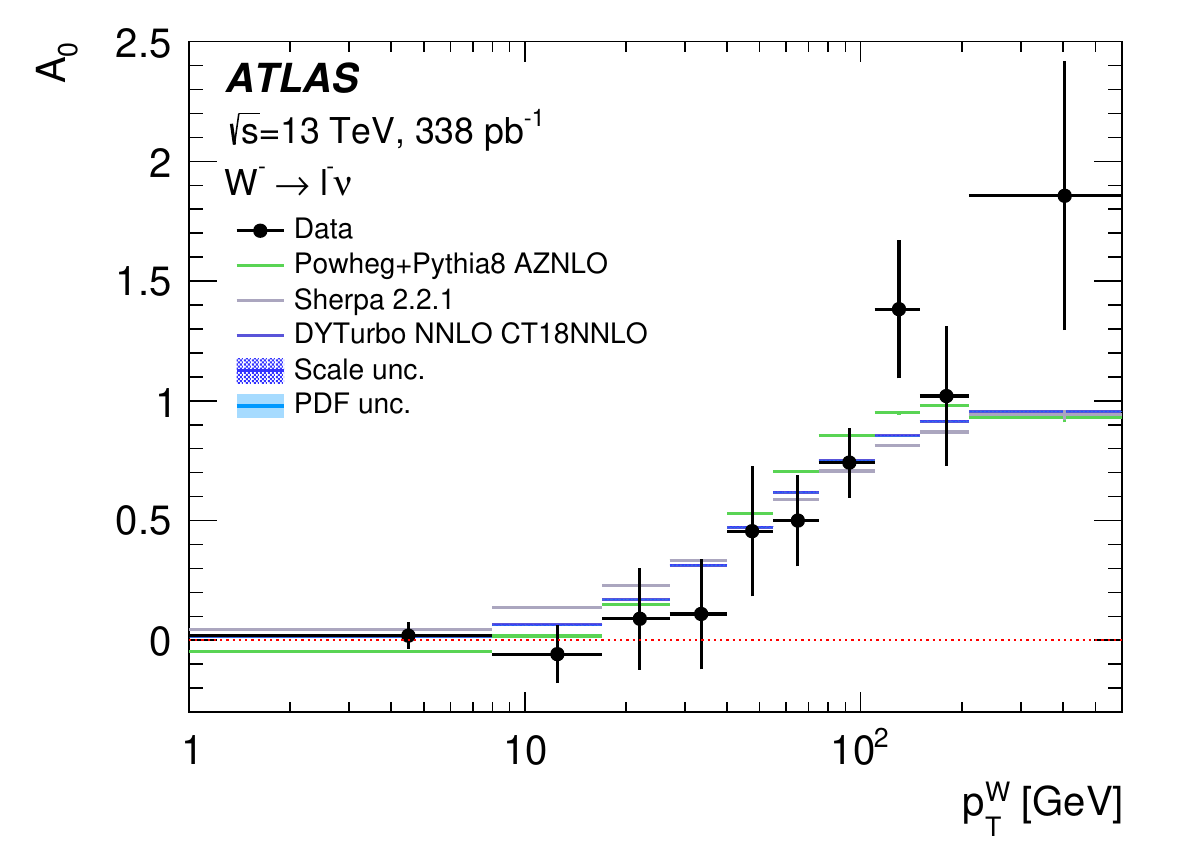}}
\subfloat[]{\includegraphics[width=\half]{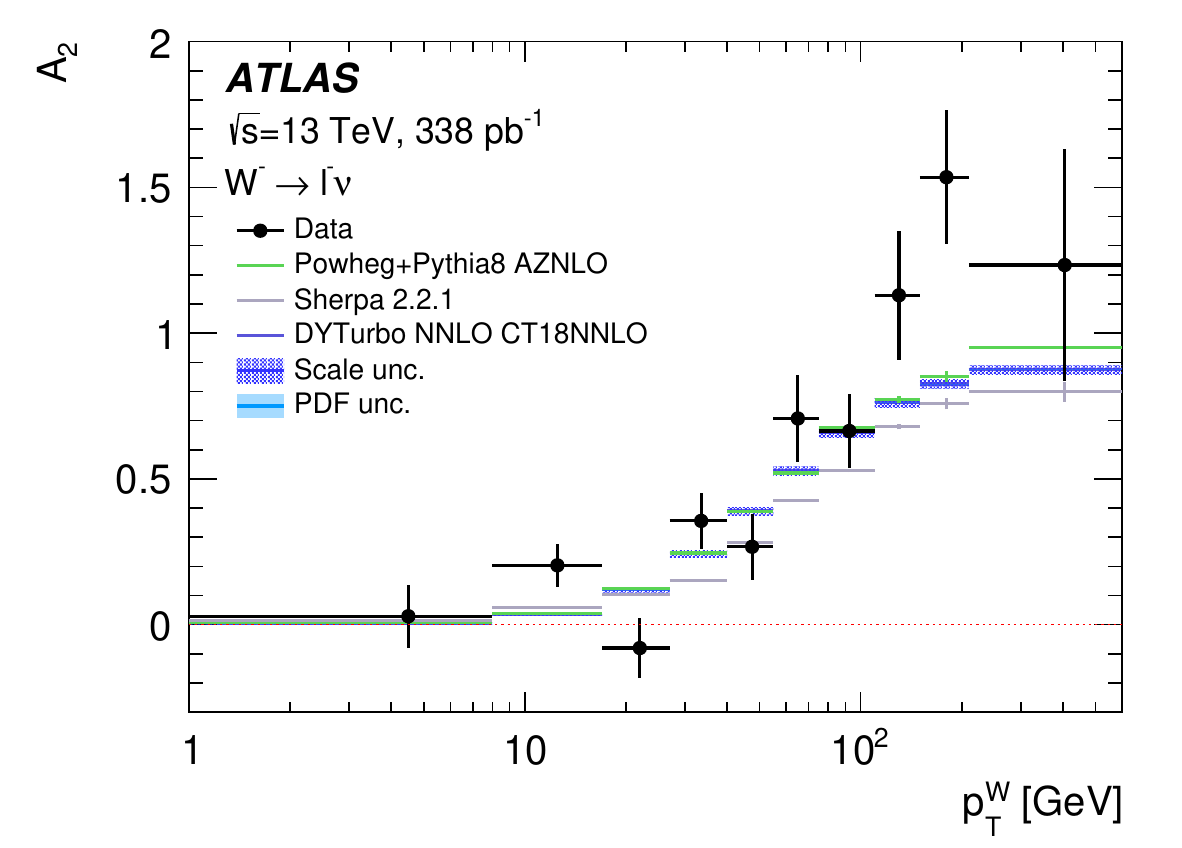}}\\
\subfloat[]{\includegraphics[width=\half]{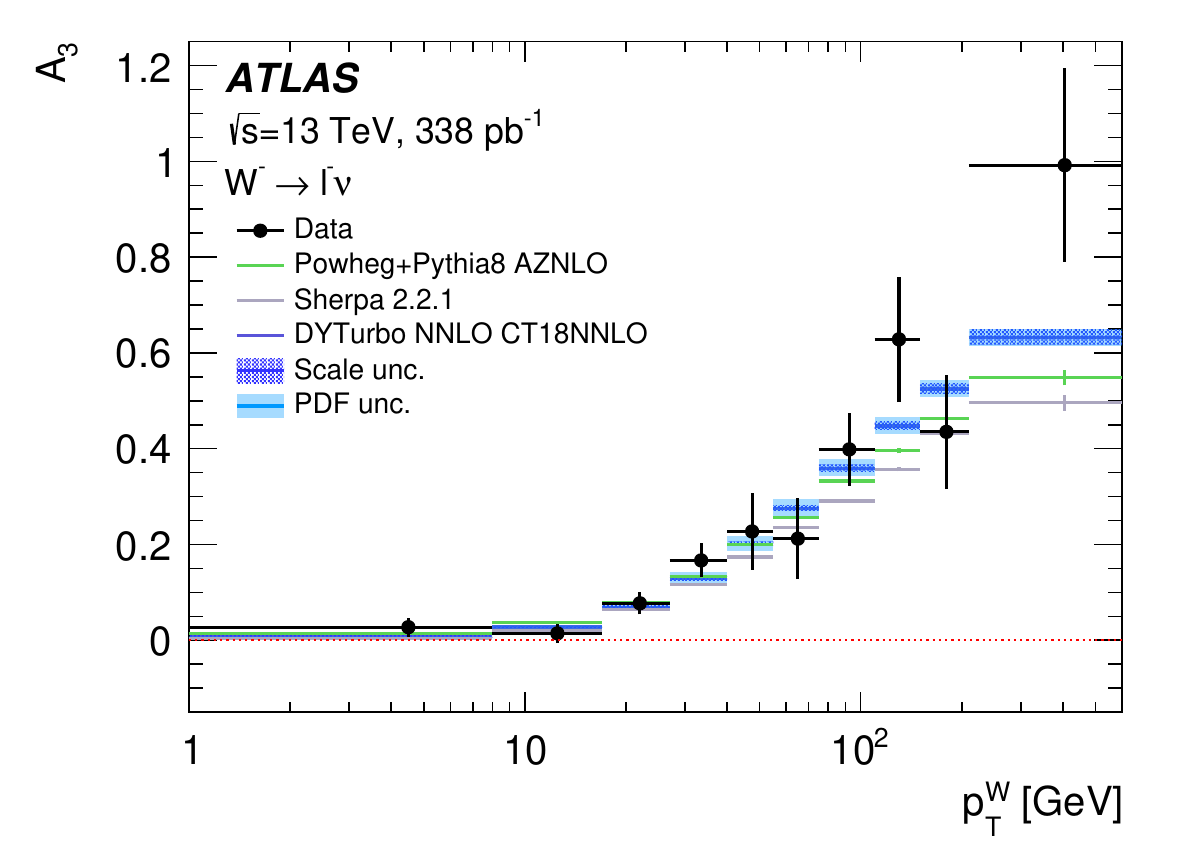}}
\subfloat[]{\includegraphics[width=\half]{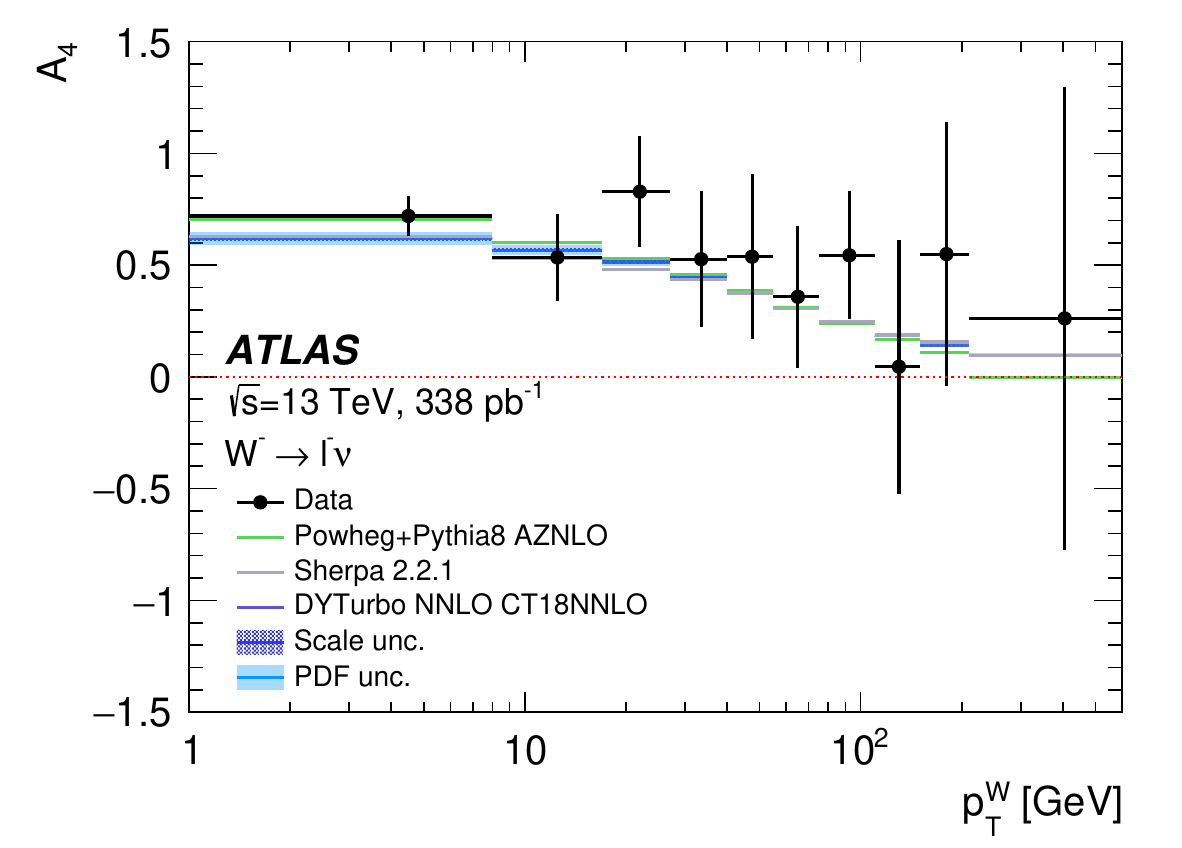}}
\caption{The measured (a) \Azero, (b) \Atwo, (c) \Athree and (d) \Afour coefficients extracted for the \Wmlnu channel as a function of \ptw with comparison to the predictions from \textsc{DYTurbo}, \Powheg, and \Sherpa. The errors on the data points represent the total uncertainty in the measurement. Theory uncertainties in the  \textsc{DYTurbo}  predictions include variations of the renormalisation and factorisation scales, as well as PDF uncertainties for the CT18NNLO PDF set. }
\label{fig:WM_pt}.
\end{figure}

\begin{figure}[ht]
\centering
\subfloat[]{\includegraphics[width=\half]{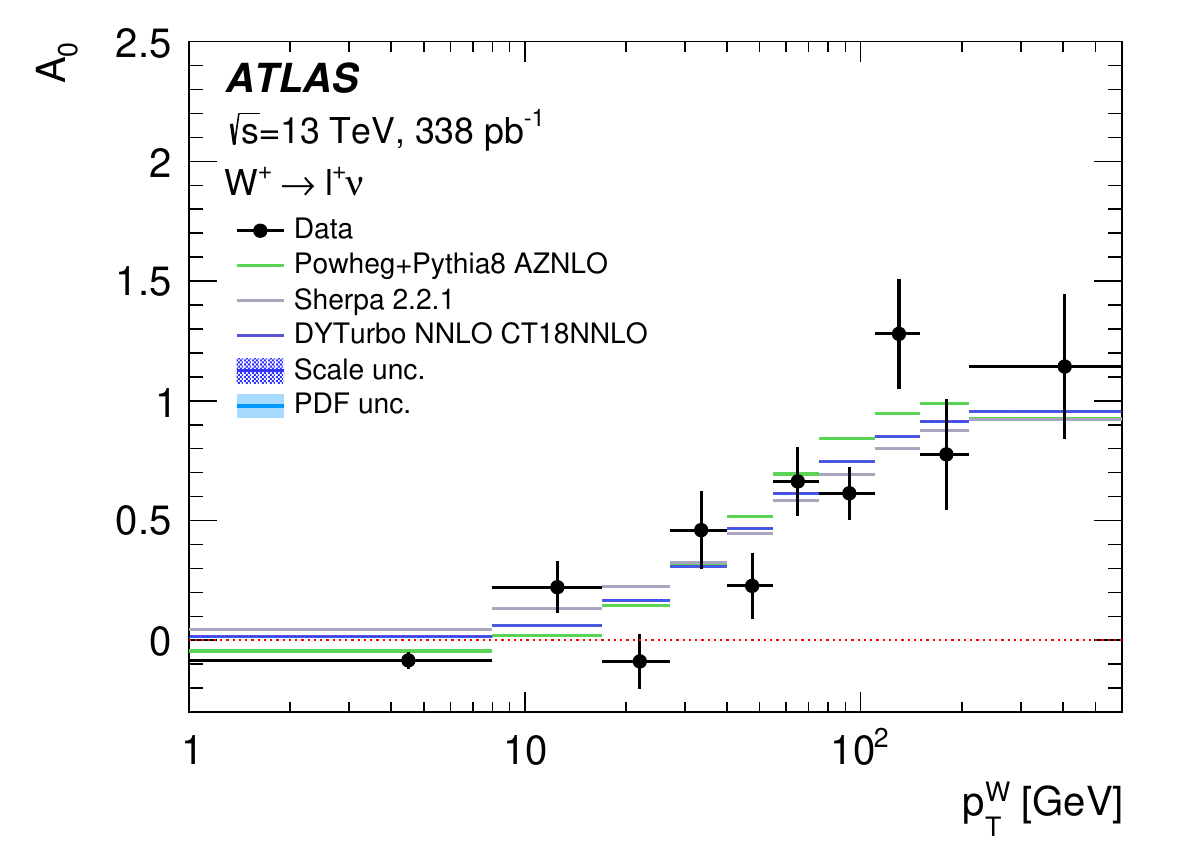}}
\subfloat[]{\includegraphics[width=\half]{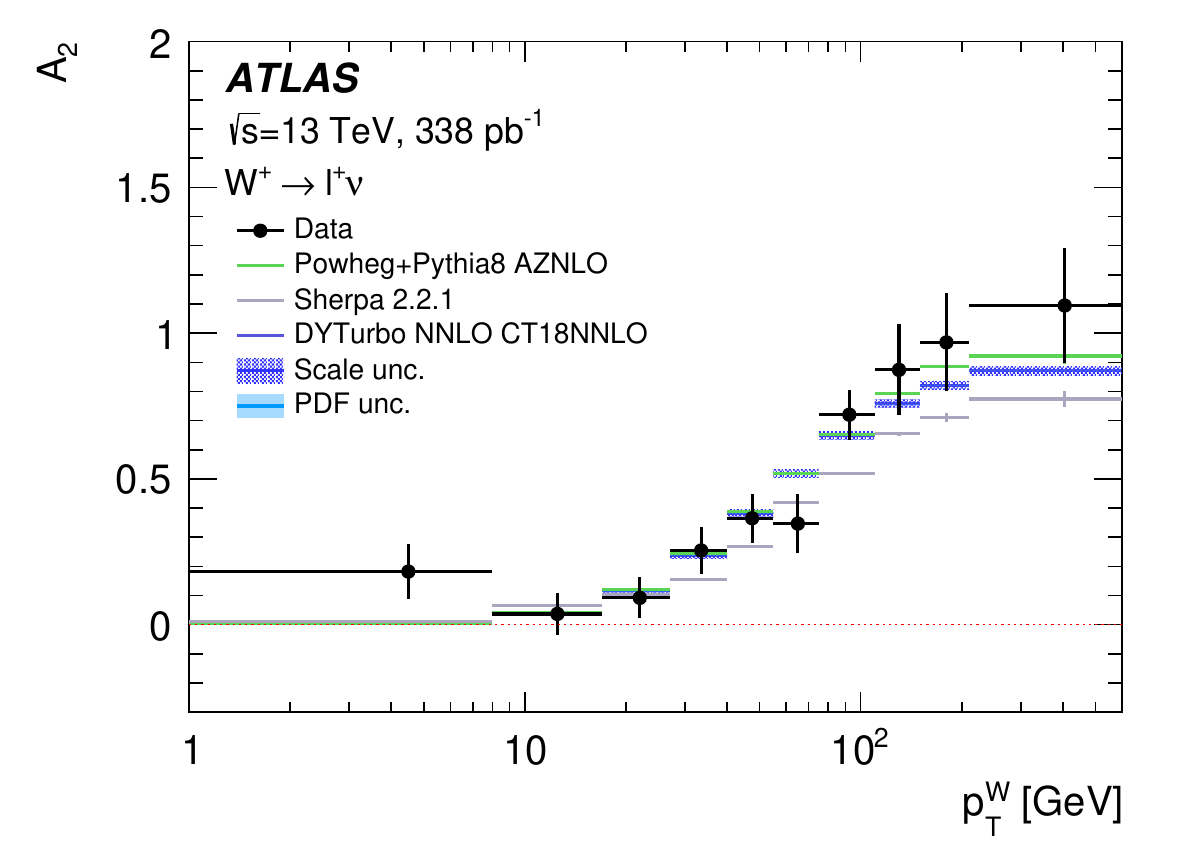}}\\
\subfloat[]{\includegraphics[width=\half]{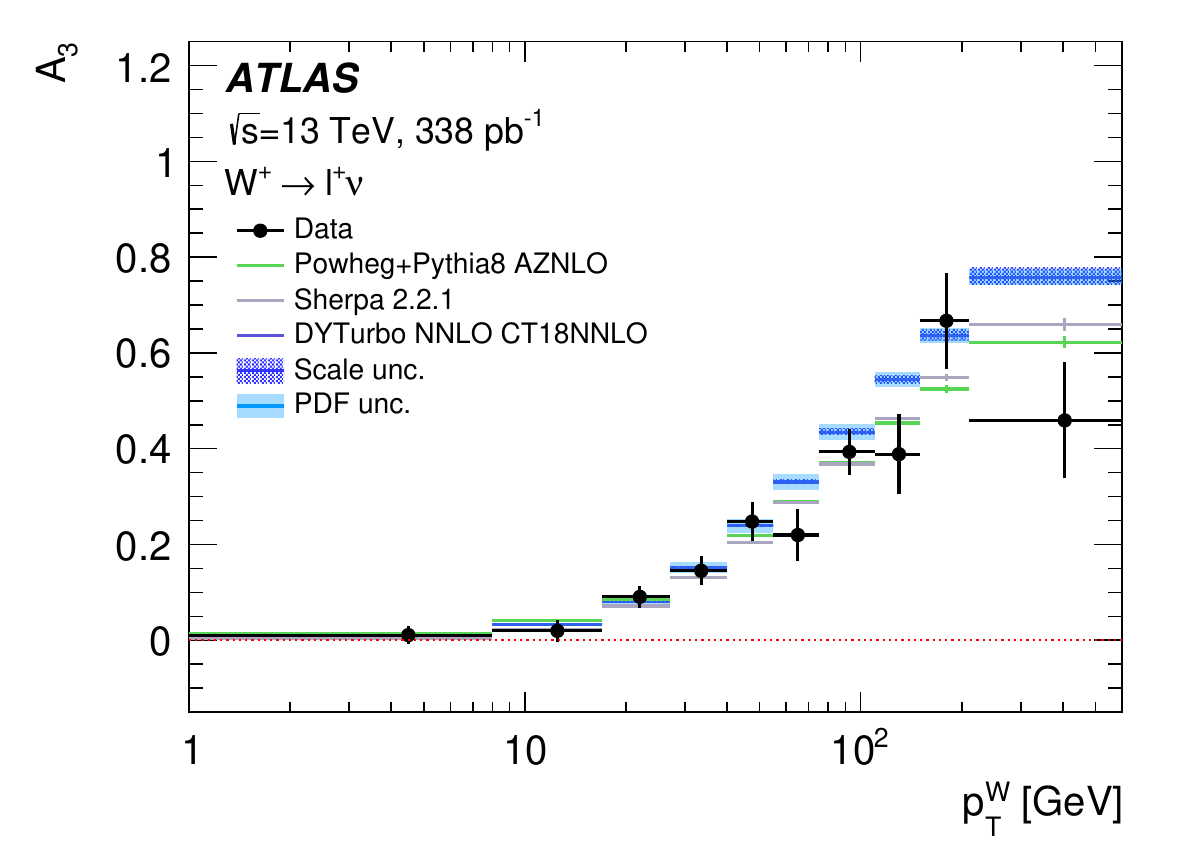}}
\subfloat[]{\includegraphics[width=\half]{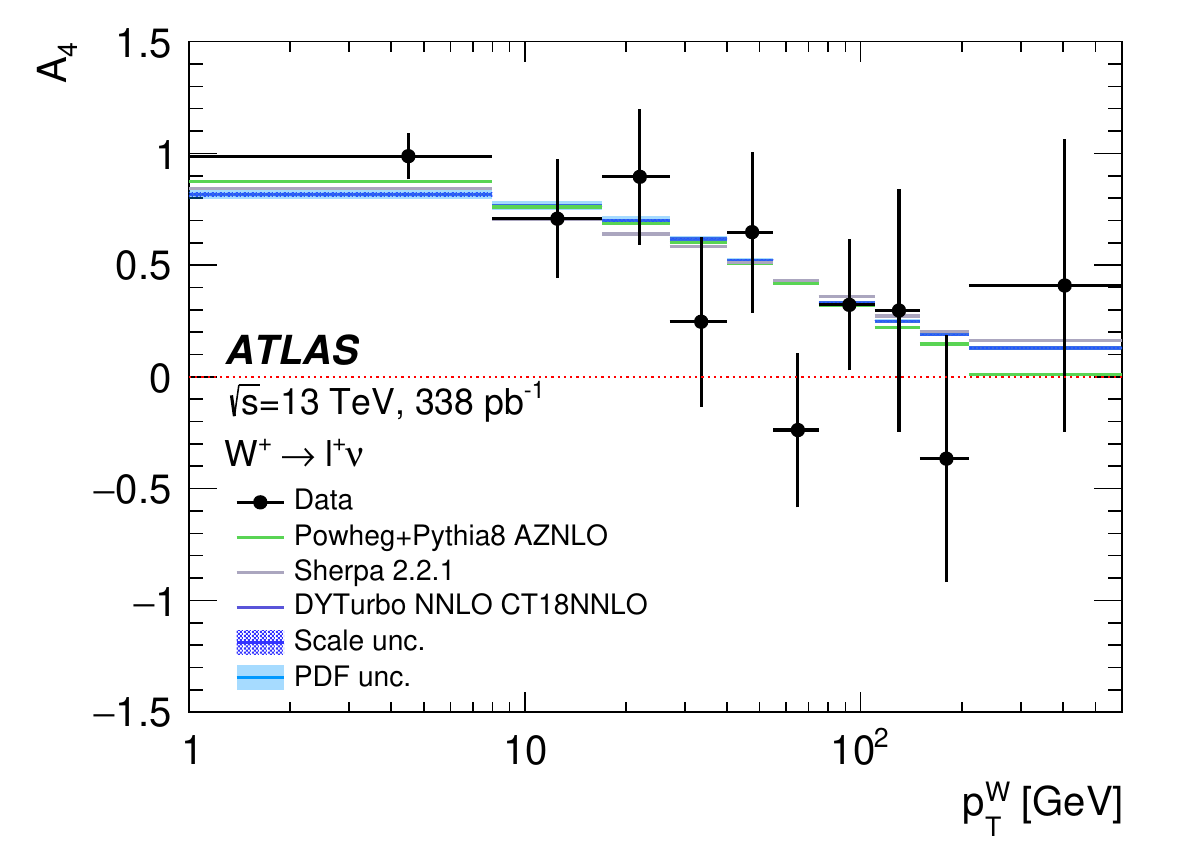}}
\caption{The measured (a) \Azero, (b) \Atwo, (c) \Athree and (d) \Afour coefficients extracted for the  \Wplnu channel as a function of \ptw with comparison to the predictions from \textsc{DYTurbo}, \Powheg, and \Sherpa. The errors on the data points represent the total uncertainty in the measurement. Theory uncertainties in the  \textsc{DYTurbo}  predictions include variations of the renormalisation and factorisation scales, as well as PDF uncertainties for the CT18NNLO PDF set. }
\label{fig:WP_pt}
\end{figure}

\begin{figure}[ht]
\centering
\subfloat[]{\includegraphics[width=\half]{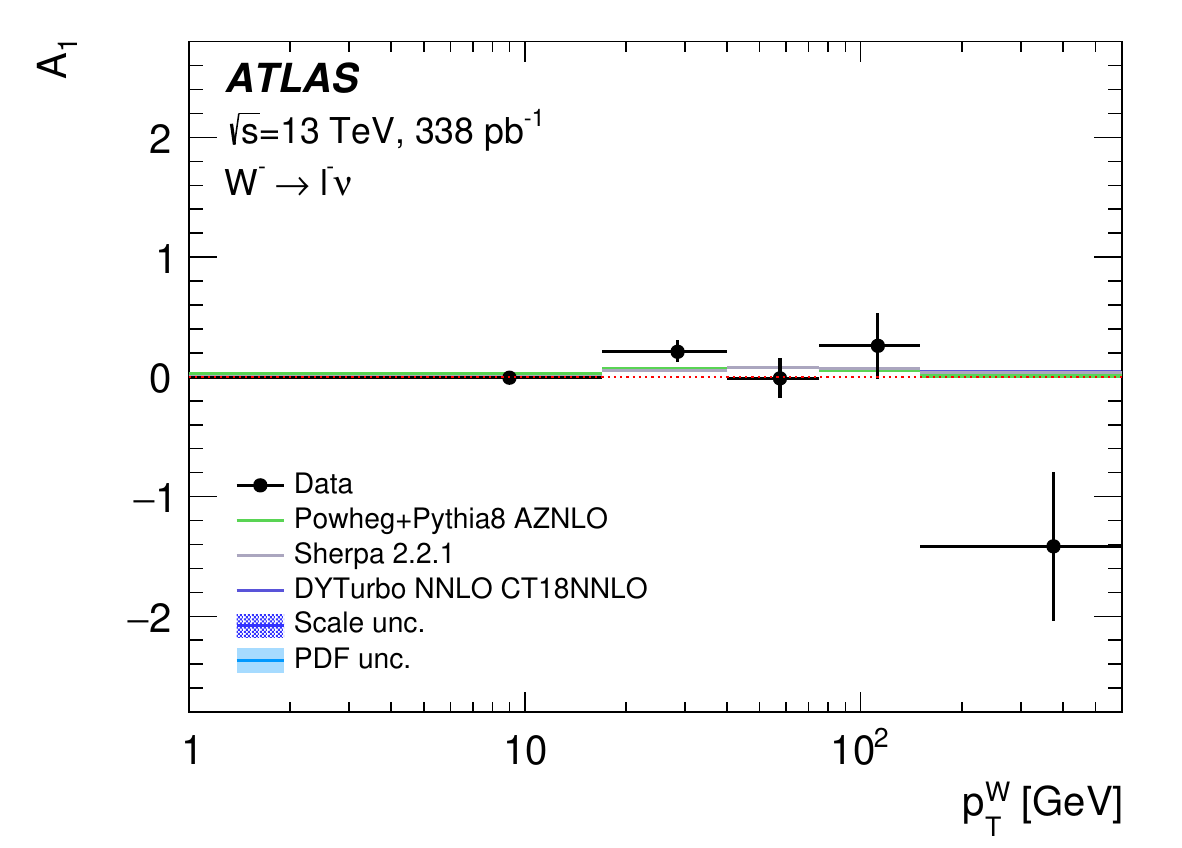}}
\subfloat[]{\includegraphics[width=\half]{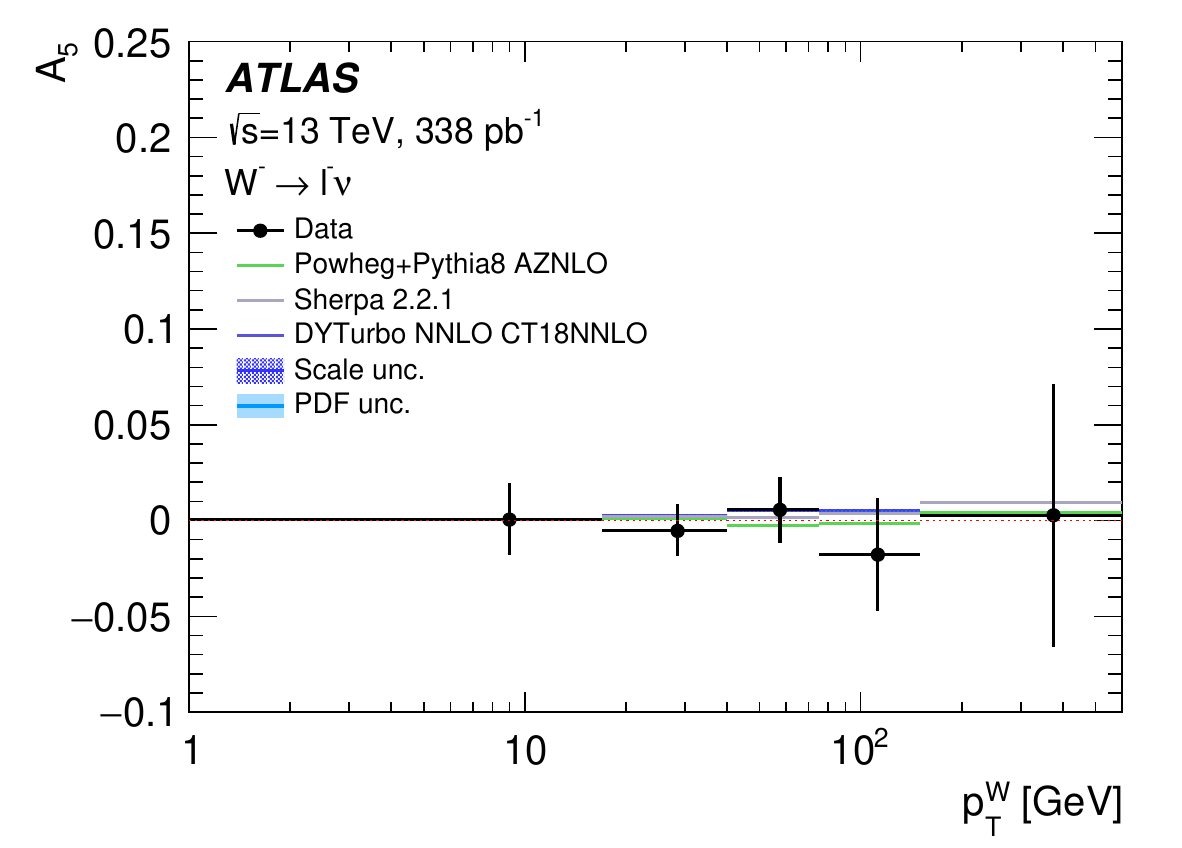}}\\
\subfloat[]{\includegraphics[width=\half]{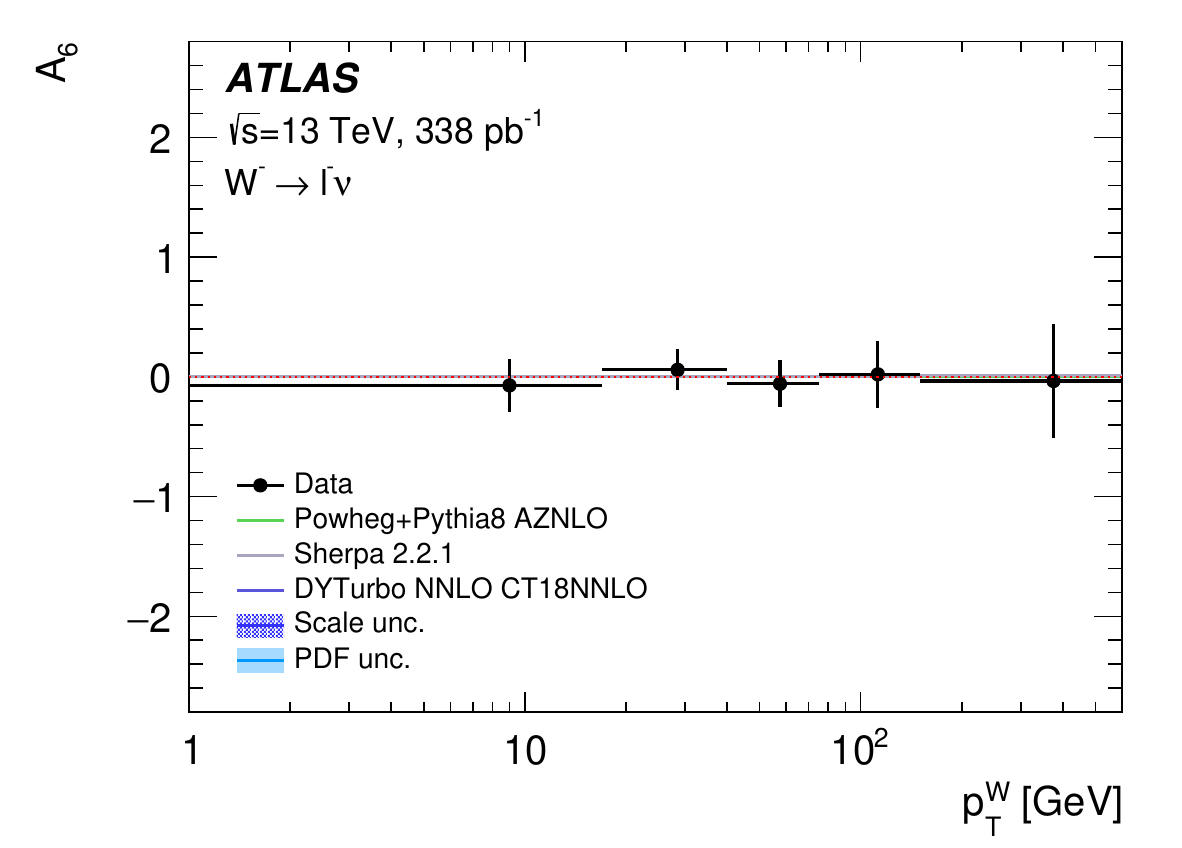}}
\subfloat[]{\includegraphics[width=\half]{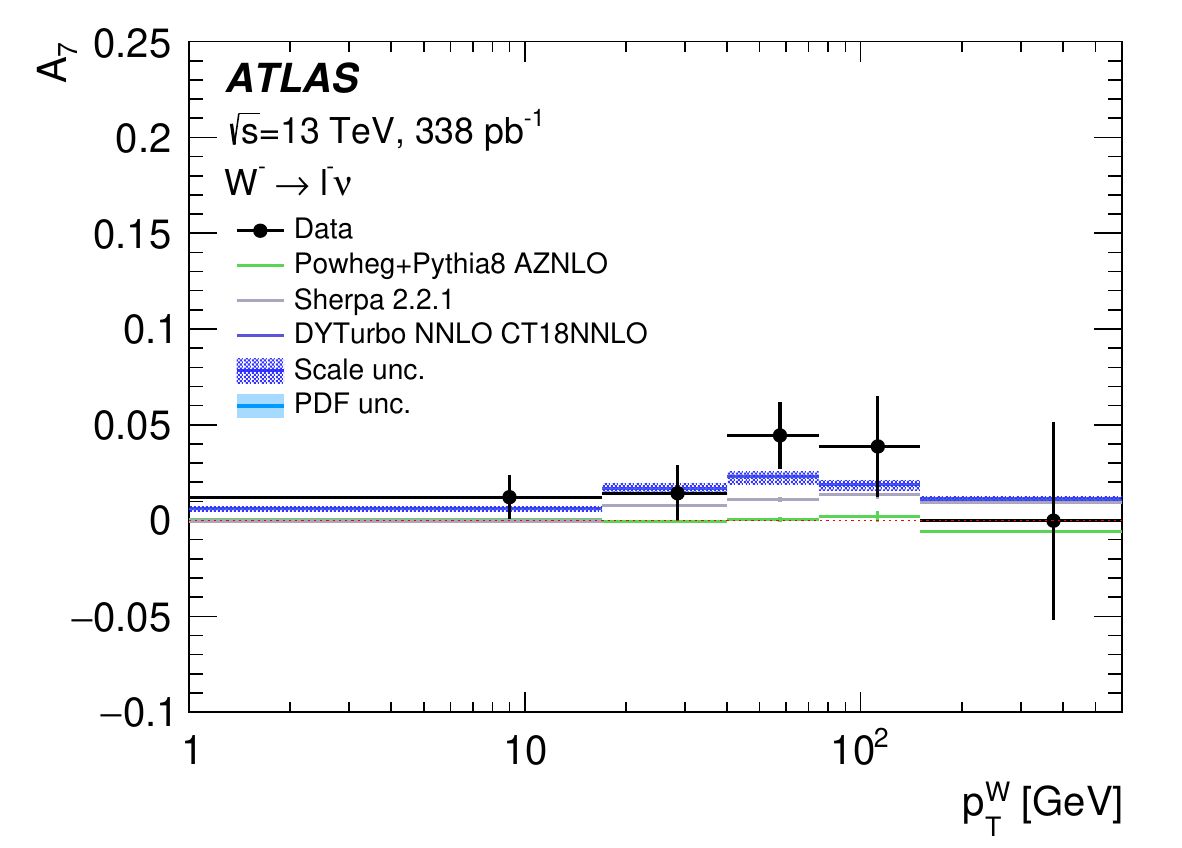}}
\caption{The (a) \Aone, (b) \Afive, (c) \Asix and (d) \Aseven coefficients extracted for the channel \Wmlnu as a function of \ptw with comparison to the predictions from \textsc{DYTurbo}, \Powheg, and \Sherpa.  The errors represent the total uncertainty in the measurement. As described in the text, a post-fit rebinning is applied in which two adjacent bins are averaged, effectively reducing the number of bins by a factor of two to mitigate strong bin-to-bin anticorrelations induced by statistical fluctuations.
The error bars for the predictions show the total uncertainty for  \textsc{DYTurbo}, but only the statistical uncertainties for    \Powheg+\Pythia~and \Sherpa~MC generators.}
\label{AUXfig:WM_pt}
\end{figure}

\begin{figure}[ht]
\centering
\subfloat[]{\includegraphics[width=\half]{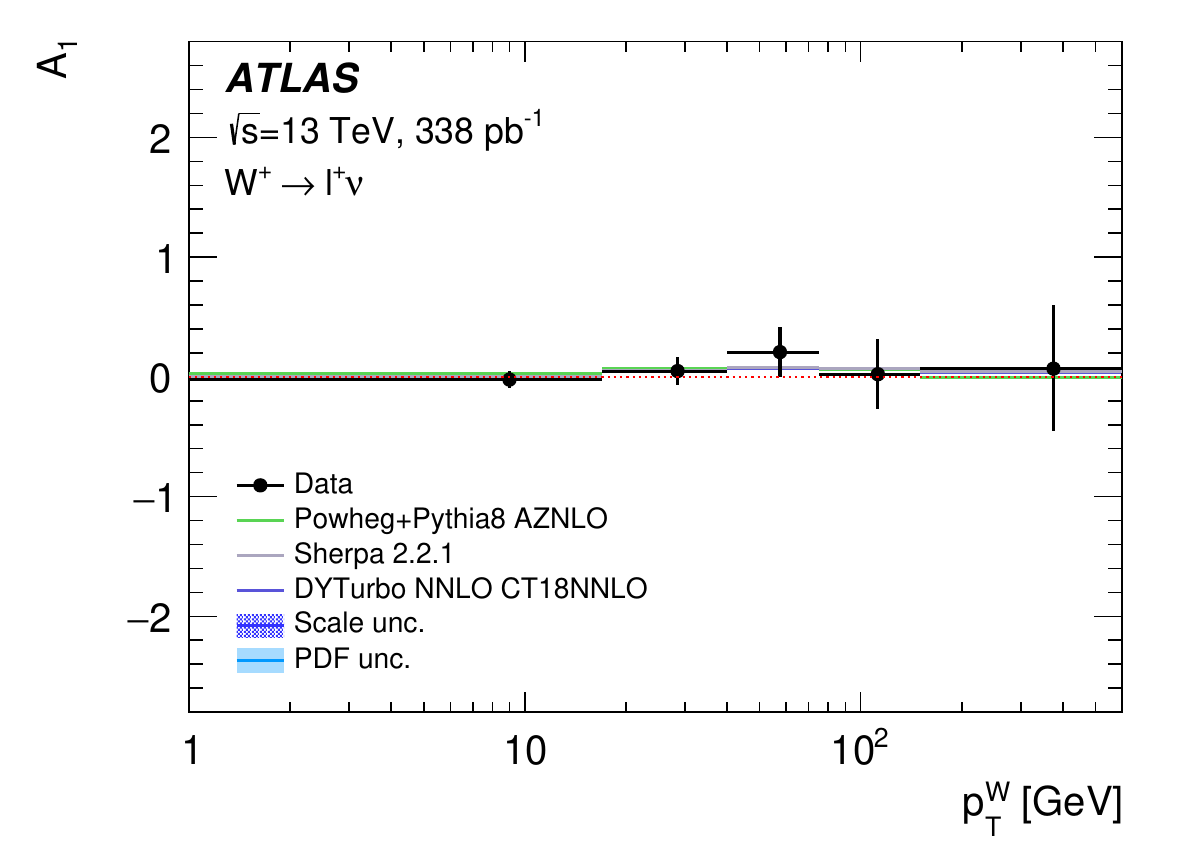}}
\subfloat[]{\includegraphics[width=\half]{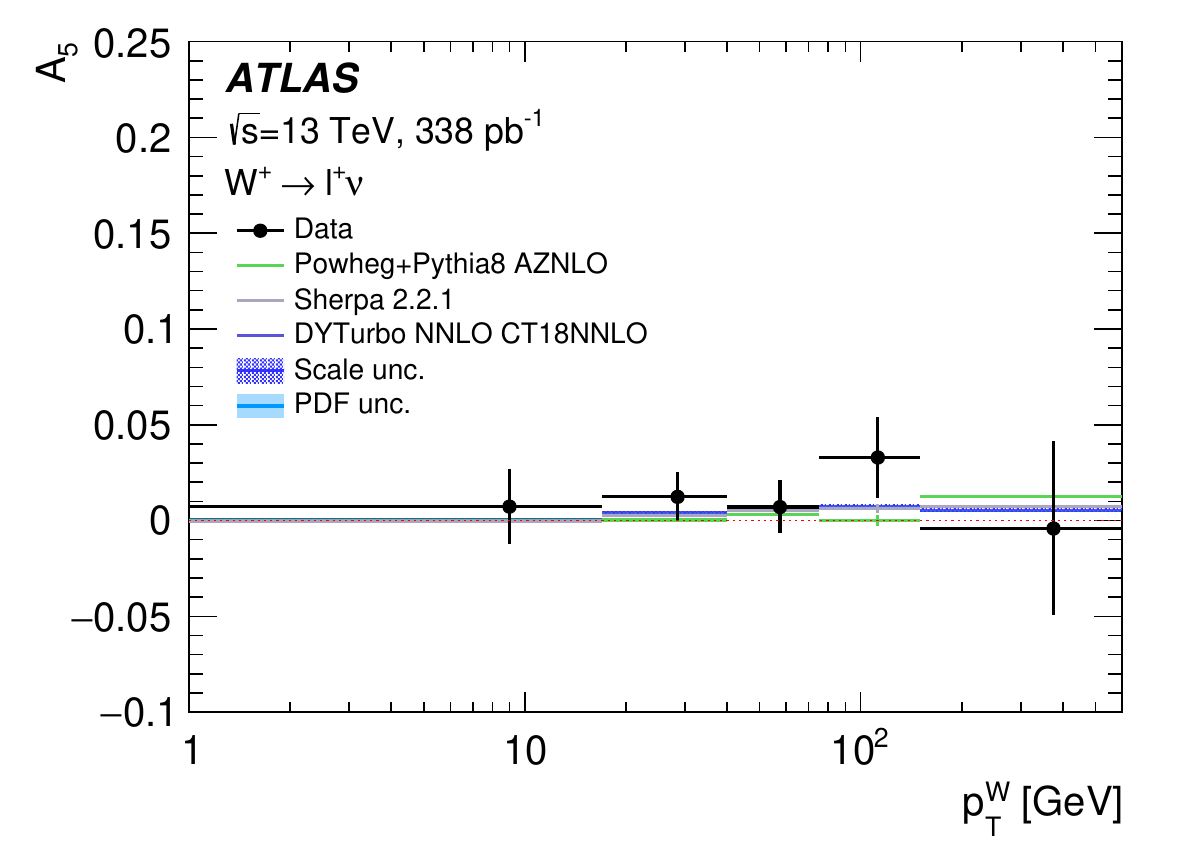}}\\
\subfloat[]{\includegraphics[width=\half]{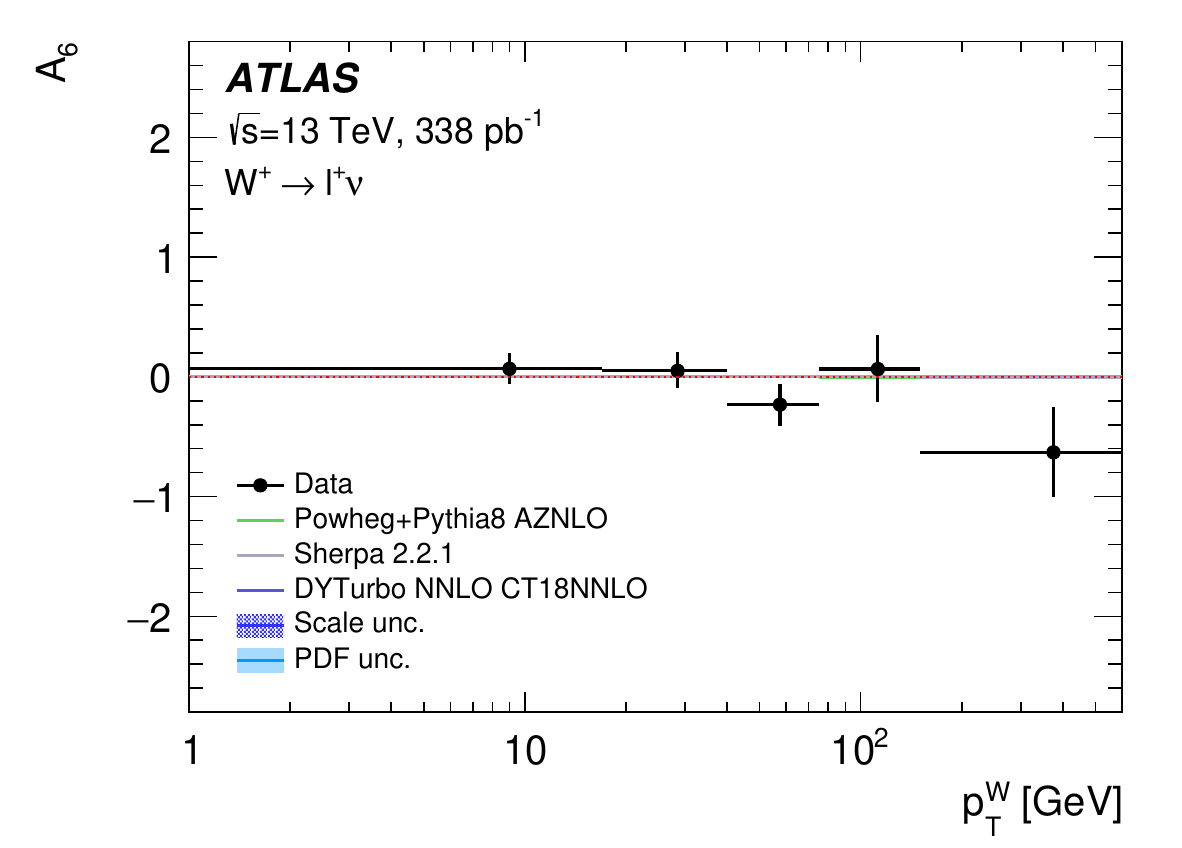}}
\subfloat[]{\includegraphics[width=\half]{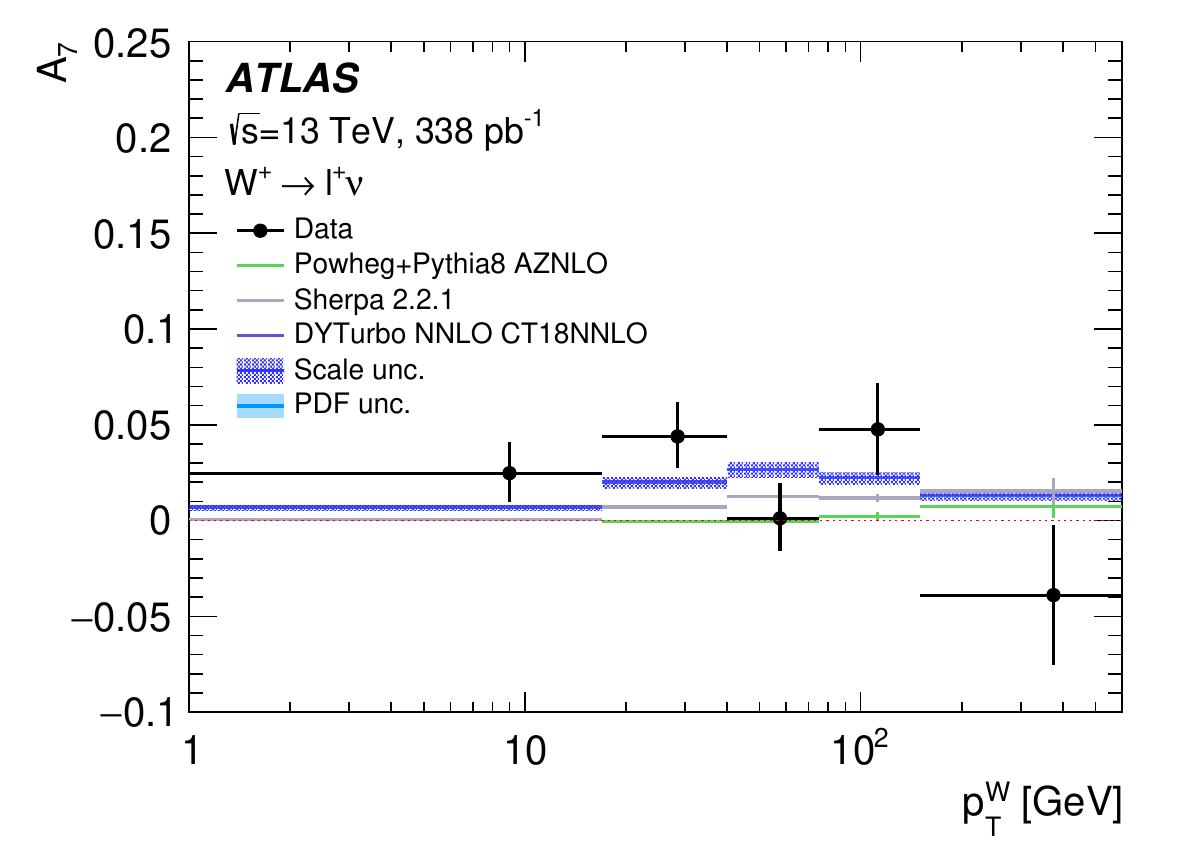}}
\caption{The (a) \Aone, (b) \Afive, (c) \Asix and (d) \Aseven coefficients extracted for the channel \Wplnu as a function of \ptw with comparison to the predictions from \textsc{DYTurbo}, \Powheg, and \Sherpa. The errors represent the total uncertainty in the measurement. As described in the text, a post-fit rebinning is applied in which two adjacent bins are averaged, effectively reducing the number of bins by a factor of two to mitigate strong bin-to-bin anticorrelations induced by statistical fluctuations. The error bars for the predictions show the total uncertainty for  \textsc{DYTurbo}, but only the statistical uncertainties for    \Powheg+\Pythia~and \Sherpa~MC generators. }
\label{AUXfig:WP_pt}
\end{figure}

\begin{figure}[ht]
\centering
\subfloat[]{\includegraphics[width=\half]{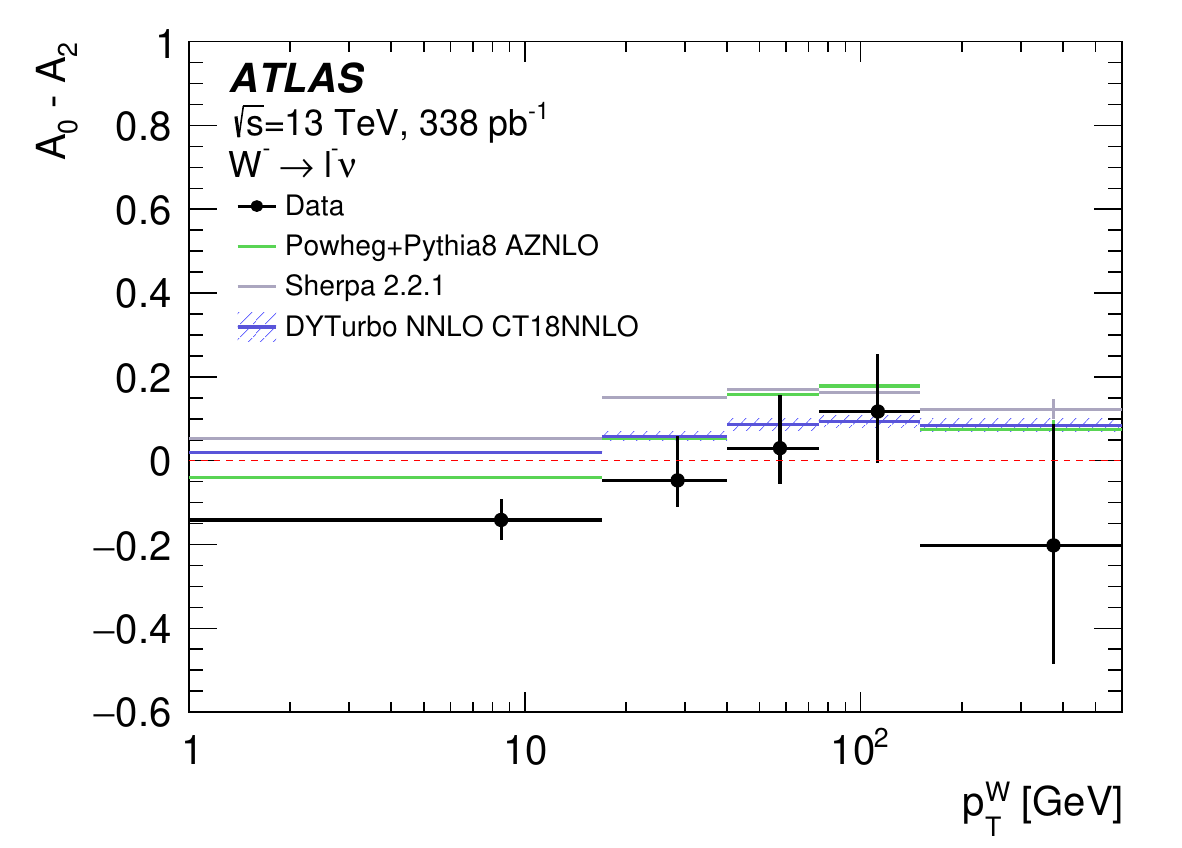}}
\subfloat[]{\includegraphics[width=\half]{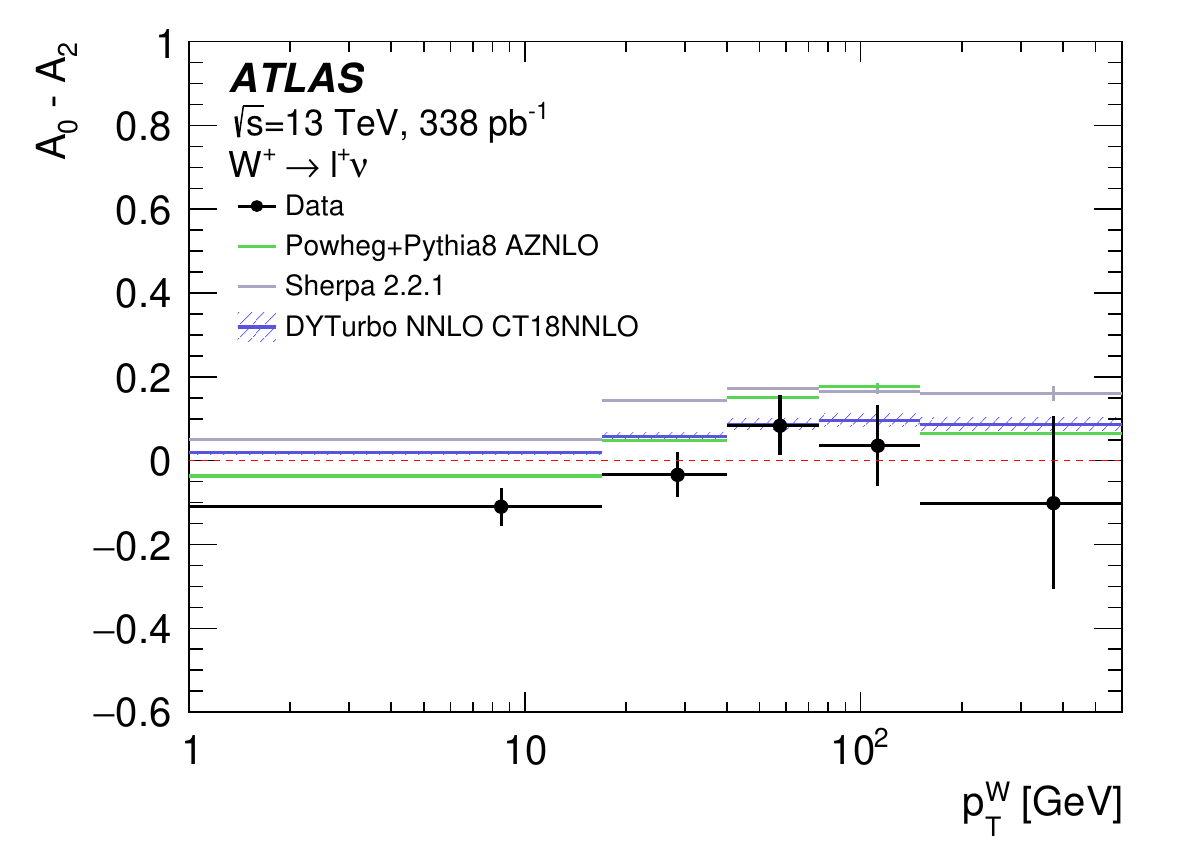}}
\caption{Comparison between the measured values of $\Azero - \Atwo$ as function of \ptw and predictions from the \Powheg\ and \Sherpa\ event generators, as well as from \textsc{DYTurbo} for (a) \Wminus and (b) \Wplus. The error bars for the \textsc{DYTurbo} predictions show the total theory uncertainty.   The data points are shown with their total uncertainties. As described in the text, a post-fit rebinning is applied in which two adjacent bins are averaged, effectively reducing the number of bins by a factor of two to mitigate strong bin-to-bin anticorrelations induced by statistical fluctuations.}
\label{fig:Ai_LTRelation_pt}
\end{figure}

\begin{figure}[ht]
\centering
\subfloat[]{  \includegraphics[width=0.5\linewidth]{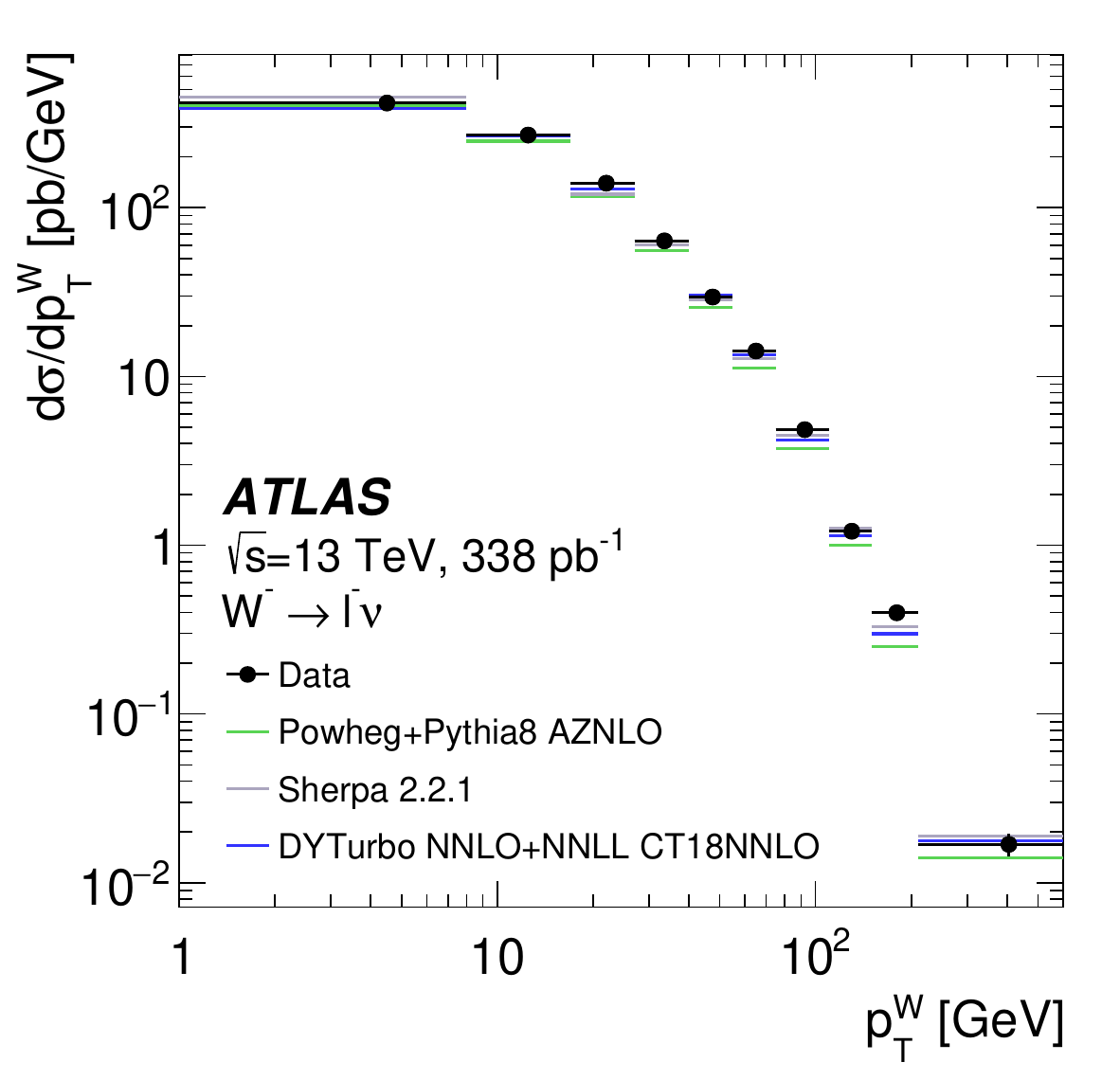}}
\subfloat[]{  \includegraphics[width=0.5\linewidth]{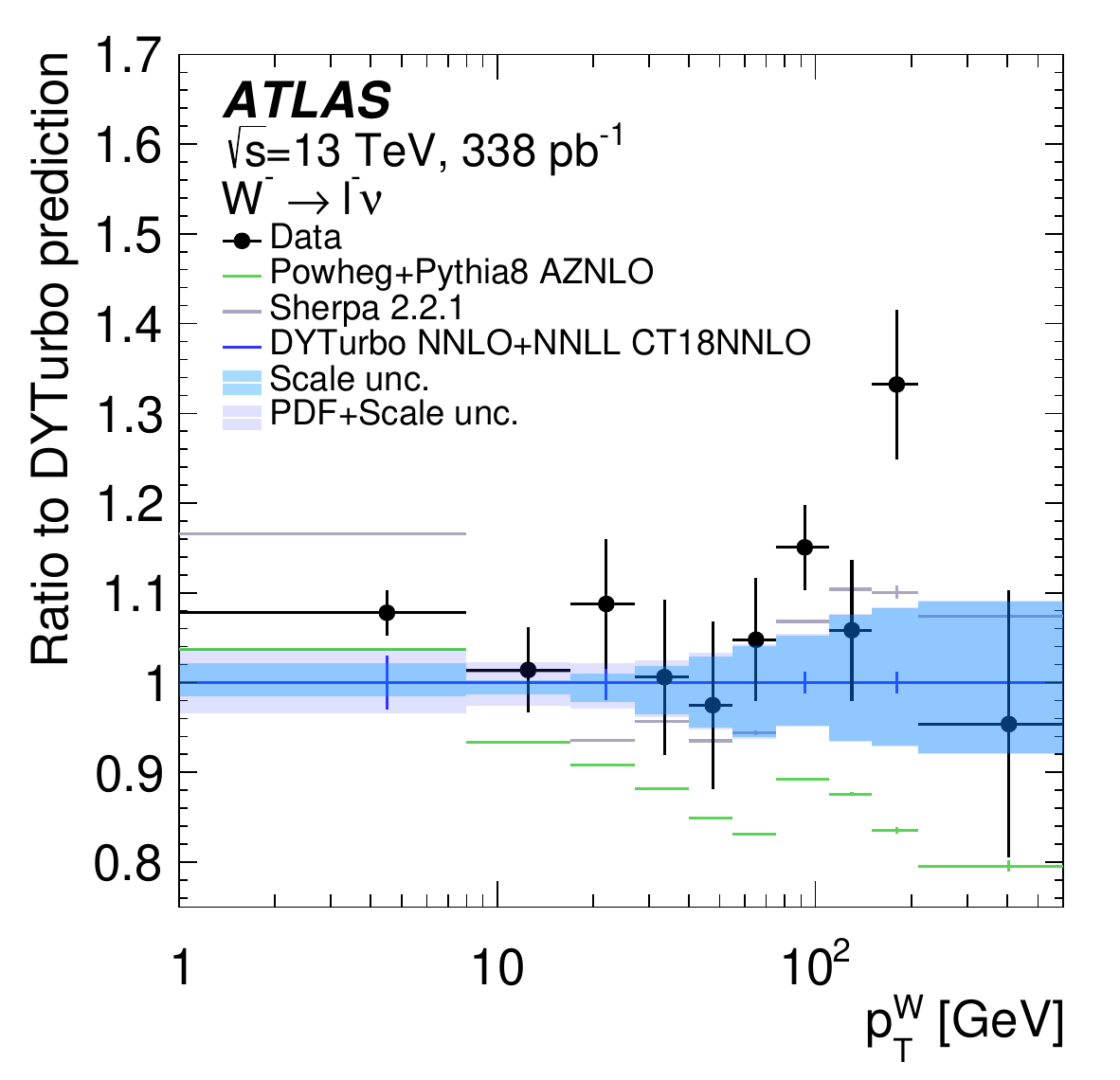}}

\subfloat[]{  \includegraphics[width=0.5\linewidth]{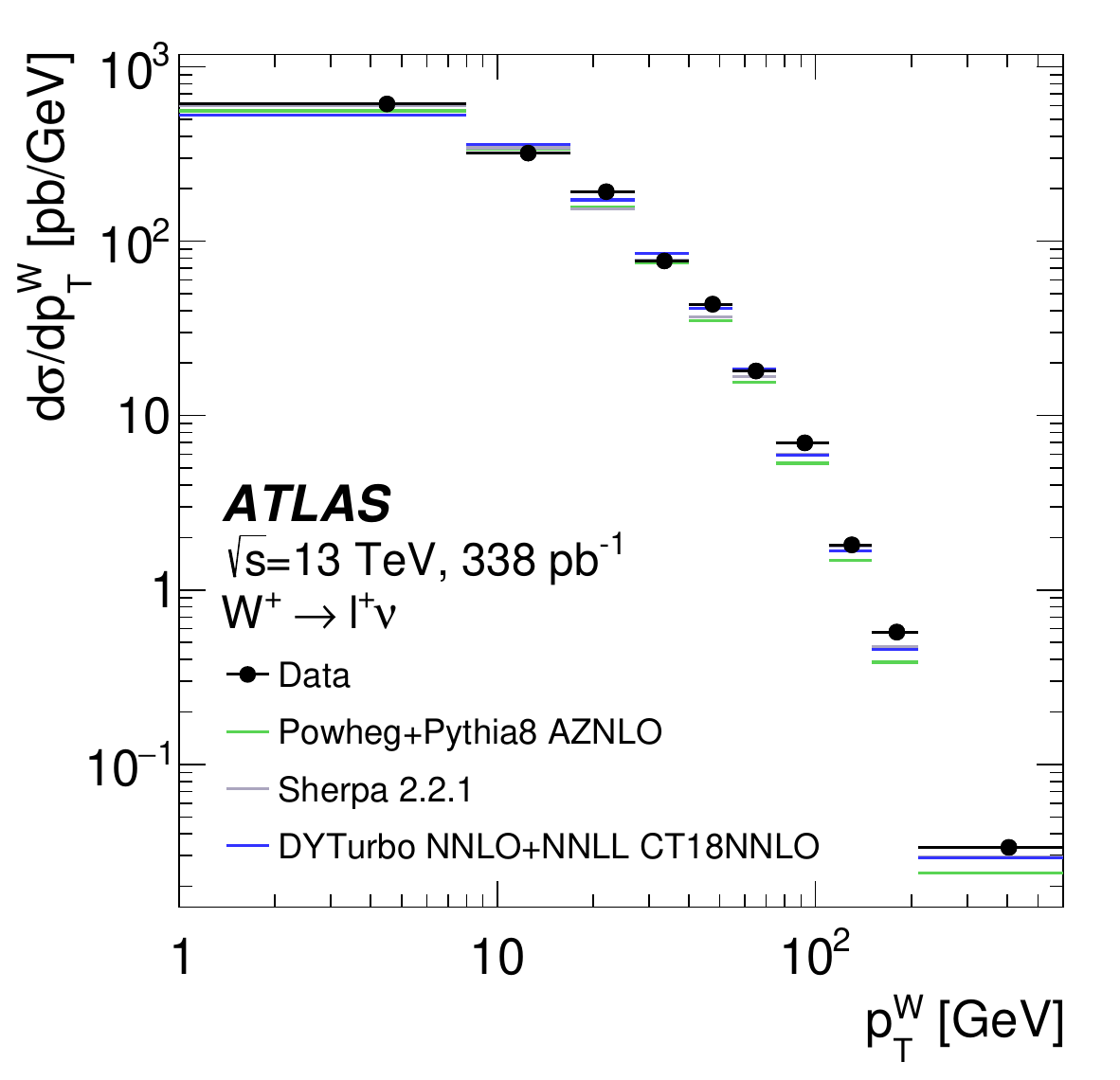} }
\subfloat[]{  \includegraphics[width=0.5\linewidth]{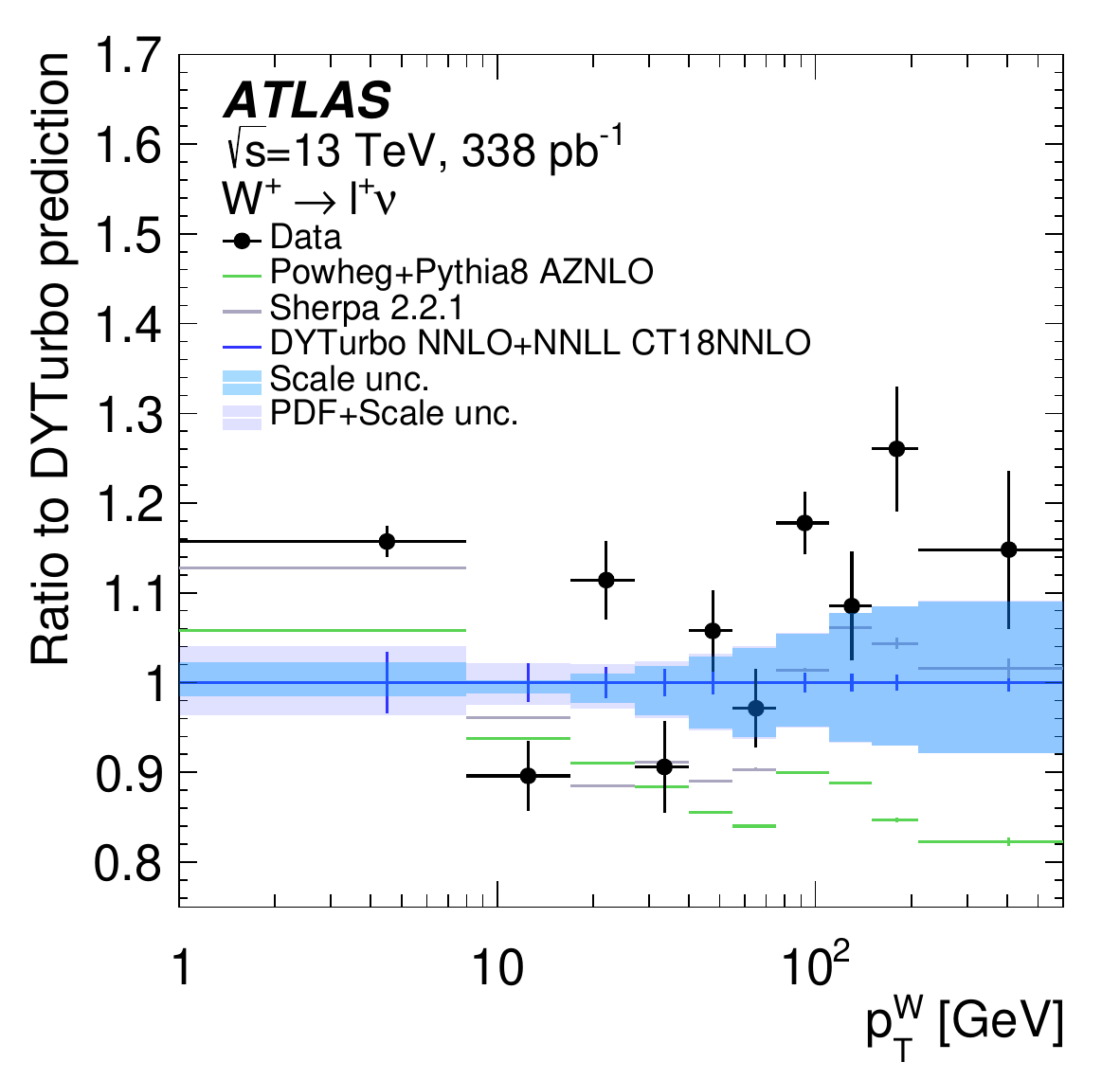} }
\caption{
Comparison of the measured absolute differential cross-sections \dsdpT\ for (a) \Wminus and (c) \Wplus bosons with predictions from \Powheg\ and \Sherpa\ event generators, as well as from \textsc{DYTurbo}. The quoted cross-sections include the branching ratio to a single lepton flavour.
Ratio of the measured \dsdpT\  and the MC predictions to \textsc{DYTurbo} for (b) \Wminus and (d) \Wplus.  Theory uncertainties in the  \textsc{DYTurbo}  predictions include variations of the renormalisation, factorisation, and resummation scales, as well as PDF uncertainties. The data points are shown with their total uncertainties. }
\label{fig:W_A9DY}
\end{figure}

This section presents the measurements of the full set of angular coefficients, \Arangesub, together with the differential cross-sections as functions of \ptw  in the full phase space of
the leptons from \Wboson-boson decays, shown separately for $W^-$ and $W^+$-boson production.

The results are first measured separately in the electron and muon channels with the methodology detailed in Section~\ref{sec:methodology}. The compatibility of the two channels is assessed with a $\chi^2$ test. The results are found to agree within their statistical uncertainties, with compatibility p-values of $0.92$ for the $W^-$ channel and $0.98$ for the  $W^+$ channel. The electron and muon channels are subsequently combined.

The measured angular coefficients, obtained from the combination of the electron and muon channels, are compared with theory predictions from the \textsc{Powheg}+\textsc{Pythia} and \textsc{Sherpa} MC generators, as well as to fixed-order predictions at $\mathcal{O}(\alpha_s^2)$ in perturbative QCD, calculated with the \textsc{DYTurbo} program.
Theory uncertainties in the \textsc{DYTurbo} predictions are evaluated by varying the nominal factorisation and renormalisation scales independently up and down by a factor of two, excluding variations in opposite directions. In this analysis the theory uncertainties associated with the choice of these scales are varied simultaneously in the numerator and denominator of the cross-section ratios used to define the angular coefficients. The resulting envelope of all variations is taken as the uncertainty. Alternative treatments may yield larger uncertainties as described in Ref.~\cite{gauld_precise_2017}.
PDF uncertainties for the CT18NNLO~\cite{Hou:2019efy}  set are evaluated using the standard eigenvector method and combined accordingly.

The results for the leading four angular coefficients, $A_0$ and $A_{2-4}$ are shown in Figures~\ref{fig:WM_pt} and~\ref{fig:WP_pt}.  These coefficients are clearly determined  and show good agreement with theory predictions. This analysis provides the first direct measurement of the  $A_0$ and $A_4$ coefficients as a function of \ptw in both $W^-$ and $W^+$ boson production. The precision achieved in the measurement of the $A_3$ coefficients exceeds that of any previous measurement.
The measurement is intrinsically more sensitive to coefficients that do not depend on $\cos\theta$, as they are less affected by the twofold sign ambiguity in the angular reconstruction of the neutrino momentum. Owing to the statistics-limited nature of the measurement, strong bin-to-bin anticorrelations are present, reaching about $80\%$ negative correlation in the low-\ptw region.

Although the data statistics are not enough to distinguish overall non-zero values for the smaller angular coefficients, \Asamll, these results are included for completeness. Despite their limited sensitivity, the extraction of these coefficients and of the Lam-Tung relation provides a useful consistency check and may inform future analyses with larger datasets.
To improve the readability of the results, a post-fit rebinning procedure is applied, reducing the number of measurement bins by a factor of two. This step is motivated by the presence of strong bin-to-bin anticorrelations observed during the extraction of the \Ai. The rebinning enhances the stability of the results and allows for a clearer interpretation of potential trends, even in regions dominated by statistical uncertainties.
The measured \Asamll coefficients are shown in Figures~\ref{AUXfig:WM_pt} and~\ref{AUXfig:WP_pt}.  The $A_7$ coefficient is expected to be larger in \Wboson-boson production compared with \Zboson-boson production due to the parity-violating nature of the weak interaction (see Refs.~\cite{Frederix:2014cba,Hagiwara:1984hi,Hagiwara:2006qe}). A slight deviation from zero is visible in the intermediate \pt range, though still limited by statistical uncertainties.
The Lam-Tung relation, $\Azero - \Atwo=0$, is studied using the combined-fit results, and shown in Figure~\ref{fig:Ai_LTRelation_pt} for $W^{-}$ and $W^{+}$ production. Uncertainties in the measurements  are derived from the full covariance matrix to account for correlations between $\Azero$ and $\Atwo$. The data measurements lack the precision needed to probe potential deviations from zero, which are expected from calculations including higher-order diagrams.

As described in Section~\ref{sec:methodology}, the profile likelihood ratio fit is used to simultaneously extract the angular coefficients and the differential cross-sections in each measurement bin.
The resulting \dsdpT within the full phase space of the decay leptons are shown in Figure~\ref{fig:W_A9DY}  for $W^-$ and $W^+$ channels,  together with the predicted values from \Powheg+\Pythia\ and \Sherpa~2.2.1 MC generators and the analytical predictions computed with the \textsc{DYTurbo} program.
The predictions are computed by performing the resummation of
logarithmically-enhanced contributions in the small transverse momentum region of the
lepton pairs at  NNLL accuracy
and also include the corresponding finite-order QCD contributions at
NNLO (corrections up to $\mathcal{O}(\alpha_s^2)$ ).
Non-perturbative contributions to the initial-state partons transverse momentum are included in \textsc{DYTurbo} as in Ref.~\cite{STDM-2016-10}.
Theory uncertainties
are estimated through up and down variations of the nominal factorisation,
renormalisation, and resummation scales by a factor of two, excluding
variations in opposite directions, and taking the envelope of all
variations. The PDF uncertainties for the CT18NNLO PDF sets
are estimated by calculating
the variations for all eigenvectors or replicas and combining
the results following the prescriptions.

The differential cross-sections are measured with a precision of $3\%$ at low  \ptw. Owing to the threefold ($\cos\theta$,$\phi$,\ptw) nature of the measurement and the statistical nature of its limitations, strong bin-to-bin anticorrelations of up to about $70\%$ are observed.
The \textsc{DYTurbo} resummed predictions show the best agreement across the full
spectrum.
The theory uncertainties at very low \ptw are dominated by the PDF uncertainty but at higher \ptw are dominated by the scale variations.
The MC predictions describe
the low-\ptw region reasonably well. At higher transverse momenta,  \Sherpa~2.2.1 provides an overall good description, whereas \Powheg+\Pythia\ shows worse agreement for $\ptw > 40\,\GeV$, as expected.
Overall, the $W^+$ measurements show the largest
deviation from the predictions.

\FloatBarrier


%

\section{Conclusions}

This paper presents for the first time the measurement of the full set of angular coefficient for the \Wboson-boson and a differential measurement in \ptw of absolute cross-sections within the full phase space of the decay leptons.
Such a measurement is achieved by exploiting the methodology already developed and published for the extraction of the $Z$-boson angular coefficients. The  twofold ambiguity in the reconstruction of the neutrino momenta in the CS frame  and the corresponding ambiguity on the sign of \coscs is solved on a statistical basis by imposing a simple \mw constraint.
The measurement is performed using the data samples collected in dedicated low pile-up runs with the ATLAS detector at the LHC, corresponding to an integrated luminosity of 338\,pb$^{-1}$ at $\sqrt{s} =13$\,TeV.
These low pile-up datasets offer a much improved resolution of the hadronic recoil compared with that from the nominal high pile-up Run~2 LHC proton--proton data, enabling the success of the challenging task of measuring neutrino-related observables.
A profile likelihood fit extracts the eight angular coefficients and the corresponding unpolarised cross-section simultaneously as parameters of interest in each measurement bin in \ptw~space. The uncertainties in these measurements are mostly statistical in nature and the experimental and theory systematic uncertainties are at the percent level over most of the range.

The angular coefficients
\Asubrange   measured for both $W^-$ and $W^+$ bosons, are found to be significantly different from zero. These measurements are compared with \textsc{DYTurbo} predictions computed at  $\mathcal{O}(\alpha_s^2)$  (NNLO). The  \dsdpT  distributions, measured within the full phase space of the decay leptons, are also compared with \textsc{DYTurbo} predictions at NNLO, including transverse momentum resummation at NNLL accuracy in QCD and using various PDF sets. Overall, good agreement between the measurements and the theory predictions is observed.


%

%
%
%
%
%
\section*{Acknowledgements}
%
%

%
%

%
%

We thank CERN for the very successful operation of the LHC and its injectors, as well as the support staff at
CERN and at our institutions worldwide without whom ATLAS could not be operated efficiently.

The crucial computing support from all WLCG partners is acknowledged gratefully, in particular from CERN, the ATLAS Tier-1 facilities at TRIUMF/SFU (Canada), NDGF (Denmark, Norway, Sweden), CC-IN2P3 (France), KIT/GridKA (Germany), INFN-CNAF (Italy), NL-T1 (Netherlands), PIC (Spain), RAL (UK) and BNL (USA), the Tier-2 facilities worldwide and large non-WLCG resource providers. Major contributors of computing resources are listed in Ref.~\cite{ATL-SOFT-PUB-2025-001}.

We gratefully acknowledge the support of ANPCyT, Argentina; YerPhI, Armenia; ARC, Australia; BMWFW and FWF, Austria; ANAS, Azerbaijan; CNPq and FAPESP, Brazil; NSERC, NRC and CFI, Canada; CERN; ANID, Chile; CAS, MOST and NSFC, China; Minciencias, Colombia; MEYS CR, Czech Republic; DNRF and DNSRC, Denmark; IN2P3-CNRS and CEA-DRF/IRFU, France; SRNSFG, Georgia; BMFTR, HGF and MPG, Germany; GSRI, Greece; RGC and Hong Kong SAR, China; ICHEP and Academy of Sciences and Humanities, Israel; INFN, Italy; MEXT and JSPS, Japan; CNRST, Morocco; NWO, Netherlands; RCN, Norway; MNiSW, Poland; FCT, Portugal; MNE/IFA, Romania; MSTDI, Serbia; MSSR, Slovakia; ARIS and MVZI, Slovenia; DSI/NRF, South Africa; MICIU/AEI, Spain; SRC and Wallenberg Foundation, Sweden; SERI, SNSF and Cantons of Bern and Geneva, Switzerland; NSTC, Taipei; TENMAK, T\"urkiye; STFC/UKRI, United Kingdom; DOE and NSF, United States of America.

Individual groups and members have received support from BCKDF, CANARIE, CRC and DRAC, Canada; CERN-CZ, FORTE and PRIMUS, Czech Republic; COST, ERC, ERDF, Horizon 2020, ICSC-NextGenerationEU and Marie Sk{\l}odowska-Curie Actions, European Union; Investissements d'Avenir Labex, Investissements d'Avenir Idex and ANR, France; DFG and AvH Foundation, Germany; Herakleitos, Thales and Aristeia programmes co-financed by EU-ESF and the Greek NSRF, Greece; BSF-NSF and MINERVA, Israel; NCN and NAWA, Poland; La Caixa Banking Foundation, CERCA Programme Generalitat de Catalunya and PROMETEO and GenT Programmes Generalitat Valenciana, Spain; G\"{o}ran Gustafssons Stiftelse, Sweden; The Royal Society and Leverhulme Trust, United Kingdom.

In addition, individual members wish to acknowledge support from CERN: European Organization for Nuclear Research (CERN DOCT); Chile: Agencia Nacional de Investigaci\'on y Desarrollo (FONDECYT 1230812, FONDECYT 1240864, Fondecyt 3240661, Fondecyt Regular 1240721); China: Chinese Ministry of Science and Technology (MOST-2023YFA1605700, MOST-2023YFA1609300), National Natural Science Foundation of China (NSFC - 12175119, NSFC 12275265); Czech Republic: Czech Science Foundation (GACR - 24-11373S), Ministry of Education Youth and Sports (ERC-CZ-LL2327, FORTE CZ.02.01.01/00/22\_008/0004632), PRIMUS Research Programme (PRIMUS/21/SCI/017); EU: H2020 European Research Council (ERC - 101002463); European Union: European Research Council (BARD No. 101116429, ERC - 948254, ERC 101089007), European Regional Development Fund (HE COFUND GA No.101081355, ERDF), European Union, Future Artificial Intelligence Research (FAIR-NextGenerationEU PE00000013), Italian Center for High Performance Computing, Big Data and Quantum Computing (ICSC, NextGenerationEU); France: Agence Nationale de la Recherche (ANR-21-CE31-0013, ANR-21-CE31-0022, ANR-22-EDIR-0002, ANR-24-CE31-0504-01); Germany: Baden-Württemberg Stiftung (BW Stiftung-Postdoc Eliteprogramme), Deutsche Forschungsgemeinschaft (DFG - 469666862, DFG - CR 312/5-2); China: Research Grants Council (GRF); Italy: Istituto Nazionale di Fisica Nucleare (ICSC, NextGenerationEU), Ministero dell'Università e della Ricerca (NextGenEU 153D23001490006 M4C2.1.1, NextGenEU I53D23000820006 M4C2.1.1, NextGenEU I53D23001490006 M4C2.1.1, SOE2024\_0000023); Japan: Japan Society for the Promotion of Science (JSPS KAKENHI JP22H01227, JSPS KAKENHI JP22H04944, JSPS KAKENHI JP22KK0227, JSPS KAKENHI JP24K23939, JSPS KAKENHI JP24KK0251, JSPS KAKENHI JP25H00650, JSPS KAKENHI JP25H01291, JSPS KAKENHI JP25K01023); Norway: Research Council of Norway (RCN-314472); Poland: Ministry of Science and Higher Education (IDUB AGH, POB8, D4 no 9722), Polish National Science Centre (NCN 2021/42/E/ST2/00350, NCN OPUS 2023/51/B/ST2/02507, NCN UMO-2019/34/E/ST2/00393, UMO-2022/47/O/ST2/00148, UMO-2023/49/B/ST2/04085, UMO-2023/51/B/ST2/00920, UMO-2024/53/N/ST2/00869); Portugal: Foundation for Science and Technology (FCT); Spain: Generalitat Valenciana (ASFAE/2022/008), Ministry of Science and Innovation (MCIN \& NextGenEU PCI2022-135018-2, MICIN \& FEDER PID2021-125273NB, RYC2019-028510-I, RYC2020-030254-I, RYC2021-031273-I, RYC2022-038164-I), Ministerio de Ciencia, Innovación y Universidades/Agencia Estatal de Investigaci\'on (PID2022-142604OB-C22); Sweden: Carl Trygger Foundation (Carl Trygger Foundation CTS 22:2312), Swedish Research Council (Swedish Research Council 2023-04654, VR 2021-03651, VR 2022-03845, VR 2022-04683, VR 2023-03403, VR 2024-05451), Knut and Alice Wallenberg Foundation (KAW 2018.0458, KAW 2022.0358, KAW 2023.0366); Switzerland: Swiss National Science Foundation (SNSF - PCEFP2\_194658); United Kingdom: Royal Society (NIF-R1-231091); United States of America: U.S. Department of Energy (ECA DE-AC02-76SF00515), Neubauer Family Foundation.

%
%


%
%
%
%
%
\printbibliography
\clearpage
 
\begin{flushleft}
\hypersetup{urlcolor=black}
{\Large The ATLAS Collaboration}

\bigskip

\AtlasOrcid[0000-0002-6665-4934]{G.~Aad}$^\textrm{\scriptsize 104}$,
\AtlasOrcid[0000-0001-7616-1554]{E.~Aakvaag}$^\textrm{\scriptsize 17}$,
\AtlasOrcid[0000-0002-5888-2734]{B.~Abbott}$^\textrm{\scriptsize 123}$,
\AtlasOrcid[0000-0002-0287-5869]{S.~Abdelhameed}$^\textrm{\scriptsize 119a}$,
\AtlasOrcid[0000-0002-1002-1652]{K.~Abeling}$^\textrm{\scriptsize 55}$,
\AtlasOrcid[0000-0001-5763-2760]{N.J.~Abicht}$^\textrm{\scriptsize 49}$,
\AtlasOrcid[0000-0002-8496-9294]{S.H.~Abidi}$^\textrm{\scriptsize 30}$,
\AtlasOrcid[0009-0003-6578-220X]{M.~Aboelela}$^\textrm{\scriptsize 45}$,
\AtlasOrcid[0000-0002-9987-2292]{A.~Aboulhorma}$^\textrm{\scriptsize 36e}$,
\AtlasOrcid[0000-0001-5329-6640]{H.~Abramowicz}$^\textrm{\scriptsize 157}$
% ...
\bigskip
\\

% ...
$^{17}$Department for Physics and Technology, University of Bergen, Bergen; Norway.\\
% ...
$^{30}$Physics Department, Brookhaven National Laboratory, Upton NY; United States of America.\\
% ...
$^{36}$$^{(a)}$Facult\'e des Sciences Ain Chock, Universit\'e Hassan II de Casablanca;$^{(b)}$Facult\'{e} des Sciences, Universit\'{e} Ibn-Tofail, K\'{e}nitra;$^{(c)}$Facult\'e des Sciences Semlalia, Universit\'e Cadi Ayyad, LPHEA-Marrakech;$^{(d)}$LPMR, Facult\'e des Sciences, Universit\'e Mohamed Premier, Oujda;$^{(e)}$Facult\'e des sciences, Universit\'e Mohammed V, Rabat;$^{(f)}$Institute of Applied Physics, Mohammed VI Polytechnic University, Ben Guerir; Morocco.\\
% ...
$^{45}$Physics Department, Southern Methodist University, Dallas TX; United States of America.\\
% ...
$^{49}$Fakult\"{a}t Physik, Technische Universit{\"a}t Dortmund, Dortmund; Germany.\\
% ...
$^{55}$II. Physikalisches Institut, Georg-August-Universit\"{a}t G\"ottingen, G\"ottingen; Germany.\\
% ...
$^{104}$CPPM, Aix-Marseille Universit\'e, CNRS/IN2P3, Marseille; France.\\
% ...
$^{119}$$^{(a)}$New York University Abu Dhabi, Abu Dhabi;$^{(b)}$United Arab Emirates University, Al Ain; United Arab Emirates.\\
% ...
$^{123}$Homer L. Dodge Department of Physics and Astronomy, University of Oklahoma, Norman OK; United States of America.\\
% ...
$^{157}$Raymond and Beverly Sackler School of Physics and Astronomy, Tel Aviv University, Tel Aviv; Israel.\\
% ...
$^{a}$ Also at Affiliated with an institute formerly covered by a cooperation agreement with CERN.\\
$^{b}$ Also at An-Najah National University, Nablus; Palestine.\\
$^{c}$ Also at Borough of Manhattan Community College, City University of New York, New York NY; United States of America.\\
$^{d}$ Also at Center for Interdisciplinary Research and Innovation (CIRI-AUTH), Thessaloniki; Greece.\\
$^{e}$ Also at Centre of Physics of the Universities of Minho and Porto (CF-UM-UP); Portugal.\\
$^{f}$ Also at CERN, Geneva; Switzerland.\\
% ...

\end{flushleft}


%

%
%

%

%
%
%
%

\end{document}